 \documentclass[final,authoryear,3p,times]{elsarticle}

\usepackage[normalem]{ulem}

\usepackage{amsmath}
\usepackage{amssymb}
\usepackage{amsthm}
\usepackage{bm}
\usepackage{graphicx}		
\graphicspath{{./figures/}}		

\usepackage{tikz}
\usetikzlibrary{shapes.geometric,arrows,decorations.pathmorphing,backgrounds,positioning,fit,calc,3d,er,calc,math}
\tikzstyle{process} = [rectangle, minimum width=3cm, minimum height=1cm, text width=13 cm, text centered, draw=black] 
\tikzstyle{arrow} = [thick,->,>=stealth]
\usepackage{overpic}

\usepackage{float}
\usepackage{caption}
\usepackage{subcaption}
\usepackage[edges]{forest}
\usepackage{tikz-qtree}
\usepackage{adjustbox}
\usepackage{footmisc}
\usepackage{color,soul}
\usepackage[framemethod=tikz]{mdframed}
\usepackage{lipsum}

\journal{}

\usepackage[]{hyperref}
\hypersetup{
    colorlinks,%
    citecolor=blue,%
    filecolor=black,%
    linkcolor=magenta,%
    urlcolor=black
}

\makeatletter
\newcommand*{\rom}[1]{\expandafter\ \romannumeral #1}
\makeatother
\newcommand\misfitSnapWidth{0.22}
\newcommand\misfitSnapWidthInside{0.9}

\usepackage{nicefrac}
\begin{document}

\begin{frontmatter}



\title{A discrete dislocation dynamics study of precipitate bypass mechanisms in nickel-based superalloys}


\author[miami]{Sabyasachi Chatterjee}
\author[mae]{Yang Li}
\author[miami]{Giacomo Po}
\ead{gpo@miami.edu}
\address[miami]{Department of Mechanical and Aerospace Engineering, University Miami, Coral Gables, FL 33146}
\address[mae]{Department of Mechanical and Aerospace Engineering, University of California Los Angeles, Los Angeles, CA 90095}

\address{}

\begin{abstract}
Order strengthening in nickel-based superalloys is associated with the extra stress required for dislocations to bypass the $\gamma'$ precipitates distributed in the $\gamma$ matrix. Depending on the operating conditions and microstructure, a rich variety of bypass mechanism has been identified, with various shearing and Orowan looping processes gradually giving way to  climb bypass as the operating conditions change from the low/intermediate temperatures and high stress regime, to the high temperature and low stress regime. When  anti phase boundary (APB) shearing and Orowan looping mechanisms operate, the classical picture  is that, at for a given volume fraction, the  bypass mechanism changes  from shearing to looping with increased particle size and within  a broad coexistence size window. Another possibility, which is supported by indirect experimental evidence, is that a third ``hybrid" transition mechanism may operate. In this paper we  use discrete dislocation dynamics (DDD) simulations to study dislocation bypass mechanisms in Ni-based superalloys. We develop a new method to compute generalized stacking fault forces in DDD simulations, based on a concept borrowed from complex analysis and known as the winding number of a closed curve about a point. We use this method to study the mechanisms of bypass of  a square lattice of spherical $\gamma'$ precipitates  by $a/2\langle110\rangle\{111\}$ edge dislocations, as a function of the precipitates volume fraction and size. We show that not only the hybrid mechanism is possible, but also that it is operates as a transition mechanism between the shearing and looping regimes over a wide range of precipitates volume fraction and radii. Based on our simulation results, we propose a simple model for the strength of this mechanism. We also consider the effects of a $\gamma/\gamma'$ lattice misfit on the bypass mechanisms, which we approximate by an additional precipitate stress computed according to Eshelby's inclusion theory. We show that in the shearing and hybrid looping-shearing regimes, a lattice misfit generally results in an increased bypass stress. For sufficiently high lattice misfit, the critical bypass configuration in attractive dislocation-precipitates interactions  changes dramatically, and the bypass stress is controlled by the pinning of the trailing dislocation on the exit side of the precipitates, similar to what has been reported in the high-temperature creep literature.
\end{abstract}


\end{frontmatter}


\section{Introduction\label{introduction}}

Since their introduction in the 1950s, nickel (Ni)-based superalloys have emerged as the dominant class of structural materials in aerospace applications requiring mechanical and chemical resistance at high temperatures \citep{sims1984history,schafrik2008superalloy,pollock2006nickel}. Despite continuous progress in competing classes of materials \citep{zhao2003ultrahigh}, Ni-based superalloys remain the reference class of materials for these applications. 
Although gas turbine engines have historically represented the archetypal target of Ni-based superalloys,  other industries have benefited from their usage, including the automotive, the gas and oil exploration, the metal processing, and the nuclear industries \citep{matsuo1987strengthening,sommitsch2012microstructure}. Ni-based superalloys also find applications as materials for boilers, heat exchangers and tubomachinery in advanced ultra-supercritical (A-USC) fossil- and nuclear-based power generation \citep{rakowski2006nickel,stein2013nickel,gandy2014advances}.  

Ni-based superalloys owe their high-temperature mechanical properties to several factors \citep{reed2008superalloys}. First, the face-centered cubic (fcc) structure of the Ni matrix guarantees that the alloys are both tough and ductile over the entire operative temperature window without suffering from a ductile to brittle transition. Second, among the fcc metals, Ni possesses relatively high activation energy of self-diffusion, thus endowing an intrinsic resistance to diffusional creep. Third, the presence of secondary phases, such as the coherent, ordered, intermetallic ($\text{Ni}_3\text{Al}$) $\gamma'$ precipitates, and carbides/borides typically residing on grain boundaries provide the high-temperature creep resistance by hindering dislocation motion and grain boundary sliding. Fourth, a rich composition of alloy elements is tailored to stabilize and enhance the various phases of the alloy and further improve the resistance to diffusional creep.

Order strengthening by $\gamma'$ precipitation is a consequence of the hindering effect that precipitates have on dislocation motion. The additional stress required for dislocations to bypass  precipitates depends on the  type of the underlying escape mechanism. Four types of  mechanisms are possible. Two of them are essentially athermal, namely particle shearing and   Orowan looping, while two mechanisms are mediated by mass diffusion, namely dislocation climb around particles, and to a lesser extent particle dragging  \citep{mclean1985threshold}.  Which specific mechanism is dominant depends on both microstructural parameters (e.g. alloy composition, lattice misfit,  precipitates size, morphology, and volume fraction) and external conditions (e.g. applied stress and operating temperature). 
At low and intermediate temperatures mass diffusion is slow, and therefore the athermal mechanisms of particle shearing and looping by dislocations are dominant.   Together with the grain size, the precipitate size  distribution and volume fraction are essential microstructural parameters influencing both monotonic and cyclic deformation of the superalloys \citep{ghonem1993elevated,alexandre2004modelling,pineau2009high,gv2012influence,antolovich2015microstructural,graverend_fatigue}. 
Of particular importance is the role of the athermal mechanisms  in controlling different regimes of creep deformation. For example,  in the low temperature creep regime, approximately defined as $T/T_m<0.6$  \citep{pollock2002dislocations}, several single crystal superalloys with high precipitates volume fraction exhibit primary creep mediated by shearing of the $\gamma'$ precipitates by paired $a/2\langle110\rangle\{111\}$ dislocations and $a\langle112\rangle\{111\}$ dislocations ribbons \citep{kear1974deformation,matan1999creep}. A transition to secondary creep eventually takes place as  $a/2\langle110\rangle\{111\}$ dislocations percolate through the $\gamma$ channels and form a geometrically-necessary and non statically-recoverable network that reduces the misfit energy and simultaneously inhibit further $a\langle112\rangle\{111\}$ slip \citep{rae2000primary}. At intermediate temperature, and correspondingly lower stresses, shearing becomes more difficult, and dislocations are typically confined to glide in the $\gamma$ phase with local climb rearrangements, although precipitates shearing by paired $a/2\langle110\rangle\{111\}$ dislocations is observed in the later stages of secondary creep \citep{pollock1992creep,pollock2002dislocations}. In low volume fraction polycrystalline superalloys, proper looping around individual particles becomes possible, and the competition between shearing and looping is more evident \citep[e.g.][]{reppich1982some2,lerch1984study}, especially when a bimodal size distribution of secondary and tertiary $\gamma'$ precipitates is present \citep{viswanathan2005investigation}. For these alloys, a rich map of deformation mechanism emerges, with various shearing processes gradually giving way to  climb bypass as the operating conditions change from the low/intermediate temperatures and higher stress regime to the high temperature and low stress regime \citep{smith2016determination,smith2016creep}. The question of which is the dominant mechanism in a given creep regime emerged early in relation to an anomaly observed in the creep behavior of precipitation and dispersion hardened materials: standard power-law creep predicts abnormally high stress-dependence and activation energy of their secondary creep rate.  This anomaly was eventually reconciled with theory by introducing the concept of a \textit{threshold stress} in the  power-law creep equation \citep{lund1976high}, although different explanations of the origin of such term have been proposed \citep{mclean1985threshold,arzt1986threshold,rosler1990new,mishra1994threshold,arzt2001interface}.

Order strengthening with $\gamma'$ precipitates bypassed  by pairs of $a/2\langle110\rangle\{111\}$ dislocations is particularly relevant for superalloys operating in the low/intermediate temperature regime. The classical picture that emerges  from the vast literature dedicated to this subject \citep[e.g.][]{reppich1982some1,ardell1985precipitation,martin2012precipitation} is that the mechanism of bypass depends on  microstructural parameters such as precipitates volume fraction  and average  size. For fixed (and not too high) volume fraction and increasing particle size, it is well-established that the  bypass mechanism transitions from  from \textit{weakly-coupled} sharing, through \textit{strongly-coupled} shearing, to Orowan looping. Understanding how these mechanisms affect  strength has been instrumental in the design of superalloys. For example, several commercial alloys have a microstructure corresponding to the transition between the weak and strong shearing mechanisms, which is the condition of ``critical dispersion" yielding the maximum strength for a given volume fraction \citep{reppich1982some1}. The shearing to looping transition is arguably less understood. \cite{reppich1982some2} point out that the transition takes place over a wide range of particle diameters, with both processes being distinct and observable side by side, probably due to the local variability of the precipitates microstructure. \cite{nabarro1995physics}, however, suggest another possibility, where the transition may also take place by means of a \textit{hybrid looping-shearing process}: under sufficiently high stress, the leading dislocation in the pair may form a loop which is stable around the precipitate until the approaching trailing dislocation drives the loop into the precipitate, forming  APB, and then quickly shears it removing the APB. It should be noted that, this hybrid process is difficult to observe experimentally, because the shearing event is unstable. However, the hybrid process leaves a  signature characterized by a planar slip deformation mode, and by the presence of single loops around  precipitates and lack of hardening because the ``latent" shearing events prevent accumulation of additional loops. Such  signature was in fact recognized in secondary $\gamma'$ precipitates ($f=0.32$ and $d=50-100$nm) in superalloy Ren\'e 88 DT under $972$MPa stress at $650^\circ$C by \cite{viswanathan2005investigation} and  \cite{unocic2008mechanisms}.

Several modeling techniques have been developed to predict the high-temperature strength and creep resistance of precipitation-hardened alloys (\cite{kim_superalloy, kim_creep,graverend_highT,graverend_misfit, collins_superalloy}).
Discrete Dislocation Dynamics (DDD) simulations have been employed extensively in the literature to shed light on the mechanisms of interaction between $a/2\langle110\rangle\{111\}$ dislocations and $\gamma'$ precipitates \citep{mohles_2004,rao_ddd,yashiro_ddd,vattre_DDD,queyreau_Orowan,huang_misfit,climb_effect_raabe,li_wang_ppt,hussein_uchic_2017,ddd_cyclic_lin,gao_sizeEffect, ngan_ddd}. However, the hybrid mechanism described above was not reported in these studies, despite its strong, albeit indirect, experimental evidence. This fact may partially be attributed to the fact that  APB forces on dislocations have so far been introduced ad-hoc in DDD simulations, possibly preventing the observation of this more complex hybrid bypass mode. The objective of this paper is to investigate the transition between the shearing and looping processes of $\gamma'$ precipitates by $a/2\langle110\rangle\{111\}$ dislocation pairs, and understand if and under which conditions the hybrid looping-shearing mechanisms takes place.  We begin by reviewing the classical theory of athermal precipitation strengthening in section \ref{theory}. In section \ref{framework} we introduce a new method to compute generalized stacking fault forces in DDD simulations, based on a concept borrowed from complex analysis and known as the winding number of a closed curve about a point. We use this method to study the mechanisms of bypass  of spherical $\gamma'$ precipitates arranged in a square lattice by $a/2\langle110\rangle\{111\}$ edge dislocations, as a function of the precipitates volume fraction and average radius. We show that not only the hybrid looping-shearing mechanism is possible, but also that it operates as the transition mechanism between the shearing and looping regimes over a wide range of precipitates volume fraction and radii. Based on our simulation results, we propose a simple model for the strength of this mechanism. In the same section, we consider the effects of a $\gamma/\gamma'$ lattice misfit on the bypass mechanisms. The lattice misfit is approximated by an additional precipitate stress computed according to Eshelby's inclusion theory. We show that in the shearing and hybrid looping-shearing regimes, a lattice misfit generally results in an increased bypass stress. For sufficiently high lattice misfit, the critical bypass configuration in attractive dislocation-precipitates interactions  changes dramatically, and the bypass stress is controlled by the pinning of the trailing dislocation on the \textit{exit} side of the precipitates, similar to what has been reported in the high-temperature creep literature \citep[see][]{arzt2001interface}. The summary in section \ref{summary} concludes our work.

\section{Theory of athermal precipitation strengthening\label{theory}}

The theory of  precipitation strengthening is vast, and detailed reviews can be found  elsewhere \citep{Kelly1963,BrownHam1971,reppich1982some1,ardell1985precipitation,martin2012precipitation}. The objective of this section is to briefly recall selected concepts that serve as the theoretical framework to interpret our  simulation results. 

\subsection{Particle shearing}
\begin{figure}[t!]
    \centering
    \begin{subfigure}{.48\textwidth}
  \centering
  \begin{tikzpicture}
  \def\R{1.5cm}
  \def\h{0.619*\R}
  \def\r{0.7854*\R}
    \shade[ball color = black!20!white, opacity = 1] (0,0) circle (\R);
    \draw[dashed] (-2*\R,\h) to  node[below,midway,sloped]{} (2*\R,\h);
        \draw[<->,thick] (-\r,\h) to  node[above,midway,sloped]{$2r$} (\r,\h);
    \draw[->,thick] (0,0) to  node[below,midway,sloped]{$R$}  (\r,\h);
    \draw[thick] (-2*\R,\h+4) -- (-2*\R+8,\h+4);
        \draw[thick] (-2*\R+4,\h+4) -- (-2*\R+4,\h+10);
  \end{tikzpicture}
  \caption{}
  \label{fig:Rr}
\end{subfigure}%
    \begin{subfigure}{.48\textwidth}
  \centering
  \begin{tikzpicture}
  \def\r{0.5cm}
  \def\L{2cm}
    \fill[color = black!20!white, opacity = 1] (0,0) circle (\r);
        \fill[color = black!20!white, opacity = 1] (0,\L) circle (\r);
            \fill[color = black!20!white, opacity = 1] (\L,0) circle (\r);
                \fill[color = black!20!white, opacity = 1] (\L,\L) circle (\r);
\draw[<->] (0,0) to node[left,midway]{$L$} (0,\L);
\draw[<->] (\r,0) to node[below,midway]{$L'$} (\L-\r,0);
\draw[->] (\L,0) to node[above,midway]{$r$} (\L+\r,0);

  \end{tikzpicture}
  \caption{}
  \label{fig:LL}
\end{subfigure}
    \caption{Elements of particle statistics, for spherical precipitates  \citep{ardell1985precipitation,martin2012precipitation} (\subref{fig:Rr}) The average particle radius is $R$, and the average radius  intersected by a random slip plane is $r=\pi R/4$. The  volume fraction  is defined as $f=\nicefrac{4}{3}\pi R^3 N_v$, where $N_v$ is the number of spherical particles  per unit volume. Since the number of particles per unit  area intersected by a plane  is $N_s=2RN_v$, then $f=\nicefrac{2\pi}{3}R^2N_s$. If $r_i$'s are the radii of the particles on this plane,  the  average $\langle r_i^2\rangle =\nicefrac{2}{3}R^2$ allows  to conveniently measure the volume fraction as the area fraction  $f=\pi \langle r_i^2\rangle N_s$ on any planar section of the material. In lieu of the identity $\langle r_i^2\rangle=32 \langle r_i\rangle^2/(3\pi^2)$,  the volume fraction can also be measured as $f=\nicefrac{32}{3\pi} r^2 N_s$, where we have let $r=\langle r_i\rangle$. From the fact that $\langle r_i\rangle=\pi R/4$ we also have $f=\nicefrac{2\pi}{3}R^2 N_s$. We can also measure $f$ from the number of particles per unit length $N_L=\ell^{-1}$ intersected on a random line in the material. Here  the ``mean free path" $\ell$ satisfies the  condition that the volume ``dragged" by a particle contains only one other particle, that is $\ell \pi R^2 N_v=1$. This yields $f=\nicefrac{4R}{3\ell}$, or $f=\nicefrac{16r}{3\pi\ell}$.
     (\subref{fig:LL}) The square lattice particle spacing $L$ on the slip plane, and the inter-particle spacing $L'=L-2r$.}
         \label{fig:RrL}
\end{figure}
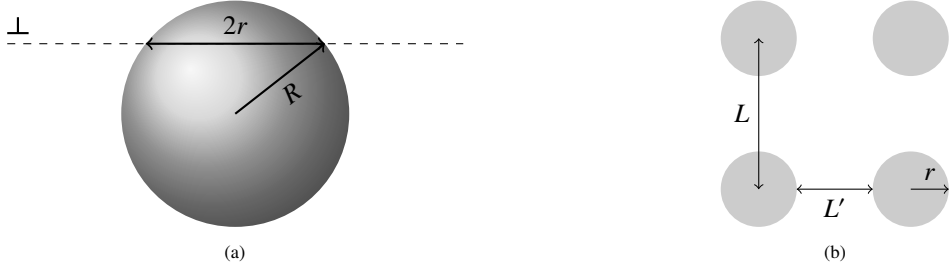

Consider a dislocation pinned by an array of point-particles with \textit{effective} spacing $\lambda$, and  subject to an applied shear stress $\tau$. The breakaway condition is satisfied when the  stress reaches a critical value $\tau_c$ corresponding the condition as  $\tau_c b\lambda = F$, where $F$ is the particle strength. Note that the effective spacing $\lambda$, is not the average particle spacing $L$ on the glide plane. The effective particle spacing depends on stress, the argument being that an increasing stress causes a more bowed out dislocation between the pinning points, and consequently a larger number of particles that come into contact with it. This dependence is  modeled by \textit{Friedel statistics} \citep{friedel1964dislocations} \citep[see also][]{fleisgher1961solution}, which assumes that the area swept by the dislocation due to an unpinning event contains only one new obstacle. For a square lattice of particles on the glide plane, this leads to the \textit{Friedel spacing} $\lambda=L\sqrt{2T/F}$, where $T$ is the dislocation line tension (typically approximated as $\approx 1/2\mu b^2$). In turn, the square lattice spacing is $L=N_s^{-\nicefrac{1}{2}}$, where $N_s$ is the number of particles per unit area of the plane, which can be expressed in terms of the volume fraction $f$ and either the average particle radius $R$, or the average radius  $r$ at the intersection between the glide plane and the particles, as summarized in Fig.~\ref{fig:RrL}. This leads to the critical resolved shear stress
\begin{align}
\tau_c=\frac{1}{bL}\sqrt{\frac{F^3}{2T}}=\frac{1}{b r}\sqrt{\frac{3\pi fF^3}{64 T}}=\frac{1}{b R}\sqrt{\frac{3fF^3}{4\pi T}}\, .
\label{friedelStrength}
\end{align}

Several mechanisms may impart the particle strength $F$, including chemical, stacking-fault, modulus, coherency, and order strengthening \citep{nembach1985precipitation,ardell1985precipitation,martin2012precipitation}. Here we  focus on order strengthening, which is the leading contribution to $F$ for  $a/2\langle110\rangle\{111\}$ dislocations interacting with $\gamma'$ precipitates in Ni-based superalloys.
 An  antiphase boundary (APB) of surface energy density $\gamma_{APB}$ is left in the precipitate in the  wake of the cutting dislocation, thus yielding a force $F=2r\gamma_{APB}$, . Replacing this value in Eq.~\eqref{friedelStrength_ham} yields the stress required for a single dislocation to shear the precipitate \citep{ham1968proceedings} 
 \begin{align}
\tau_c^{H}=\frac{\gamma_{APB}}{b }\sqrt{\frac{3\pi \gamma_{APB}\, fr}{8 T}}=\frac{\gamma_{APB}}{b }\sqrt{\frac{3\pi^2 \gamma_{APB}\, fR}{32 T}}\, .
\label{friedelStrength_ham}
\end{align}

Because the shearing process  is  complete only when a second dislocation cuts the precipitate and restores the order, it is  more significant to consider shearing by a pair of $a/2\langle110\rangle\{111\}$ dislocations. Among other scenarios, \cite{BrownHam1971} considered the case of \textit{weakly coupled} dislocations, separated by a distance larger than the average particle diameter. The force equilibrium of the pair of leading and trailing dislocations, that is $\tau_cb\left(\lambda_L+\lambda_T\right)=2\gamma_{APB}\left(r_L-r_T\right)$. This yields the  estimate $\tau_c\approx \nicefrac{\gamma_{APB}}{2b}(\nicefrac{2r_L}{\lambda_L}-\nicefrac{2r_T}{\lambda_T})$. 
If the trailing dislocation is almost straight then $\lambda_T$ is the ``mean free path" between two particles on a random straight line, hence $2r_T/\lambda_T=\nicefrac{3\pi}{8}f\approx f$. On the other hand $\lambda_L$ is given by the Friedel spacing, thus leading to the weakly-coupled strengthening
\begin{align}\label{tauWCS}
\tau_c^{BH}=\frac{\gamma_{APB}}{2b}\left(\sqrt{\frac{3\pi \gamma_{APB}fr}{8 T}}-f\right)=\frac{\gamma_{APB}}{2b}\left(\sqrt{\frac{3\pi^2 \gamma_{APB}fR}{32 T}}-f\right)\, .
\end{align}
\cite{ardell1985precipitation} has verified that for small $rf$, the scaling $\tau_c\sim \sqrt{fr/T}$ holds for several commercial alloys.
For larger overaged particles, however, the  precipitate diameter becomes comparable to the distance between the dislocation pair, and the assumptions of weak coupling does not apply. Instead the dislocation pair  becomes \textit{strongly coupled} while shearing the same precipitate. \cite{huther1978interaction} considered this situation, and after equilibrium considerations obtained the following estimate for spherical particles
\begin{align}\label{tauSCS}
\tau_c^{HR}=0.44 p \frac{T}{br}\sqrt{\frac{\pi^2\gamma_{APB}rf}{4pT}-f}\, ,
\end{align}
 where $p$ is a fitting constant of order unity.

\subsection{Orowan looping}\label{sec:Orowan_looping}
When the conditions for shearing are not met, dislocations can bypass coherent particles by the process of Orowan looping, a concept that was initially developed to explain precipitation hardening by  unshearable obstacles \citep{Orowan1948symposium}. The   idea is that a flexible dislocation with line tension $T$ can be bent by a critical stress satisfying  $\tau_c b=T/\rho$, where $\rho$ is the radius of curvature between the obstacles. Although several estimates of the appropriate line tension have been proposed \citep[see][]{Kelly1963}, a crude approximation using $T=\nicefrac{1}{2}\mu b^2$ and $\rho=L'/2$, where $L'$ is the inter-particle spacing, yields the celebrated result 
\begin{align}
    \tau^O =\frac{\mu b}{L'}\, ,
\end{align} 
where for a square lattice of particles $L'=N_s^{-\nicefrac{1}{2}}-2r=\left(\sqrt{\nicefrac{32}{3 \pi f}} - 2 \right)r$.

Bacon, Kocks, and Scattergood (BKS) improved  this estimate by considering the effect of the dislocation self-interaction on the Orowan stress \citep{bks_1973}. The argument used to derive the critical bypass stress is that the applied force must be balanced by the self energy of the bowing line, that is $L'b\tau_c=\mu b^2 A\left[\ln\left(X/r_0\right)+B\right]$, where  $r_0$ is a conventional cut-off radius, $B$ is a fitting constant of the order unity,  $X$ is an appropriate outer screening distance of the dislocation elastic fields, and $A=\nicefrac{1}{2\pi}$ for edge dislocations or $A=\nicefrac{1}{2\pi(1-\nu)}$ for screw dislocations.  Taking $r_0\approx b$, $B\approx 0.7$ and using the harmonic mean $X=(1/L'+\nicefrac{1}{2r})^{-1}$ yields
\begin{align}
    \tau_c^{BKS}= \frac{\mu b A}{ L'} \left[\ln \left(\frac{b}{L'}+\frac{b}{2r}  \right)^{-1}+ 0.7\right]\, .
    \label{tauBKS}
\end{align}
\cite{queyreau_Orowan} considered the effect of subsequent loops piling-up around the precipitates. The loop pile-up is equivalent to an increased effective radius of the precipitate, hence a simple modification of Eq.~\eqref{tauBKS} was proposed with $r$ replaced by an effective radius $r_e$. Note that \eqref{tauBKS} applies only approximately to shearable particles because partial shearing may take place during the looping process, therefore altering the nominal parameters of the precipitates distribution. 

Finally, note that deciding whether edge or screw dislocations control the bypass stress is not a trivial task. If we consider a standard dislocation line energy of the form $E(\theta)=\alpha \nicefrac{\mu b^2}{4\pi (1-\nu)}(1-\nu \cos^2\theta)$, where $\theta$ is the angle from the perfect screw orientation, then we can compute the corresponding line tension as $T=E+\nicefrac{d^2 E}{d\theta^2}$. The line tension of  screw dislocations computed in this way is a factor $(1+\nu)/(1-2\nu)$ higher than for edge dislocations, making  screw dislocations about four times stiffer than edge dislocations. Then, the bypass stress should be controlled by edge dislocations in weakly-coupled shearing, and by screw dislocations in strongly-coupled shearing and looping. However, screw dislocations may also bypass precipitates by cross-slip. Since the main  objective of the paper is to investigate the mechanisms of shearing to looping transition we will consider only the case of initially edge dislocations  in our simulation results.

\section{Computational Framework\label{framework}}

The objective of this paper is to study the mechanisms of  $\gamma'$ precipitates bypass by $a/2\langle110\rangle\{111\}$ dislocations, with particular interest in the regime of transition between shearing and Orowan looping. In order to do so within the DDD framework, we need to develop a robust method to compute the forces exerted by generalized stacking faults on dislocations. Consider a slip plane in the material. Let $\gamma(\bm s(\bm x))$ be the generalized stacking fault energy density of that plane at point $\bm x$, and $\bm s(\bm x)$ be the slip vector at that point. Then the misfit energy of the plane is
\begin{align}
    E_\text{misfit}=\int \gamma(\bm s(\bm x)) \,\text{d}A\, ,
\end{align}
where $\text{d}A$ is an elementary area element of the glide plane. Dislocations are  lines of discontinuity for the function $\bm s(\bm x)$ and consequently they are  possible lines of discontinuity for the function $\gamma(\bm s(\bm x))$. If $\bm w$ is the dislocation velocity in the slip plane, then the transport theorem for discontinuous functions yields the rate
\begin{align}
    \dot E_\text{misfit}=-\int [\gamma(\bm s(\bm x))\, \bm d]\cdot \bm w \,\text{d}l\, ,
\end{align}
where $[\gamma(\bm s(\bm x))\, \bm d]=\gamma(\bm s(\bm x^+))\bm d^++\gamma(\bm s(\bm x^-))\bm d^-$. Here the upperscripts  $+$ and $-$  simply indicate the two sides of glide  plane separated by the dislocation,  and $\bm d^+$ and $\bm d^-$ are  the vectors orthogonal to the dislocation line in the glide plane pointing inside the corresponding region ($\bm d^+=-\bm d^-$). The configurational force per unit length on the dislocation due to generalized stacking fault is therefore
\begin{align}
    \bm f=  [\gamma(\bm s(\bm x))\bm d]=\gamma(\bm s(\bm x^+))\bm d^++\gamma(\bm s(\bm x^-))\bm d^-\, .
    \label{SFforce}
\end{align}

For a certain phase in the material (e.g. the fcc matrix $\gamma$-phase, or the $\text{L}1_2$ precipitates $\gamma'$-phase), the generalized stacking fault energy density function $\gamma(\bm s)$ is typically constructed from \textit{ab-initio} calculations by shifting the upper side of the crystal relative to the lower side by the slip vector $\bm s$. A discussion on the different types of stacking faults in the $\gamma$ and $\gamma'$ phases has been provided in \cite{vorontsov_2012}. In this work, the \textit{ab-initio} data is fitted to the periodic function
\begin{align}\label{gamma_s}
\gamma(\bm s)=\sum_{i}\sum_j S_i \, \sin(\bm k_i \cdot \bm M_j\bm s) + C_i \, \cos(\bm k_i \cdot \bm M_j\bm s), 
\end{align}
where $\bm s$ is a local (two-dimensional) slip vector on the glide plane, the $\bm k_i$'s are wave vectors,  the $\bm M_j$ are matrices representing the symmetries of the lattice, and $S_i$ and $C_i$ are the corresponding fitting constants. The fitting procedure is detailed  in \ref{app:calc_gamma_surf}. After this one-time fitting procedure, the function $\gamma(\bm s)$ is then fully available analytically.

Once the force \eqref{SFforce} is obtained, it can be summed to the Peach-Koehler force routinely computed in DDD simulations. The total force is then used to obtained the dislocation velocity and evolve the dislocation configuration in time as already discussed elsewhere \citep[e.g.][]{po2014variational,cui2017influence,cui2018size}. However, it should be recognized that there is an intrinsic difficulty in the DDD framework, related to the computation of the force \eqref{SFforce}. In fact, while the Peach-Koehler force ultimately depends on the dislocation configuration, that is the collection of local Burgers vectors and  tangent vector of each dislocation line element, the force \eqref{SFforce} depends on the slip vectors on the two sides of any given dislocation during the simulation, that is  the vectors $\bm s(\bm x^+)$ and $\bm s(\bm x^-)$. The  difficulty is then how to efficiently compute these vectors. To solve this problem \cite{hussein_uchic_2017} construct a mesh of each glide plane intersecting a precipitate, with cells storing the local accumulated slip vector $\bm s(\bm x)$. In this approach a  planar mesh is created every time a new glide plane is needed, and great deal of precision must be used to update the slip vector once a dislocation sweeps its center. There is therefore a compromise between the accuracy of the force calculation, related to the planar mesh resolution, the corresponding memory requirement, and the computational cost associated  with updating a large number of cell points. Here we pursue a different method, which does not require any meshing nor pointwise storage and update of the slip vectors $\bm s(\bm x)$.  The tenet of the proposed method is that the identity of individual loops composing the dislocation network is maintained during the simulation. In our numerical implementation in the MoDELib
 (Mechanics Of Defect Evolution Library) code \citep{po2015model} this condition is satisfied, since MoDELib features a double topological representation of the dislocation network, where the identity of individual planar loops in the network is preserved  even when segments undergo discrete reactions such as glissile junctions and cross-slip.


\begin{figure}[t!]
	\centering
	\includegraphics[width=0.8\linewidth,trim=0 0 0 0,clip]{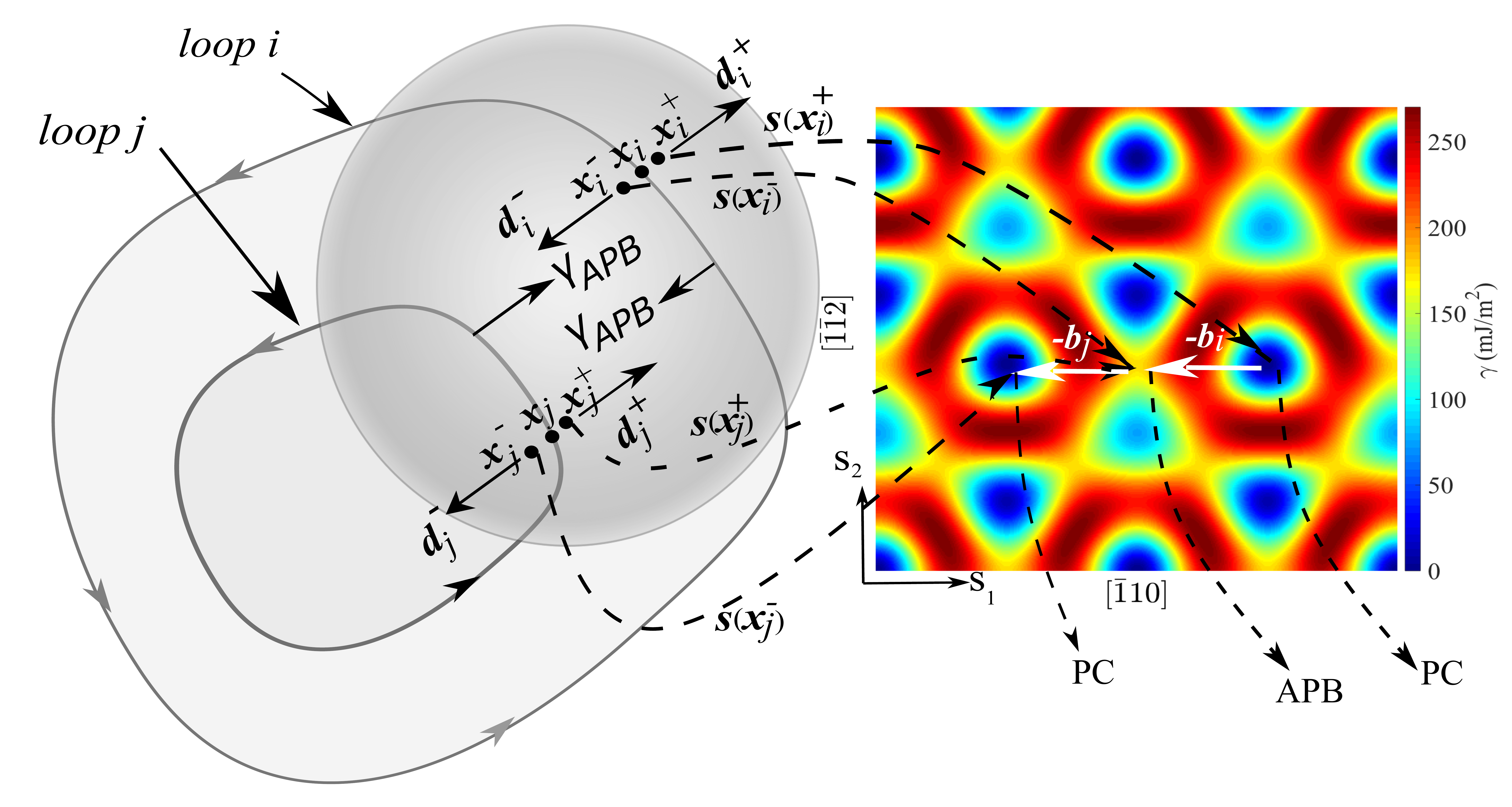}
	\caption{Method of calculation of the stacking fault force acting on $a/2\langle110\rangle\{111\}$ superpartials inside a $\gamma'$ precipitate. The superpartial dislocation loop (loop $i$) has Burgers vectors $\bm b_i = a/2[\bar{1}10]$. Minus Burgers vectors correspond to equivalent slip vectors in $\gamma$-surface. Detailed description can be found in the text. Here PC represents the perfect crystal.}
	\label{fig:SFE}
\end{figure}

In order to describe the method, let us consider Fig.~\ref{fig:SFE}. The generalized stacking fault energy surface $\gamma(\bm s)$ of the $\{111\}$ planes of the $\text{L}1_2$  $\gamma'$ phase is shown in a local coordinate system with Cartesian axes $\bm s_1$ and $\bm s_2$ along $\langle 110\rangle$ and $\langle 112\rangle$, respectively. The analytical expression of this function is given in eq.~\eqref{gamma_s}, and the fitting procedure is detailed in  \ref{app:calc_gamma_surf}. In this reference system, the vector $ \bm N = \bm s_1 \times \bm s_2$ points to the upper side of the crystal. Now consider a dislocation loop on this plane. The loop gives rise to a plastic distortion
\begin{align}\label{eq:betap}
\beta^p (\bm x) = - \int_S \delta(\bm x- \bm x') \bm b  \otimes\bm n_R\, \text{d}  A'\, ,
\end{align}
where $\mathcal{S}$ is the slip surface of the loop, $\bm b$ its Burgers vector,  $\text{d}A$  an elementary area element, and $\bm n_R$ the normal to the loop plane which is \textit{right-handed} to the loop tangent vector. It can be seen that the ``inelastic part" of the displacement field caused by this distortion is associated with the solid angle subtended by the loop  \citep{po2018non}. The slip vector $\bm s(\bm x)$ is the jump in this inelastic displacement as the field point $\bm x$ crosses the slip plane. If $\bm n_R$ is oriented in the same sense as $\bm N$, points inside  the loop and immediately above the  plane slip by an amount $-\bm b$ relative to points below the plane, so that $\bm s(x)=-\bm b$.  If the orientation is opposite $\bm s(x)=\bm b$. If the point $\bm x$ is outside the loop then $\bm s(x)=\bm 0$. 
Since $\bm n_R$ is uniquely determined by the orientation of the loop line tangent, the slip vector can efficiently be computed  in a two-dimensional reference system on the slip plane as $\bm s(\bm x)=- W_i(\bm x) \bm b$. Here $W(\bm x)$ is a quantity borrowed from complex analysis, and known as the the \textit{winding number} of the loop about $\bm x$ \citep{krantz1999handbook}. The winding number is an integer and it measures the number of times the loop wraps around $\bm x$. If $\bm n_R$ and $\bm N$ are in the same direction, then the loop is counterclockwise in the $\bm s_1$ and $\bm s_2$ plane of Fig.~\ref{fig:SFE} and it yields $W(\bm x)=1$ when it contains $\bm x$, and zero otherwise. If $\bm n_R$ and $\bm N$ are in opposite directions, then loop is clockwise it yields $W(\bm x)=-1$ if it contains $\bm x$, and zero for points outside. For more than one loop the total slip vector of any point in the plane can be obtained as 
\begin{align}
\bm s(\bm x)=-\sum_i W_i(\bm x) \bm b_i\, , 
\label{swx}
\end{align}
where the summation extends over all the dislocation loops in the plane.

 The actual calculation of the generalized stacking fault force \eqref{SFforce} along a certain dislocation segment proceeds as follows. Dislocation segments are populated with  quadrature points. For each quadrature point with position $\bm x_i$ we create two adjacent points, namely $\bm x_{i}^+=\bm x_i+\varepsilon \bm d_i^+$ and $\bm x_{i}^-=\bm x_i+\varepsilon \bm d_i^-$, where $\varepsilon$ is an infinitesimal positive scalar, as shown in  Fig.~\ref{fig:SFE}. Eq.~\eqref{swx} is then applied to these two points, thus allowing the calculation of the generalized stacking fault  \eqref{SFforce}. 
The lattice restoring force acting on dislocation segments are  obtained by integrating this stacking fault force per unit length over all quadrature points.

As an example we can consider an explicit force calculation for the two points $\bm x_i$ and $\bm x_j$ in Fig.~\ref{fig:SFE}. Suppose that these points reside on $a/2\langle110\rangle\{111\}$ loops  which are shearing a $\gamma'$ precipitate. The point $\bm x_i$ is on the outermost (leading) loop, while the point $\bm x_j$ in on the innermost (trailing) loop.  Then $\bm x_i^+$ is outside both loops so that, according to \eqref{swx}, $\bm s(\bm x_i^+)=\bm 0$. On the other hand $\bm x_i^-$ is inside the leading loop but outside the trailing loop, so that $\bm s(\bm x_i^-)=-\bm b_i$. Then the generalized stacking fault force at $\bm x_i$ is $\bm f(\bm x_i)=\gamma_\text{APB}\bm d_i^-$, which is a force that tries to expel the leading dislocation from the precipitate. On the other hand, the point  $\bm x_j^+$ is inside the leading loop and outside the trailing loop, so that $\bm s(\bm x_j^+)=-\bm b_i$, while the point $\bm x_j^-$ is inside both loops and $\bm s(\bm x_j^+)=-\bm b_i-\bm b_j$. Then the  generalized stacking fault force at $\bm x_j$ is $\bm f(\bm x_j)=\gamma_\text{APB}\bm d_j^+$. This force attracts the trailing dislocation inside the precipitate.

 It is worth noting that the method described above is completely general, and it does not require any ad-hoc ``pairing" between dislocations. The method applies seamlessly  to a variety of cases involving stacking fault forces acting on dislocations, like in the case of Shockley partials in the fcc $\gamma$ phase, superpartials in the $\text{L}1_2$ $\gamma'$ phase, and even to the more complex $a\langle112\rangle\{111\}$ dislocation ribbons mentioned in the introduction. Moreover, the winding number is a lightweight and  trivially parallelizable calculation which consumes negligible wall-clock time compared to dislocation-dislocation interactions in DDD simulations. It handles arbitrary-shaped loops, regardless of their convexity and self-intersection.

\section{DDD simulation results}

\begin{figure}[t!]
\centering
\begin{subfigure}{.4\textwidth}
\centering
\begin{tikzpicture}
\node[anchor=south west,inner sep=0] (image) at (0,0) {\includegraphics[width=\linewidth]{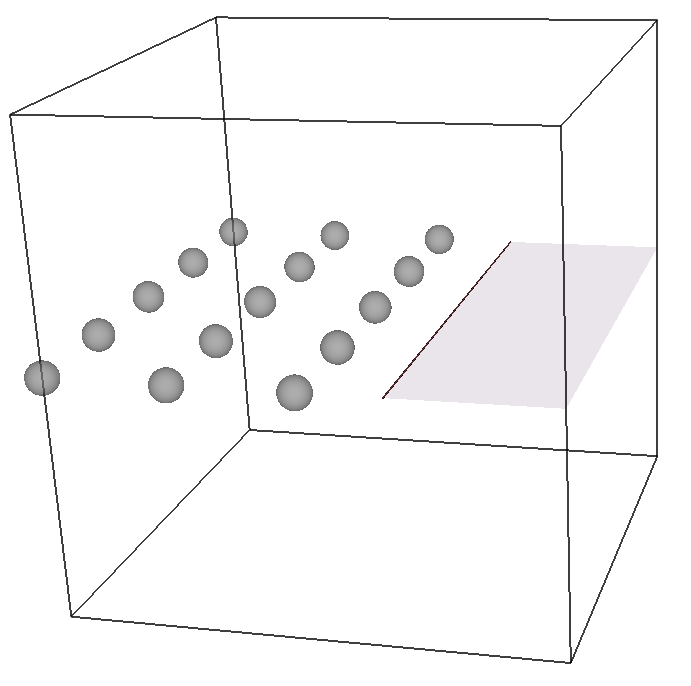}};
\begin{scope}[x={(image.south east)},y={(image.north west)}]
\draw [->, thick, blue] (0.8,0.6)  -- (0.8,0.7) node[above,black]{\small  $\{111\}$}; 
\draw [->, thick, brown] (0.7,0.55)  -- (0.6,0.55) node[below,midway,black]{\footnotesize  $\langle 110 \rangle$};
\draw [->, thick, black] (0.55,0.41)  -- (0.5,0.31) node[below,black]{\small  D}; 
\end{scope}
\end{tikzpicture}
\caption{}
\label{fig:singledis_setup}
\end{subfigure}
\hspace{1cm}
\begin{subfigure}{.5\textwidth}
\centering
\includegraphics[width=\linewidth]{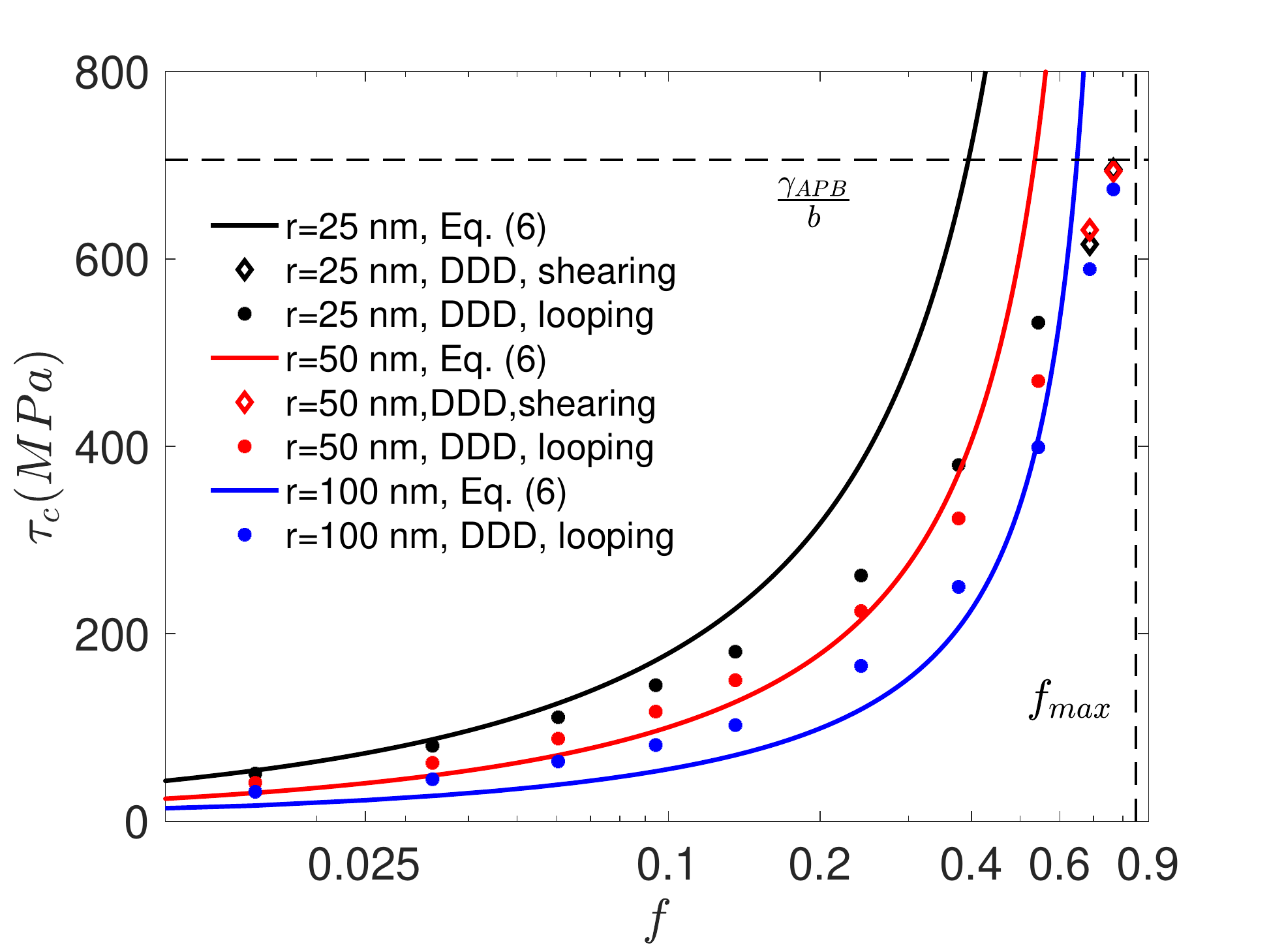}
\caption{}
\label{fig:tc_vs_f_single}
\end{subfigure}

\caption{(\subref{fig:singledis_setup}) Simulation setup for single dislocation interaction with a square lattice of precipitates. The glide plane is $\{111\}$ and the slip direction is $\langle110\rangle$. An external shear strain rate is applied on the glide plane along the slip direction. The single dislocation is marked as $D$. (\subref{fig:tc_vs_f_single}) Bypass stress as a function of volume fraction and different precipitate sizes. The maximum value of $f$ is marked as $f_{max}$ while the maximum theoretical value of the bypass stress is marked as $\nicefrac{\gamma_{APB}}{b}$. Bypass by the looping mechanism is marked by dots, while bypass by shearing mechanism is marked by diamonds.}
\end{figure}
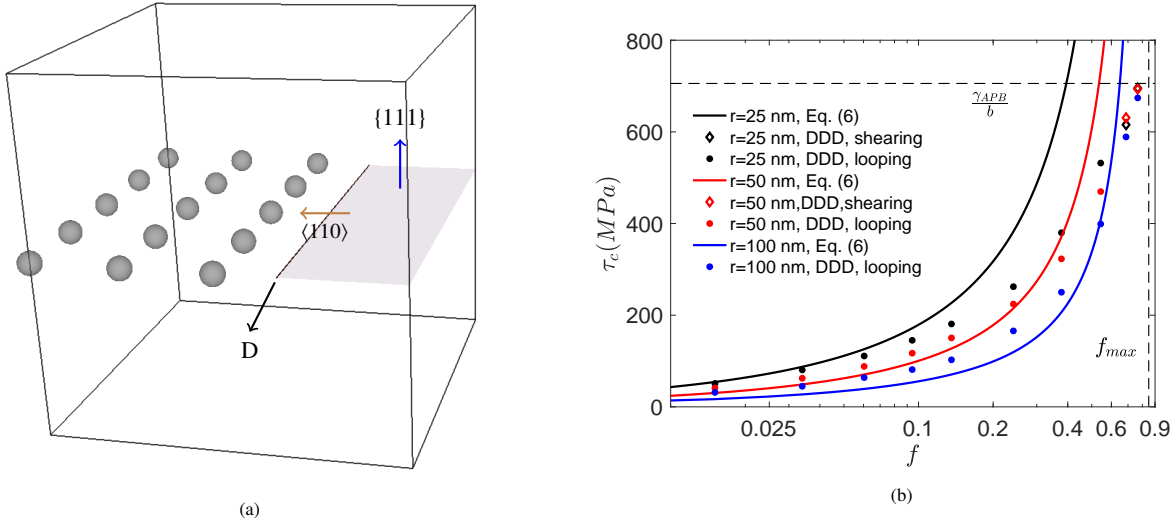


In this section we explore the mechanisms by which $a/2\langle110\rangle\{111\}$ edge dislocations overcome a square lattice of $\gamma'$ precipitates.
The radius of the precipitates intersected by the glide plane is $r$, the square lattice spacing is $L$, and the corresponding volume fraction is $f=\nicefrac{32}{3\pi}r^2/L^2$. The maximum value of $f$ occurs when $2r=L$  and it is $f_\text{max}=\nicefrac{8}{3 \pi}\approx0.85$. 
As an initial validation phase, we first consider  the interaction of a  single dislocation with the square lattice of precipitates. The interaction of a pair of dislocations in considered next. The bypass mechanisms and their corresponding strength are reported as a function of $r$ and $f$, and they are compared to analytical models. Note that, based on the method developed in section \ref{framework}, no special rules are needed to compute the force exerted by generalized stacking faults on  dislocations, even in the case of paired dislocations. 


\subsection{Single dislocation-precipitates interaction mechanisms}\label{sec:single_dd}
With reference to Fig.~\ref{fig:singledis_setup}, a single $a/2\langle110\rangle\{111\}$ dislocation is introduced within the $\gamma$ phase of the Ni-based superalloy.  The crystal is rotated such that the $\{111\}$ glide plane normal is along the global $x_3$ axis, while the $\langle110\rangle$ slip direction is along the global $x_1$ axis. The glide plane is populated with a square lattice of spherical precipitate particles. Precipitates intersect   the glide plane with a radius $r$. The size of the box is varied from $1500b$ to $12000b$, depending on the precipitate size and volume fraction. 
An external shear strain rate of $1.21 \times 10^{2} s^{-1}$ is applied on the $\{111\}$ glide plane along the $\langle110\rangle$ slip direction.  
The shear stress at which the dislocation is able to move through the precipitate square lattice is measured from the DDD simulations, and it is reported in Fig. \ref{fig:tc_vs_f_single} as a function of the volume fraction $f$, and for three different precipitate radii $r$. The strain rate was chosen to be sufficiently small that strain-rate effects on the bypass stress are negligible. The way the bypass stress is extracted from the simulations is described in \mbox{\ref{app:bypass_stress}}. 

As expected, the observed bypass mechanism in these simulations is predominantly Orowan looping,  apart from the case of high volume fraction, where shearing is observed. At low volume fraction (in particular $f \leq 0.2$) and for smaller precipitate radii  (e.g. $r=25 nm$), the bypass stress measured in the DDD simulations is in good agreement with the BKS estimate for impenetrable particles given in Eq.~\eqref{tauBKS}. As the volume fraction increases, however, dislocations start to partially penetrate the precipitates, and the BKS model does not apply anymore. A deviation between the DDD results and the BKS model is  observed for increasing $f$. The discrepancy at intermediate $f$ is mainly due to the fact that  partial shearing occurs before the looping process is complete. Hence, the effective radius and spacing of the precipitates deviate from their nominal values. The difference between penetrable $\gamma'$ precipitates and impenetrable precipitates assumed in the BKS model is more and more evident as the volume fraction $f$ approaches its maximum value $f_\text{max}$. In this case dislocations encounter an almost continuum $\gamma'$ barrier which they are forced to shear when the applied stress reaches the critical value  $\nicefrac{\gamma_{APB}}{b}$, which is independent of the particle radius. The DDD results indeed approach  this value at large $f$, independent of the value of $r$.

\subsection{Paired dislocations-precipitates interaction mechanisms}\label{sec:result_dislocation_pair}

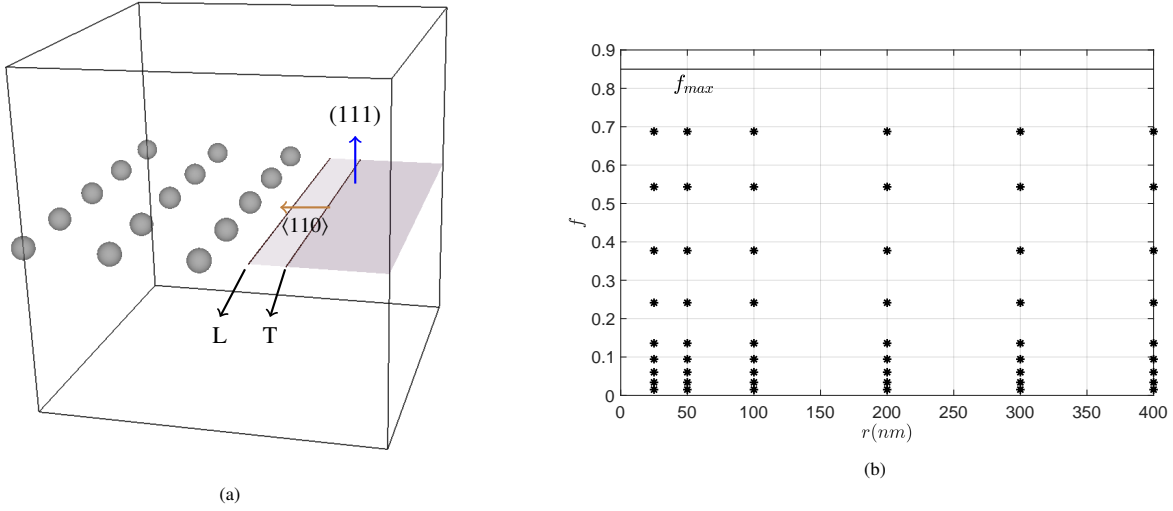
\begin{figure}[t!]
\begin{subfigure}{0.4\textwidth}
\centering
\begin{tikzpicture}
\node[anchor=south west,inner sep=0] (image) at (0,0) {\includegraphics[width=\linewidth]{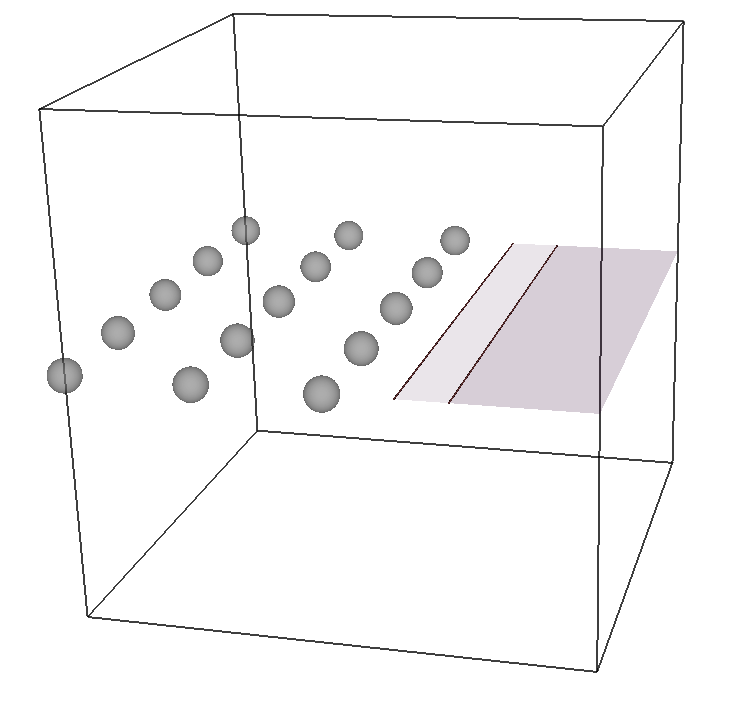}};
\begin{scope}[x={(image.south east)},y={(image.north west)}]
\draw [->, thick, blue] (0.75,0.6)  -- (0.75,0.7) node[above,black]{\small  $(111)$};  
\draw [->, thick, brown] (0.7,0.55)  -- (0.6,0.55) node[below,midway,black]{\footnotesize  $\langle 110 \rangle$};
\draw [->, thick, black] (0.53,0.42)  -- (0.48,0.32) node[below,black]{\small  L};
\draw [->, thick, black] (0.61,0.42)  -- (0.58,0.32) node[below,black]{\small  T};
\end{scope}
\end{tikzpicture}
\caption{}
\label{fig:ddd_pair_setup}
\end{subfigure}
\hspace{0.5cm}
\begin{subfigure}{0.55\textwidth}
\centering 
\includegraphics[width=\linewidth]{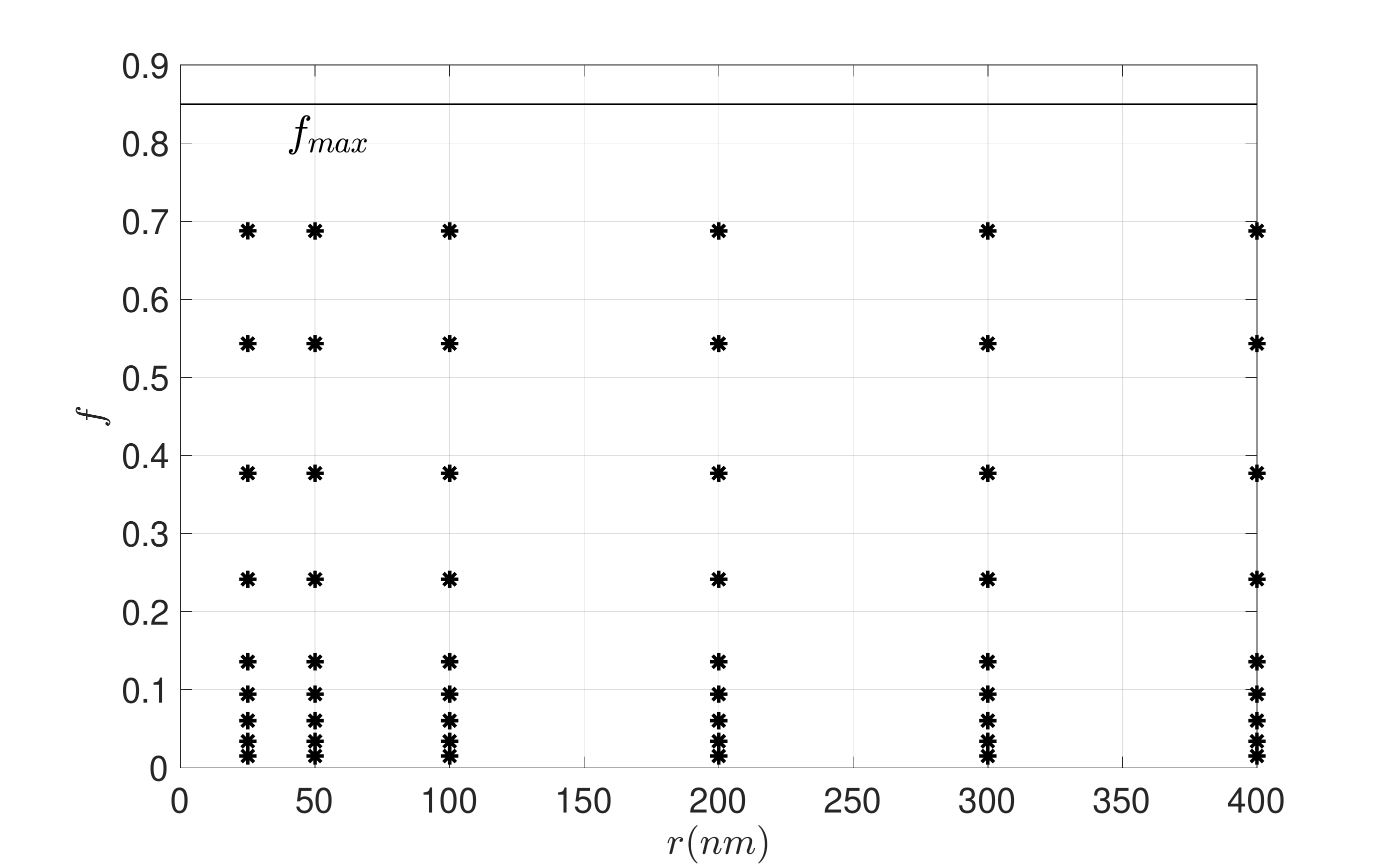}
\caption{}
\label{fig:ddd_data_points}
\end{subfigure}
\caption{(\subref{fig:ddd_pair_setup}) DDD simulation setup for paired dislocations interaction with a square lattice of precipitates. The glide plane is $\{111\}$ and the slip direction is $\langle110\rangle$. An external shear strain rate is applied on the glide plane along the slip direction. The leading and trailing dislocations are marked as L and T respectively. The slipped area is shaded.  (\subref{fig:ddd_data_points}) Set of $(r,f)$ points explored in the DDD simulations.}
\end{figure}

\tikzset{
    decision/.style={diamond, minimum height=10pt, minimum width=10pt, inner sep=1pt},
    chance/.style={circle, minimum width=10pt, draw=blue, fill=none, thick, inner sep=0pt},
    level 2/.style={sibling distance=30mm}
  }
 
\begin{figure}[t!]
  \centering   
\begin{minipage}[][14cm][t]{.85\textwidth}
\vspace*{\fill}
  \begin{subfigure}{\textwidth}
\begin{adjustbox}{width=\linewidth}
\begin{forest}
  for tree={
    parent anchor=children,
    child anchor=parent,
    l sep'+=20mm,
    where level=1{l sep'=10mm}{},
    where level=2{l sep'=15mm}{},
    where level>=3{l sep'=10mm}{},
    s sep'+=10mm,
    edge={arrow,line width=1.0pt},
  },
  [L interaction,rectangle,draw
    [{\includegraphics[width=42mm]{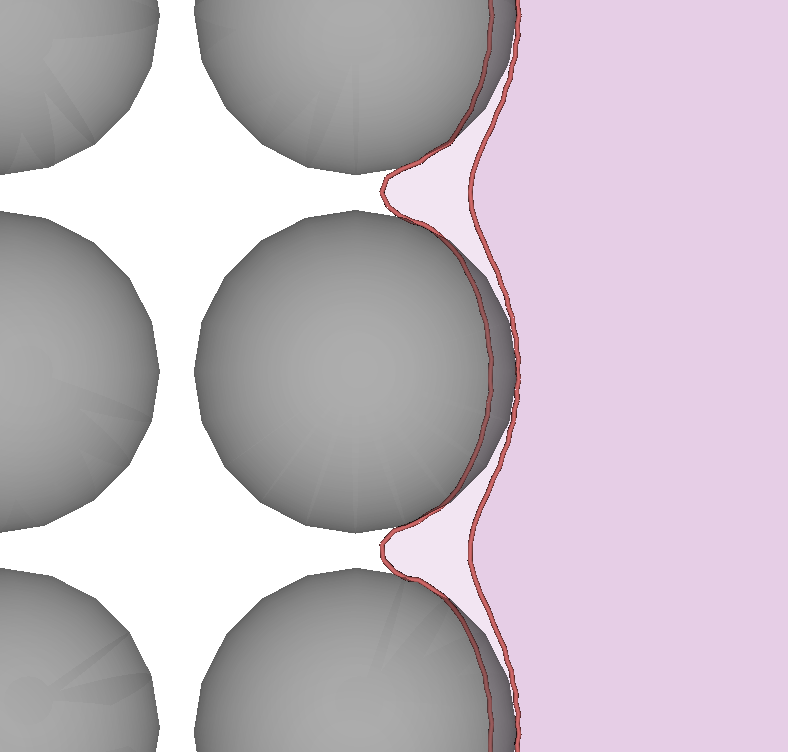}},edge label={node[midway,above,sloped] {shear}},label={above,left=1.5cm}:(A)
      [T interaction,rectangle,draw  
         [{\includegraphics[width=42mm]{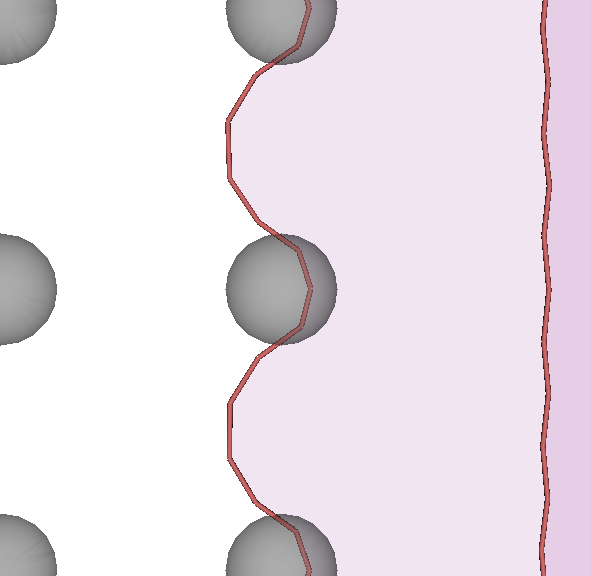}}, edge label={node[midway,above,sloped] {shear}},label={above,left=1.5cm}:(A1.1)          [{\includegraphics[width=42mm]{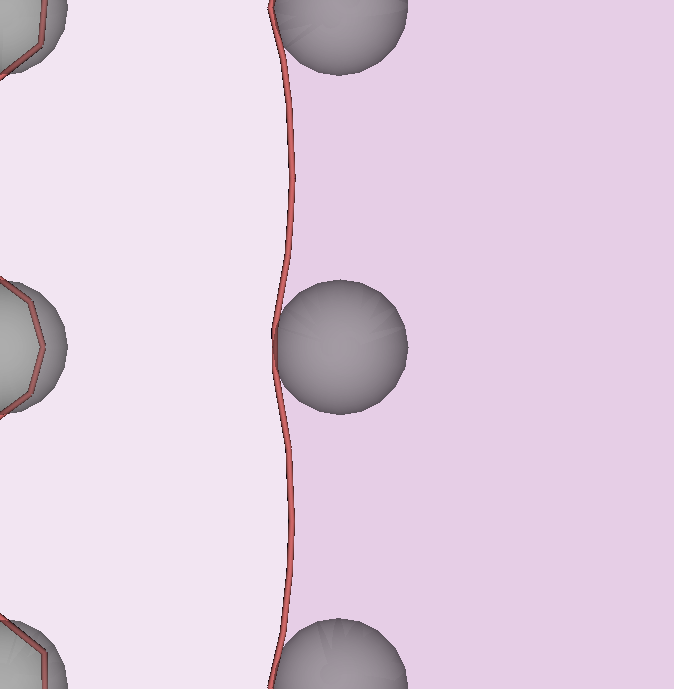}}, no edge,label={above, left=1.5cm}:(A1.2),label={below:{\large weakly-coupled shearing}}]
         ]
         [{\includegraphics[width=42mm]{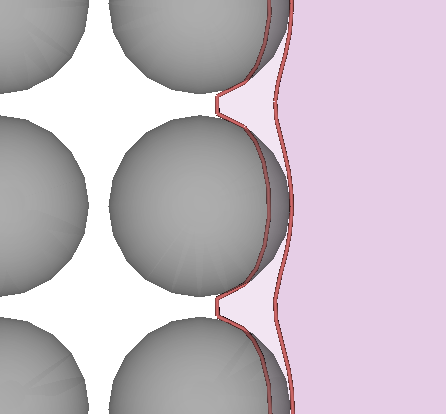}}, edge label={node[midway,above,sloped] {shear}},label={above,left=1.5cm}:(A2.1)           [{\includegraphics[width=44mm]{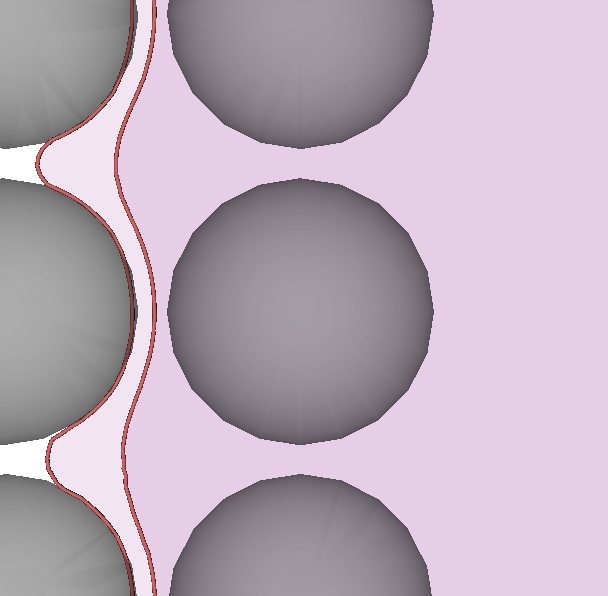}}, no edge,label={above, left=1.5cm}:(A2.2),label={below:{\large strongly-coupled shearing}}]
         ]
      ]
    ]
    [{\includegraphics[width=42mm]{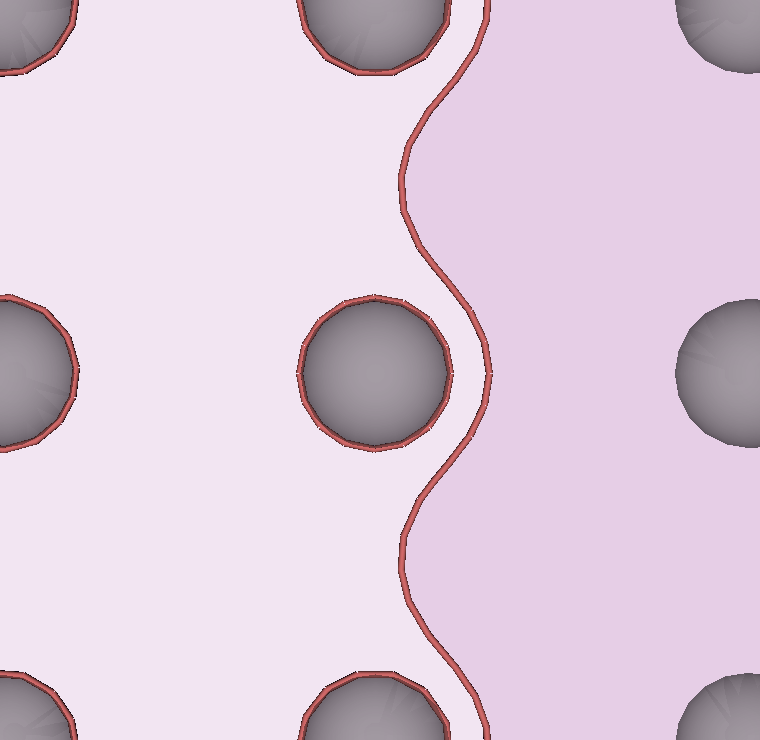}},edge label={node[midway,above,sloped] {loop}},label={above,left=1.5cm}:(B)
      [T interaction,rectangle,draw [{\includegraphics[width=42mm]{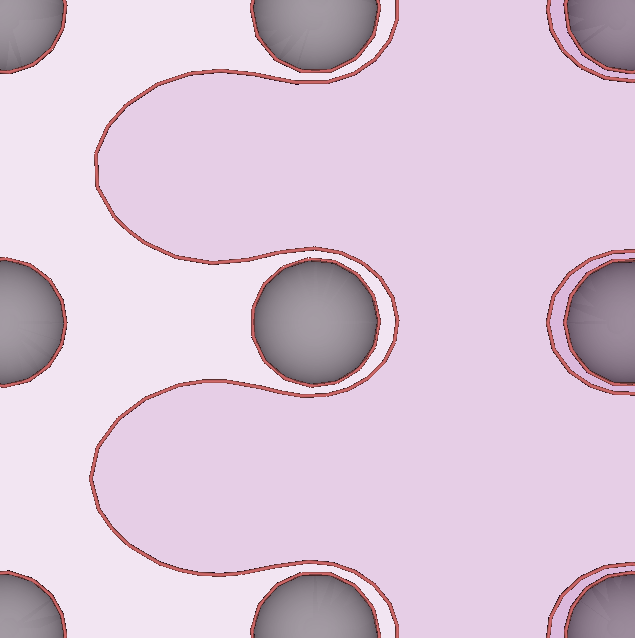}}, edge label={node[midway,above,sloped] {loop}},label={above,left=1.5cm}:(B1.1)
          [{\includegraphics[width=42mm]{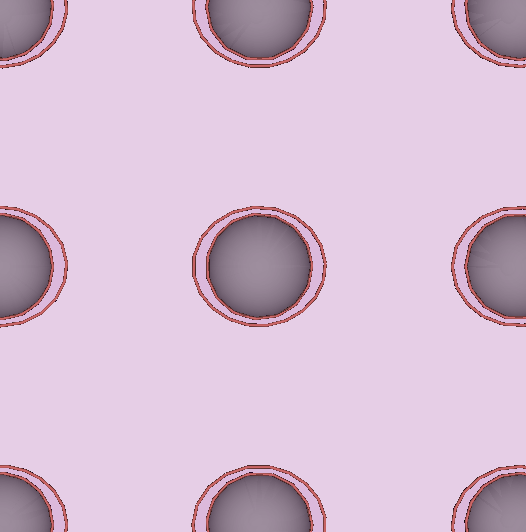}}, no edge,label={above,left=1.5cm}:(B1.2),label={below:{\large looping}}]
        ]
        [{\includegraphics[width=42mm]{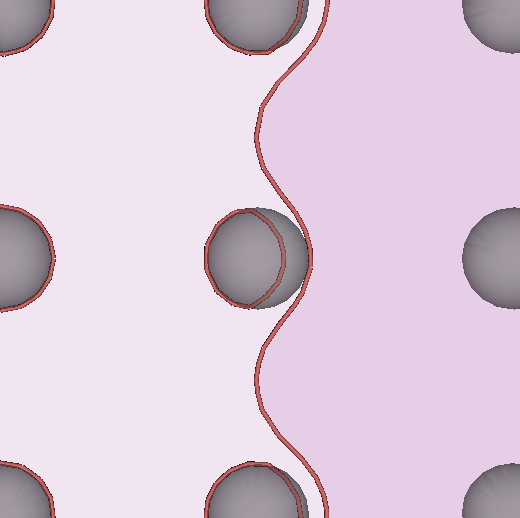}}, edge label={node[midway,above,sloped] {shear}},label={above,left=1.5cm}:(B2.1),
          [{\includegraphics[width=42mm]{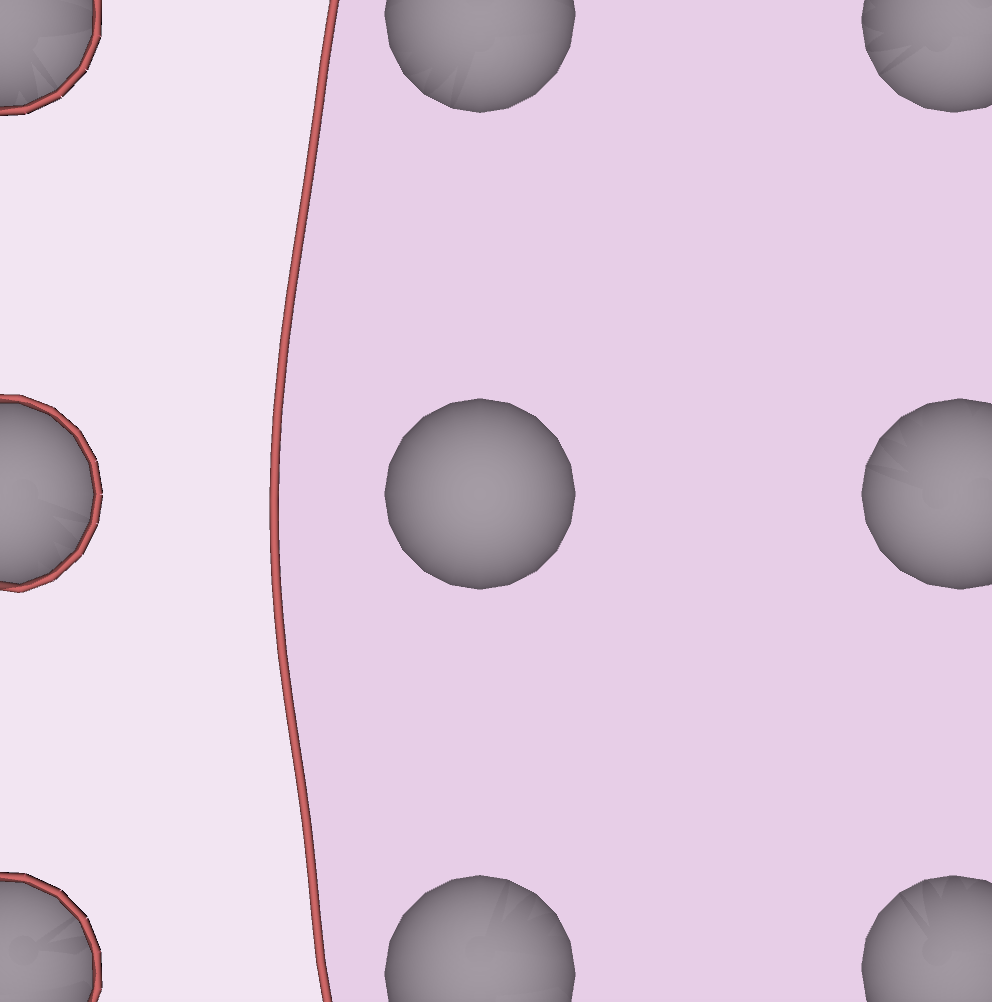}}, no edge,label={above,left=1.5cm}:(B2.2),label={below:{\large hybrid looping-shearing}}]
        ]   
      ]   
    ]
  ]
\end{forest}
\end{adjustbox}
\end{subfigure}
\end{minipage}
\caption{Tree diagram of $\gamma'$ precipitates bypass mechanisms    by paired $a/2\langle110\rangle\{111\}$ dislocations. The leading and trailing dislocations are labeled as L and T respectively. Depending on the precipitate radii and volume fraction, L either shears (A) the precipitates or loops (B) around them. If L shears, T always shears. If T is outside the precipitate and it remains almost straight during the critical configuration, the bypass mechanism satisfies the weakly-coupled shearing conditions ((A1.1) and (A1.2)). If L and T are inside the same precipitate during the critical configuration,  strongly-coupled shearing occurs ((A2.1) and (A2.2)). If L loops, T can either for a second loop around the precipitates ((B1.1) and (B1.2)), or trigger the hybrid looping-shearing bypass mechanisms, where the leading loop is  collapsed in the critical configuration (B2.1) forming APB, followed by trailing shearing which removes the APB  (B2.2). }
\label{fig:decision_tree}
\end{figure}

The same simulation setup described in section \ref{sec:single_dd} is now used in the case of a pair of $a/2\langle110\rangle\{111\}$ edge dislocations interacting with the square lattice of precipitate (Fig.~\ref{fig:ddd_pair_setup}). The set of $(r,f)$ points  explored in the DDD simulations is shown in Fig. \ref{fig:ddd_data_points}.  Depending on the precipitate radius $r$ and volume fraction $f$, three distinct types of bypass mechanisms are observed, each with different scaling of the strength as a function of $r$ and $f$. The possible outcomes of the interaction between the dislocation pair and the square lattice of precipitates is illustrated in the tree diagram of Fig.~\ref{fig:decision_tree}, which includes images extracted from the DDD simulations that are representative of the distinct bypass mechanisms. The white background in these images indicates the pristine area of the glide plane, while the lighter and darker magenta backgrounds indicate areas slipped by the leading dislocation only, and both leading and trailing dislocations, respectively. 

The first discriminant between the bypass mechanisms is the  type of interaction between the leading dislocation and the precipitates. Depending on the values of  $r$ and $f$, the leading dislocation can either shear the precipitates (image A), or loop around them (image B). In the case of shearing by the leading dislocation, the subsequent interaction of the trailing dislocation   always results in shearing of the precipitate, so that the APB fault is removed. This mode of shearing can be either weakly-coupled (Eq.~\eqref{tauWCS}) or strongly-coupled (Eq.~\eqref{tauSCS}) depending on the values of $r$ and $f$. For example, image A1.1 depicts the critical configuration of a weakly-coupled condition, with the leading dislocation inside an array of precipitates, and the trailing dislocation being almost straight outside the precipitates. Image A1.2 shows the corresponding configuration after the bypass. On the other hand, image A2.1 shows the critical configuration corresponding to the maximum shear stress, for the case of a high volume fraction. In this case the paired dislocations are strongly coupled and shear the precipitate cooperatively. Image A2.2 shows the paired dislocations after exiting the first array of precipitates, and ready to repeat the shearing process for the second array of precipitates. 

The situation is more diverse in the case that the leading dislocation initially loops around the precipitates (image B in Fig.~\ref{fig:decision_tree}). In this case, the trailing dislocation encounters an array of precipitates surrounded by repulsive  Orowan loops left in the wake of the leading dislocation. Our results indicate that two distinct bypass mechanisms may operate in this case. First, the trailing dislocation may also loop around the precipitates, creating  secondary loops concentric to the existing ones (image B1.2).  The bypass stress corresponding to the critical configuration (image B1.1) is then the stress necessary  to create  the secondary loops, which follows the BKS estimate (Eq.~\eqref{tauBKS}). However, hardening is found in this process for two reasons:  (a) because  the effective radius of the precipitates is increased by the presence of the primary loops \citep{queyreau_Orowan};  (b) because the trailing loop embryo experiences a repulsion exerted by the primary loops surrounding the subsequent array of precipitates, especially in the case of intermediate $f$ and large $r$. 

The other possible pathway revealed by the DDD simulations seems to correspond to the hybrid shear-looping mechanism already mentioned in the introduction, which is discussed in \cite{nabarro1995physics} and was indirectly observed by \cite{viswanathan2005investigation} and  \cite{unocic2008mechanisms}. In this case, the approaching trailing dislocation drives the existing loop into the precipitate on the entry side. The bypass stress corresponds to the  critical configuration in image B2.1, where the partially-collapsed primary loop reaches an unstable configuration. When this happens, the loop quickly collapses creating APB, and the trailing dislocation immediately sweeps the precipitate and removes the APB. Note that, in this process, the shearing event is difficult to observe because when the inner loop reaches the point of instability the cooperative shearing process happens almost instantaneously. 
This instability is associated with the difficulty to observe the process directly in experiments, which are more likely to capture either
the state preceding the loop collapse (single “leading” loops around precipitates, without a stacking fault) or the state after the loop collapse (no loops and no stacking faults). 
However,  indirect evidence of the process has been recognized. For example, in Ren\'e 88 DT  under high stress, \mbox{\cite{viswanathan2005investigation}} observed $a/2\langle 1 1 0 \rangle$ loops around precipitates, without obvious pairing of the dislocations, and with very few stacking faults. The authors explain this process as a secondary (trailing) $a/2\langle 1 1 0 \rangle$  dislocation  pairing with a pre-existing (leading) loop so that the precipitate can be sheared by the pair of dislocations linked by a thin APB ribbon \citep[see Fig.~15 in ][for a sketch]{viswanathan2005investigation}. The authors recognize that the annihilation of the pre-existing loop provides a strong driving force for this mechanism, and that ex-situ observations are unlikely to document it since dislocations and stacking faults are not observed inside the precipitates. 
\mbox{\cite{unocic2008mechanisms}} suggest the existence of a similar cooperative (hybrid) loop-shear mechanism in Ren\'e 88DT by inferring it from two salient features of the structures. First, the very planar slip mode, which facilitates pairing of dislocations on the same glide plane. Second, the presence of only single loops around the precipitates. Unless this hybrid loop-shear mechanism were operative, intense strain hardening would be expected due to the accumulation of multiple loops around precipitates, which was not observed.


\begin{figure}[t!]
\centering
\begin{subfigure}{0.48\textwidth}
\centering
\includegraphics[width=0.95\textwidth]{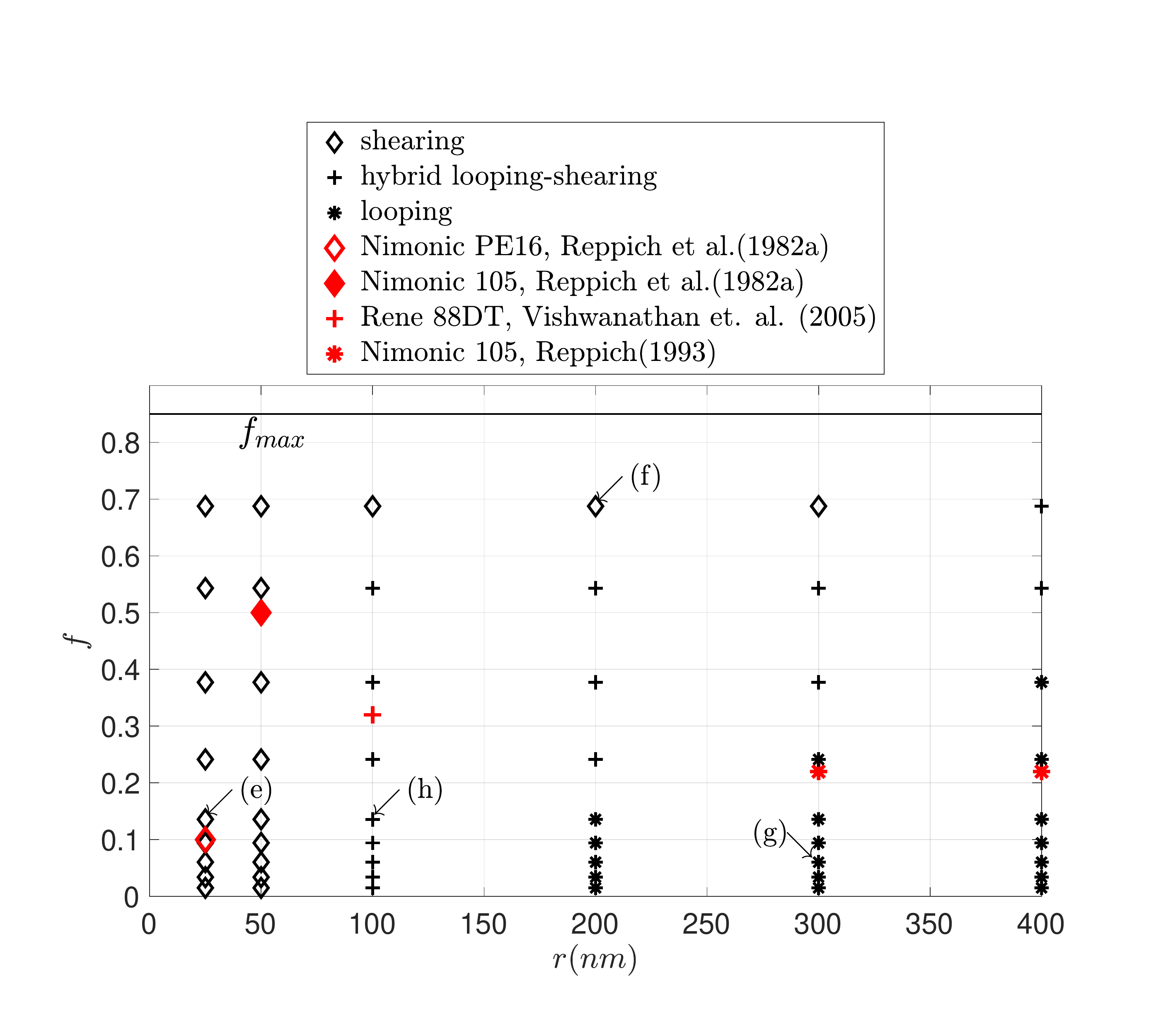}
\caption{}
\label{fig:cutting_mech_DDD}
\end{subfigure}
\begin{subfigure}{0.48\textwidth}
\includegraphics[width=0.9\textwidth]{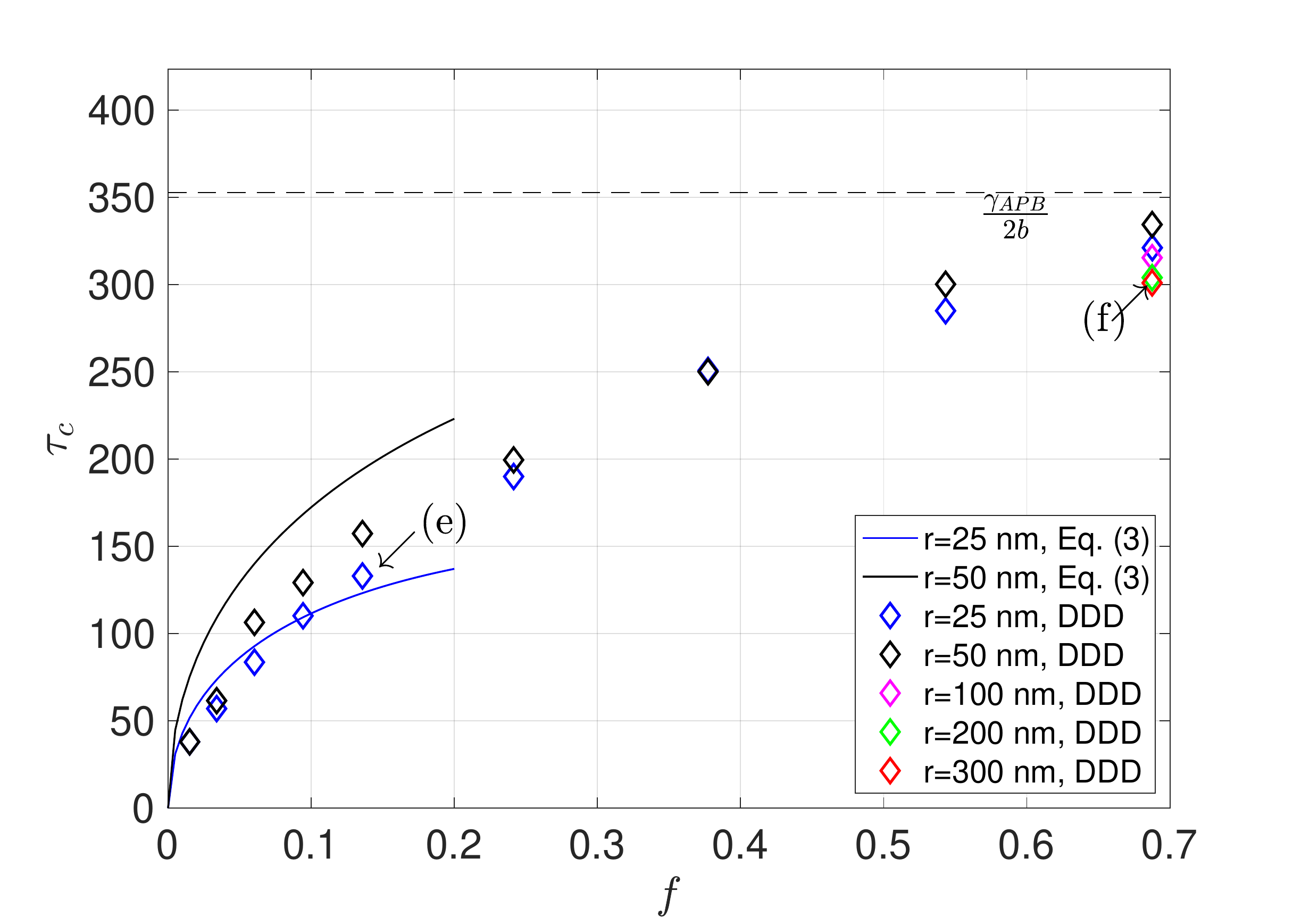}
\caption{Shearing.}
\label{fig:shearing}
\end{subfigure}
\begin{subfigure}{0.48\textwidth}
\includegraphics[width=0.9\textwidth]{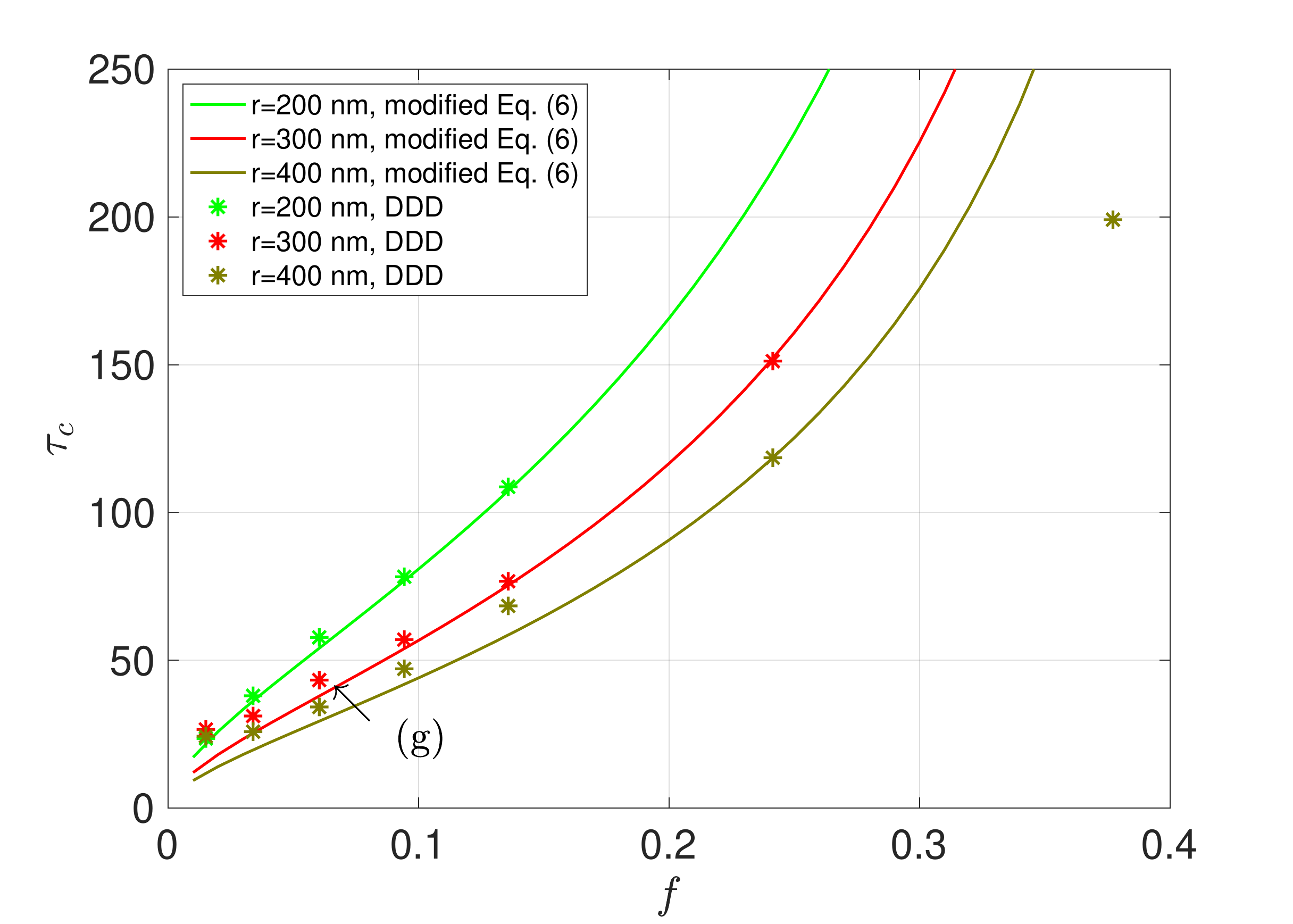}
\caption{Looping.}
\label{fig:looping}
\end{subfigure}
\begin{subfigure}{0.48\textwidth}
\includegraphics[width=0.9\textwidth]{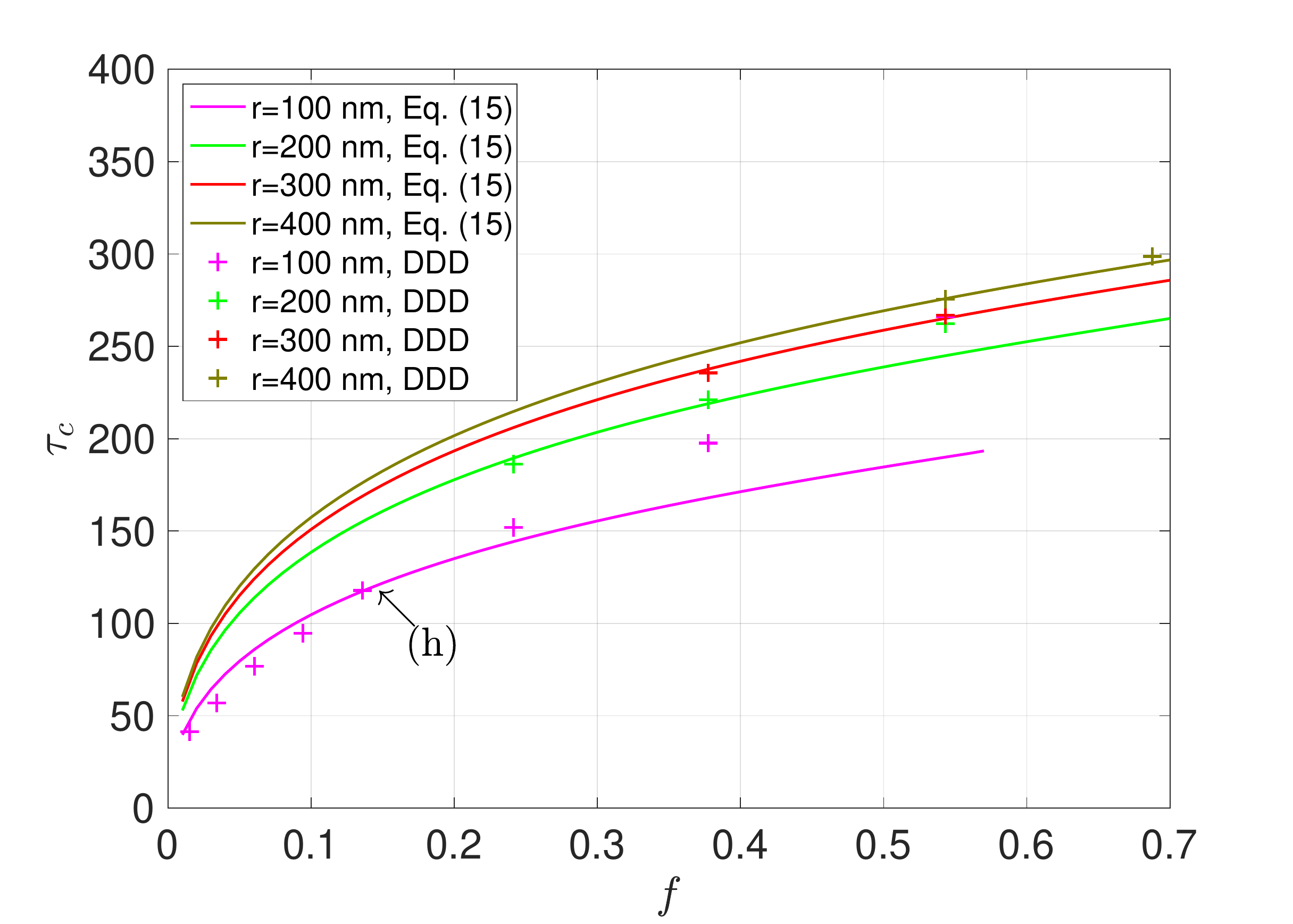}
\caption{Hybrid looping-shearing.}
\label{fig:looping_and_shearing}
\end{subfigure}
\par \bigskip
\begin{subfigure}{.22\textwidth}
  \centering
  \includegraphics[width=0.9\textwidth]{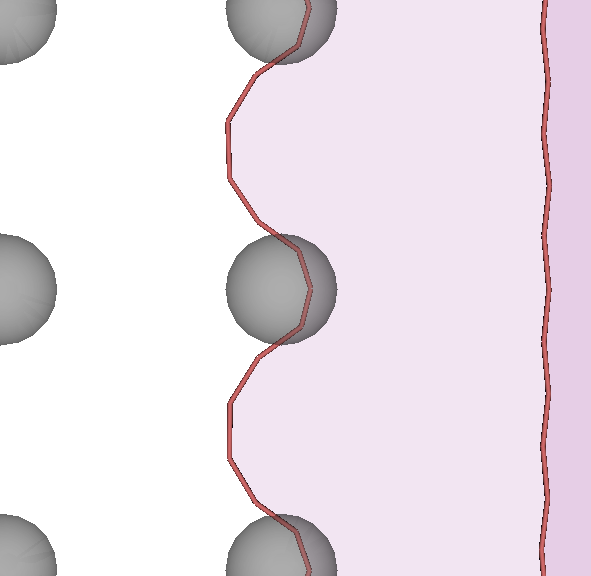}
  \caption{}
  \label{fig:Casee}
\end{subfigure}%
\hspace{0.5cm}
\begin{subfigure}{.22\textwidth}
  \centering
  \includegraphics[width=0.9\textwidth]{figures/shearShear2.png}
  \caption{}
  \label{fig:Casef}
\end{subfigure}
\hspace{0.5cm}
\begin{subfigure}{.22\textwidth}
  \centering
  \includegraphics[width=0.9\textwidth]{figures/looploop2.png}
  \caption{}
  \label{fig:Caseg}
\end{subfigure}
\hspace{0.5cm}
\begin{subfigure}{.22\textwidth}
  \centering
  \includegraphics[width=0.9\textwidth]{figures/loopShear2.png}
  \caption{}
  \label{fig:Caseh}
\end{subfigure}%
\caption{(\subref{fig:cutting_mech_DDD}) Bypass mechanism map observed in the present DDD simulations and several and experiments in the literature: the bypass mechanisms  are marked using different symbols. The mechanisms observed experimentally (marked in red) are consistently located in areas of the map occupied by the corresponding mechanisms observed in the  DDD simulations (marked in black). Four representative pairs $(f,r)$ corresponding to different bypass mechanisms are marked as $\textrm{(e)-(h)}$. We observe the large range of $f$ and $r$ over which the \textit{hybrid looping-shearing} mechanism is dominant; Comparison of theoretical models with DDD for (\subref{fig:shearing}) Shearing (\subref{fig:looping}) Looping (\subref{fig:looping_and_shearing}) Hybrid Looping-shearing;  (\subref{fig:Casee}-\subref{fig:Caseh}) Critical  configurations extracted from DDD simulations corresponding to the cases   marked as (e)-(h) in (\subref{fig:cutting_mech_DDD}). } 
\label{fig:cutting_mech_models}
\end{figure}

Fig.~\ref{fig:cutting_mech_DDD} presents a mechanism map for the three processes described above (shearing, looping, and hybrid mechanisms) as a function of volume fraction $f$ and average precipitate radius on the glide plane $r$. The shearing regime occupies the low $r$  and the  high $f$ portions of the map, while the looping regime occupies its high $r$ and low $f$ corner. Our results indicate that not only  the hybrid mechanism is possible, but it is always found as the transition mechanism between the shearing and looping regimes over the explored range  of precipitates volume fraction and radii. On the same map we also report the bypass mechanisms observed experimentally by  \mbox{\cite{reppich_1982}} in Nimonic PE16, \mbox{\cite{reppich_particle1993}} in Nimonic 105 and  \mbox{\cite{viswanathan2005investigation}} in Ren\'e 88DT. The areas of the map occupied by the three mechanisms is in general consistent with the experimental observations. However, such comparison should be considered only qualitatively, since the bypass mechanism map depends on material properties and operating conditions that may very significantly when different experiments and simulations are considered.

In order to understand how hardening depends on the coordinates $r$ and $f$, we compare the numerical simulations to appropriate analytical expressions in each regime. In the shearing regime, the DDD results are compared to the the weakly-coupled shearing model, Eq.~\eqref{tauWCS}, as shown in Fig.~\ref{fig:shearing}. Consideration of the weakly-coupled shearing mode is justified by the observation that, for small $r$ ($r=25 nm$ and $r=50 nm$), the trailing dislocation is found mostly outside the precipitates at the critical configuration. A sample configuration, corresponding to the label (e) in Figs.~\ref{fig:cutting_mech_DDD} and \ref{fig:shearing}, is illustrated in Fig.~\ref{fig:Casee}. The critical stress measured in the DDD simulations agrees with the weakly-coupled   predictions for low $f$ and small $r$ (e.g. $r=25 nm$ and $f<0.2$). Increasing $r$ and $f$ don't satisfy the weakly-coupled shearing assumptions and deviations from the theoretical model emerge. Note that at the largest volume fraction explored in the simulations ($f=0.69$), the strength measured in the simulations approaches the theoretical maximum value $\frac{\gamma_{APB}}{2b}$, with a weak dependence on $r$. This value is half of 
the already discussed theoretical maximum  for a single dislocation, and can be understood as the strength required for the dislocation pair to cut through a continuum $\gamma'$ barrier. An example of this situation, corresponding to the label (f) in Figs.~\ref{fig:cutting_mech_DDD} and \ref{fig:shearing}, is illustrated in Fig.~\ref{fig:Casef}. Note that none of the shearing models presented in section \ref{theory} (Eq. \eqref{tauWCS} or Eq. \eqref{tauSCS}) predict this limit value since they were developed under the assumption of a low volume fraction.

In the Orowan looping regime, the bypass stresses are reported in Fig.~\ref{fig:looping}. The bypass stress during the formation of an Orowan loop can be estimated by the BKS model given in Eq.~\eqref{tauBKS} if we assume that precipitates are not partially penetrated by the looping dislocations. Note that the leading dislocation loops around the precipitates at a stress that is about half of the BKS estimate, since it is assisted by the repulsion from the trailing dislocation. The critical stress is then governed by the loop formed by the trailing dislocation. The trailing dislocation encounters precipitates surrounded by a repulsive loop, which effectively act as enlarged particles.  Following \cite{queyreau_Orowan}, an increased effective radius $r_e \approx 1.33 \, r$ needs to be considered in the BKS model. However, our simulations also indicate that this correction is not sufficient in the range of radii and volume fraction considered in this regime. The difference between the effects of an augmented radius and the presence of the leading loop is that the loop exerts a repulsive stress field that further hinders the closure of the trailing loop. This difference can easily be observed in the shape of the corresponding critical configurations, as better described  in  \ref{app:comp_bks}.  
This repulsion on the exit side leads to a further increase in the bypass stress, which here is simply accounted for by using a multiplicative factor  $\alpha$ in Eq. \eqref{tauBKS}. The resulting modified BKS model, including both the effective radius $r_e$ and a multiplicative factor $\alpha=2.2$ is found to be in good agreement with the our DDD simulations for  $f<0.3$, as shown in Fig. \ref{fig:looping}.

\begin{figure}[t!]
\centering
\includegraphics[width=0.5\linewidth]{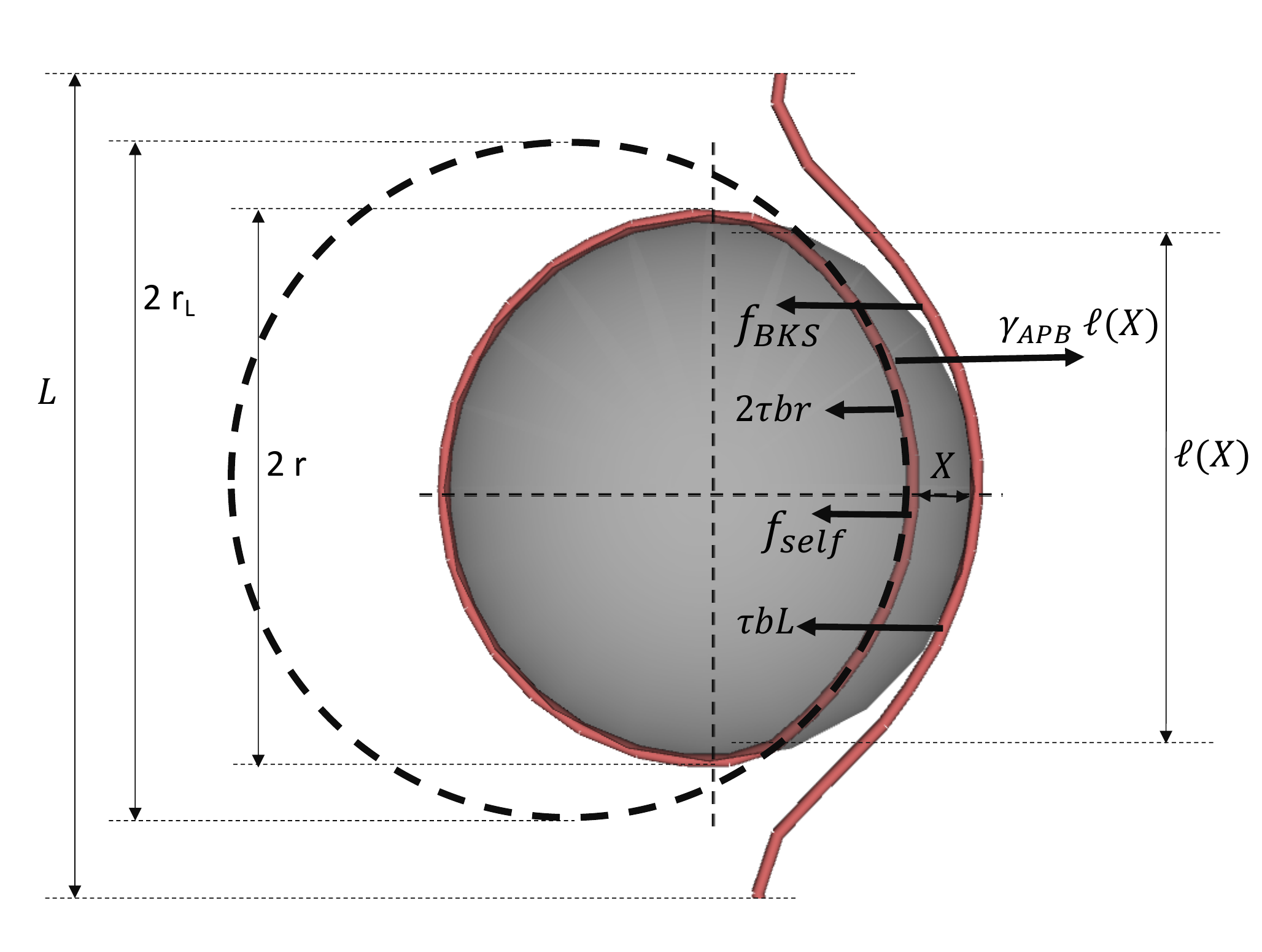}
\caption{Typical critical configuration in hybrid looping and shearing mechanism. }
\label{fig:critical_config_loopShear}
\end{figure}

Finally, let us consider the hybrid looping and shearing regime. Also in this case the leading dislocation leaves Orowan loops around the $\gamma'$ precipitates. The approaching trailing dislocation however is able to drive the enter-side of these leading loops into the precipitates, thus creating APB. The critical configuration in this regime corresponds to the collapse of the leading loops. After the leading loops collapse, the trailing dislocation  quickly shears the precipitates removing the APB. 
The critical stress for this process can be estimated with the help of the following model. Based on our results, we assume that the leading loop collapses when the trailing dislocation enters the precipitates. This condition is sketched in Fig.~\ref{fig:critical_config_loopShear}, which shows the periodic portion of length L between the bowed out parts of the trailing dislocation, and its interaction with a precipitate surrounded by the loop left behind by the leading dislocation. In the critical configuration, the enter side of the leading loop has partially been pushed into the precipitate creating APB. 
The enter-side of the leading loop remains approximately ``parallel" to the approaching curved trailing dislocation, at a separation distance  $X$ which can be estimated by the the same logic used in the case of strongly-coupled shearing. This yields  $X=2w \mu b^2/(2\pi \gamma_\text{APB})$ \citep{reed_2006}. However, the chord $\ell(X)$ subtended by loop inside the precipitate is different from the  strongly-coupled shearing case, because the partially-collapsed loop cannot be approximated by a straight line. Instead,  we assume that the shape of this partially-collapsed loop is approximately an arc of radius $r_L(X)$. Under this assumption,  simple geometry allows to determine the expressions of  $r_L(X)$ and $\ell(X)$ which are shown in  \ref{app:lx} along with their derivation. With this geometry we can consider force equilibrium in the direction of motion of the trailing dislocation, for the system composed of the trailing dislocation and the right-half of the leading loop closer to the trailing dislocation, which is shown as partially collapsed in Fig.~\ref{fig:critical_config_loopShear}. The applied stress exerts a total force $\tau b (L+2r)$ on this system. Given the self energy of the loop as $W= \frac{(2-\nu) \mu  b^2}{4 \, (1-\nu)}  r\left(\ln \frac{4r}{b} -2 \right)$ \citep{hirth_lothe}, the right-half of the loop  also experiences a force due to the loop self interaction that is
\begin{align}
    f_\text{self}= \frac{2r}{2\pi r}\frac{d W}{dr}=\frac{(2-\nu)\mu b^2}{4 \pi \, (1-\nu)}  \left[ \ln\left(\frac{4r}{b}\right) - 1 \right]\, .
\end{align}   
The trailing dislocation interacts with the left-half of the loop with a force similar to this contribution. Hence we will consider an effective self force $ w f_\text{self}$, where $1<w<2$ is a fitting parameter. 
Moreover, the self-energy of the curved trailing dislocation, which we estimate using the BKS model, results in a force
\begin{align}
    f_\text{BKS}=\frac{\mu b^2}{2 \pi} \left\{ \ln \left[\left(\frac{b}{L-2r} + \frac{b}{2r}\right)^{-1}\right] + 0.7 \right\}\, .
\end{align}
Finally, these forces, which all act to collapse the loop, are balanced by the stacking fault force $\gamma_\text{APB}\ell(X)$. Force equilibrium then yields the following critical stress for loop collapse
\begin{align}
\tau_{LS} = {\frac{1}{b (L + 2r)}} \left\{\gamma_{APB} \, \ell(X) - \frac{\mu b^2}{2 \pi} \left[ \ln \left[\left(\frac{b}{L-2r} + \frac{b}{2r}\right)^{-1}\right] + 0.7 \right] - \frac{w(2-\nu)\mu b^2}{4 \pi  \, (1-\nu)}    \left[ \ln\left(\frac{4r}{b}\right) - 1 \right] \right\}\, .
\label{tauC_LS}
\end{align}
This equation models strengthening by the hybrid looping and shearing bypass mechanisms. It is plotted in Fig.~\ref{fig:looping_and_shearing} as a function of $f$ and for different values of $r$ chosen in the DDD simulation, and using $w=1.6$.  There is good agreement with the critical stress measured in our DDD simulations, and eq.~\eqref{tauC_LS}.

\subsection{Effects of a lattice misfit}
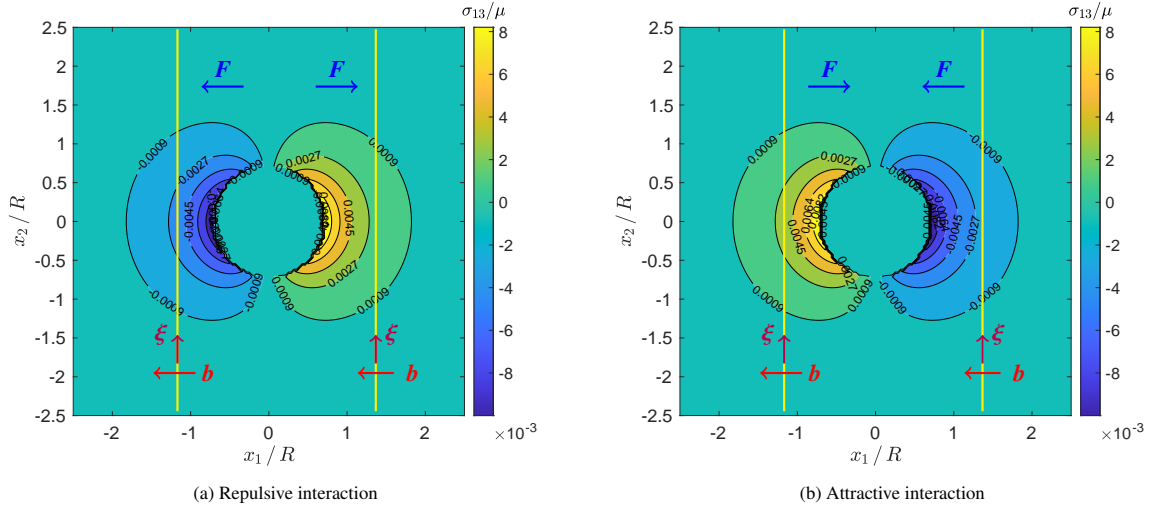
\begin{figure}[t!]
\centering
\begin{subfigure}{0.48\textwidth}
\begin{tikzpicture}
\node[anchor=south west,inner sep=0] (image) at (0,0) {\includegraphics[width=\linewidth]{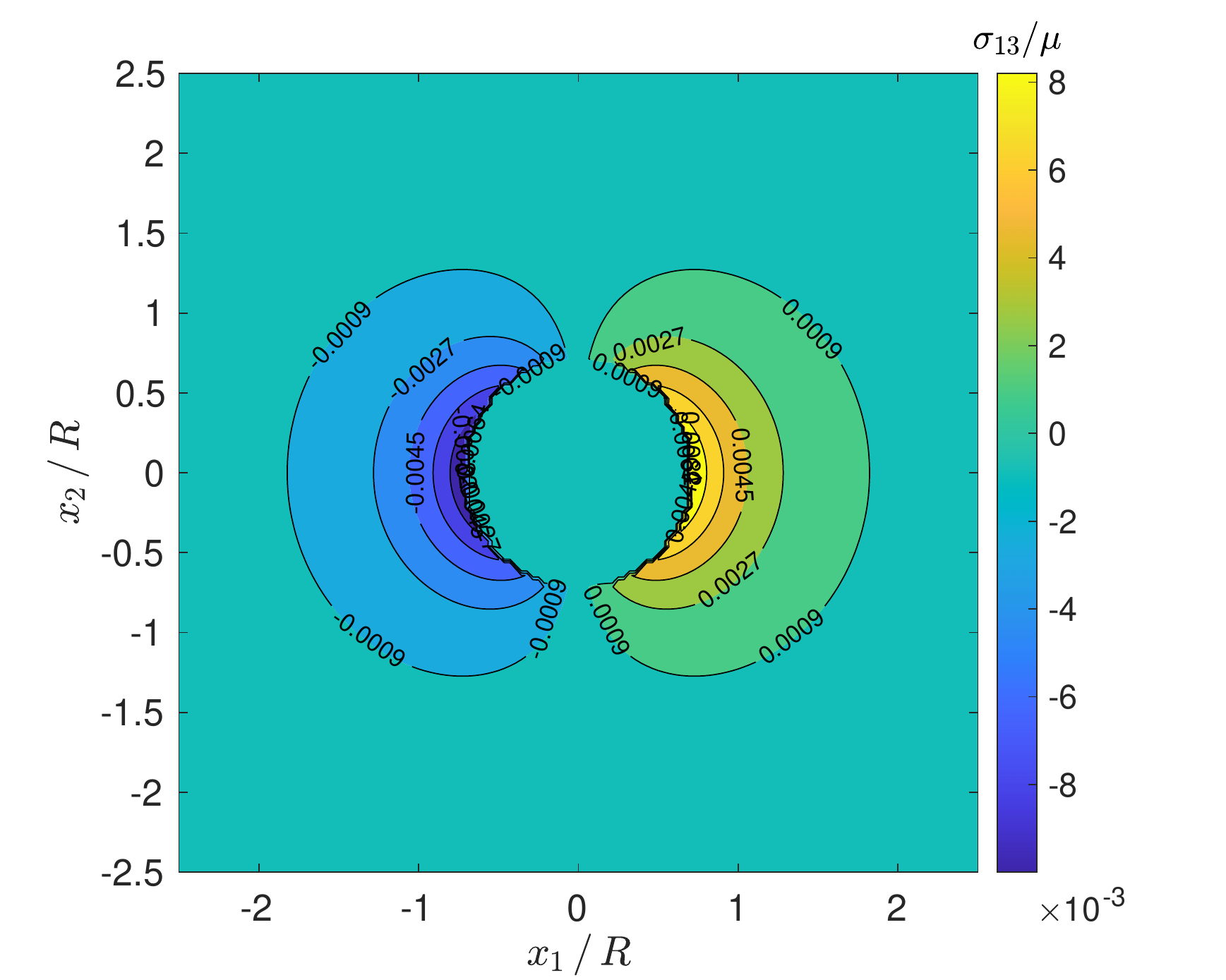}};
\begin{scope}[x={(image.south east)},y={(image.north west)}]
\draw [thick, yellow] (0.65,0.12)  -- (0.65,0.92); 
\draw [->,thick,purple] (0.65,0.22) -- (0.65,0.28) 
node[right,purple]{\small  $\bm \xi$}; 
\draw [->,thick,red] (0.68,0.2) -- (0.62,0.2) 
node[right,red,xshift=0.5cm]{\small  $\bm b$}; 
\draw [->,thick,blue] (0.55,0.8) -- (0.62,0.8) 
node[above,midway,blue]{\small  $\bm F$};
\draw [thick, yellow] (0.32,0.12)  -- (0.32,0.92); 
\draw [->,thick,purple] (0.32,0.22) -- (0.32,0.28) 
node[left,purple]{\small  $\bm \xi$}; 
\draw [->,thick,red] (0.35,0.2) -- (0.28,0.2) 
node[right,red,xshift=0.5cm]{\small  $\bm b$}; 
\draw [->,thick,blue] (0.43,0.8) -- (0.36,0.8) 
node[above,midway,blue]{\small  $\bm F$};
\end{scope}
\end{tikzpicture}
\caption{Repulsive interaction}
\label{fig:stress_e5}
\end{subfigure}
\begin{subfigure}{0.48\textwidth}
\begin{tikzpicture}
\node[anchor=south west,inner sep=0] (image) at (0,0) {\includegraphics[width=\linewidth]{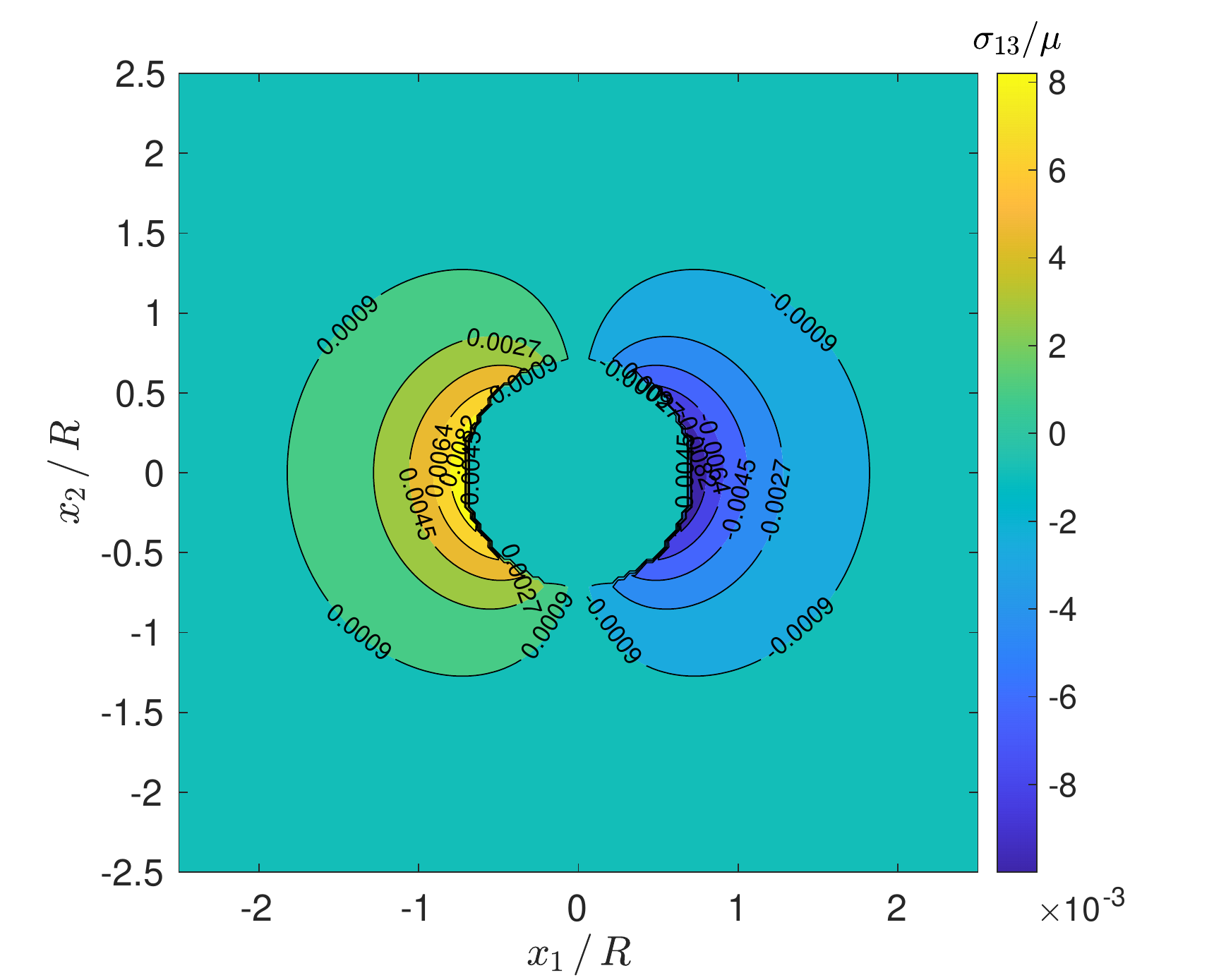}};
\begin{scope}[x={(image.south east)},y={(image.north west)}]
\draw [thick, yellow] (0.65,0.12)  -- (0.65,0.92); 
\draw [->,thick,purple] (0.65,0.22) -- (0.65,0.28) 
node[right,purple]{\small  $\bm \xi$}; 
\draw [->,thick,red] (0.68,0.2) -- (0.62,0.2) 
node[right,red,xshift=0.5cm]{\small  $\bm b$}; 
\draw [->,thick,blue] (0.62,0.8) -- (0.55,0.8) 
node[above,midway,blue]{\small  $\bm F$}; 
\draw [thick, yellow] (0.32,0.12)  -- (0.32,0.92); 
\draw [->,thick,purple] (0.32,0.22) -- (0.32,0.28) 
node[left,purple]{\small  $\bm \xi$}; 
\draw [->,thick,red] (0.35,0.2) -- (0.28,0.2) 
node[right,red,xshift=0.5cm]{\small  $\bm b$}; 
\draw [<-,thick,blue] (0.43,0.8) -- (0.36,0.8) 
node[above,midway,blue]{\small  $\bm F$};
\end{scope}
\end{tikzpicture}
\caption{Attractive interaction}
\label{fig:stress_e-5}
\end{subfigure}
\caption{Stress field of a  spherical precipitate with lattice misfit $|\delta|=0.5\%$  on the average intersection plane with height $h=\nicefrac{R}{4}\sqrt{16-\pi^2}\approx 0.62 R$ above the equatorial plane, where $R$ is the radius of the precipitate. The radius of the precipitate intersected by this plane is  $r=\nicefrac{\pi R}{4}$.  Contour lines of $\sigma_{13}$ component (in units of $\mu$) are shown.
 (\subref{fig:stress_e5}) Repulsive interaction with a dislocation.  The dislocation line is represented by the yellow vertical line. Its Burgers vector and line direction are marked as $\bm b$ and $\bm \xi$ respectively. In the given precipitate stress field, the Peach-Koehler force $\bm F$ repels the dislocation from the precipitate .  (\subref{fig:stress_e-5}) In the opposite stress field $\bm F$ attracts the dislocation to the precipitate.}
\label{fig:stressField_misfit}
\end{figure}

The bypass mechanisms considered so far are valid when precipitates possess negligible lattice and elastic misfit. The lattice misfit 
\begin{align}
\delta = 2\frac{(a_{\gamma'}-a_{\gamma})}{(a_{\gamma'}+a_{\gamma})}
\end{align}
is an important property in alloy design, since it controls the $\gamma/\gamma'$ interface energy and drives the precipitates coarsening kinetics \citep{reed_2006}. It also controls the precipitate morphology, with spherical precipitates typically observed for low misfit gradually giving way to cuboidal precipitates as $|\delta|$ approaches  $\approx 0.3\%$ \citep{pollock2006nickel}. In this section we consider the effects of a lattice misfit on the three types of bypass mechanisms described in section \ref{sec:result_dislocation_pair}. In our analysis we make the assumption that the elastic misfit is negligible, and that precipitates remain spherical. Under these assumptions, the presence of a lattice misfit results in an additional stress field that can can be calculated using the theory of Eshelby's spherical inclusion \citep{mura2013micromechanics}, as detailed in \ref{EshelbyInclusion}.
Note that the resolved shear stress on any equatorial plane of the spherical precipitates vanishes identically. However, dislocations encounter precipitates at a average height $h=\nicefrac{R}{4}\sqrt{16-\pi^2}\approx 0.62 R$ from the equatorial plane, which corresponds to the average intersection radius $r=\nicefrac{\pi R}{4}$. The resolved shear stress on this plane is shown in Fig.~\ref{fig:stressField_misfit}. The sign of the misfit stress depends on the sign of $\delta$, and it flips both across the precipitate on the glide plane, and depending on whether the glide plane is  above or below the equatorial plane. In any case, a given dislocation may experience either a repulsive or attractive interaction. In our analysis, we choose to vary $\delta$ in  the range  $-0.5\%\le\delta\le0.5 \%$, and keep the height and side of the intersection, and the ``sign" of the dislocation fixed. With this choice, in our analysis a positive misfit corresponds to a repulsive interaction, while a  negative misfit corresponds to an attractive interaction. The misfit stress inside the precipitate is constant and equal to zero. Let us denote the bypass stress required for the dislocation pair to overcome the precipitate square lattice with zero misfit as $\sigma_{0}$. For each of the three types of bypass mechanisms described in section \ref{sec:result_dislocation_pair}, we compute the additional stress $\Delta\sigma$ necessary to bypass the same precipitate square lattice, and analyze the changes in the bypass mechanism as a function of the misfit $\delta$. 

\begin{figure}[t!]
\centering
\begin{subfigure}{.45\textwidth}
   \centering
    \scalebox{1.6}
    {\includegraphics[width=0.65\textwidth]{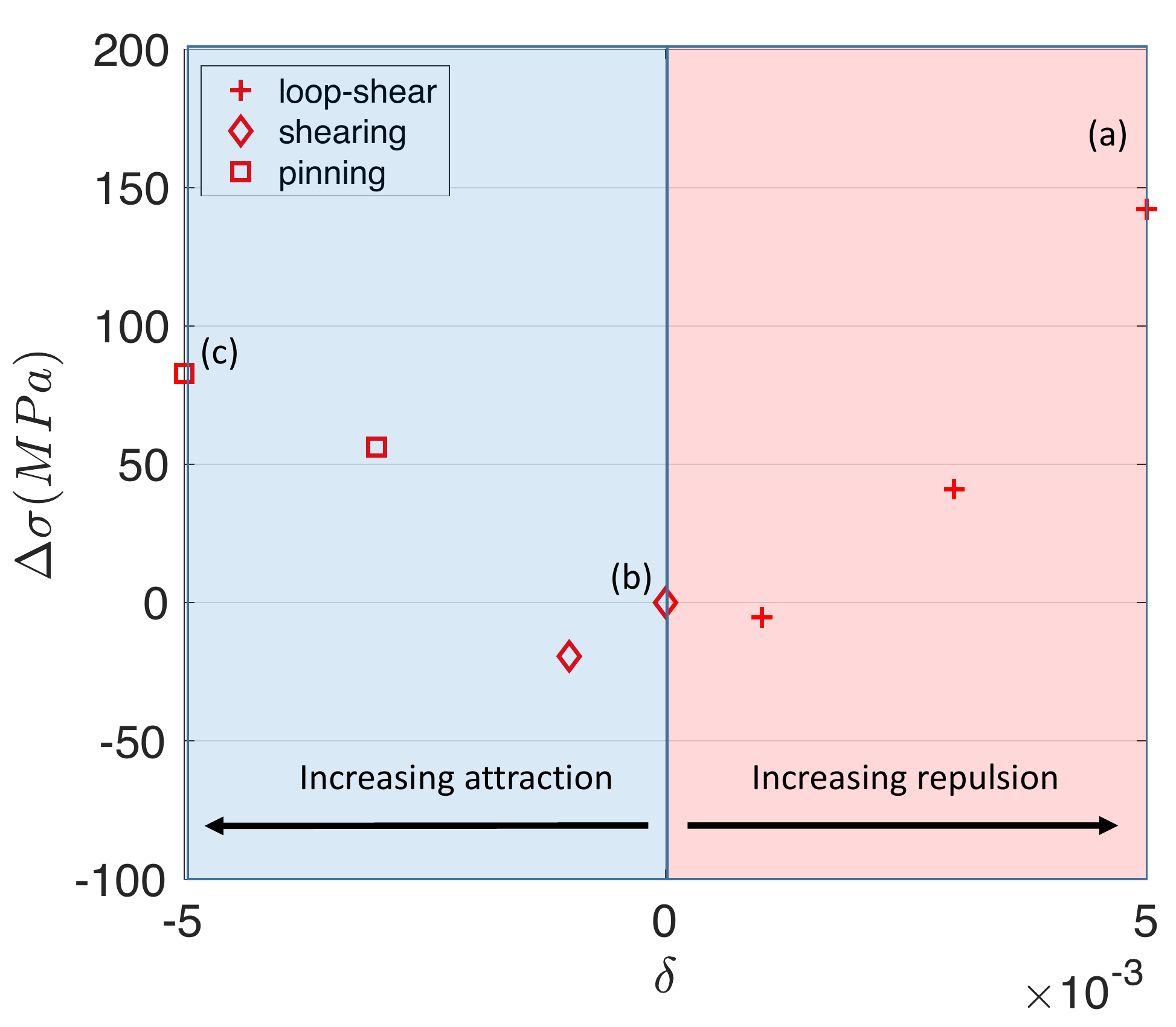}}
\end{subfigure}

\vspace{0.5cm}

\begin{subfigure}{\misfitSnapWidth\textwidth}
   \centering
    \includegraphics[width=\misfitSnapWidthInside\textwidth]{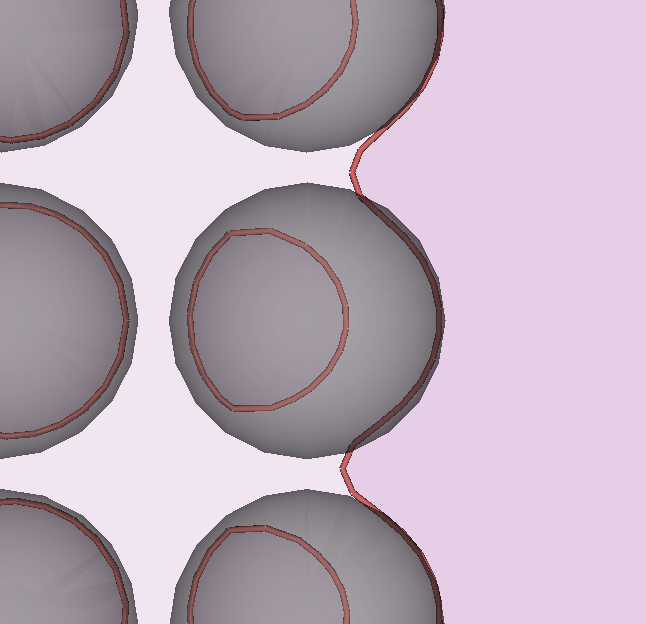}
     \caption{}
     \label{fig:ss_r100_e5}
\end{subfigure}
\hspace{0.5cm}
\begin{subfigure}{\misfitSnapWidth\textwidth}
   \centering
    \includegraphics[width=\misfitSnapWidthInside\textwidth]{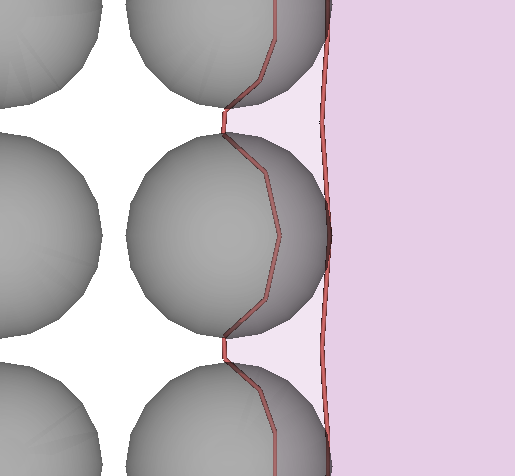}
     \caption{}
     \label{fig:ss_r100_e0}
\end{subfigure}
\hspace{0.5cm}
\begin{subfigure}{\misfitSnapWidth\textwidth}
   \centering
    \includegraphics[width=\misfitSnapWidthInside\textwidth]{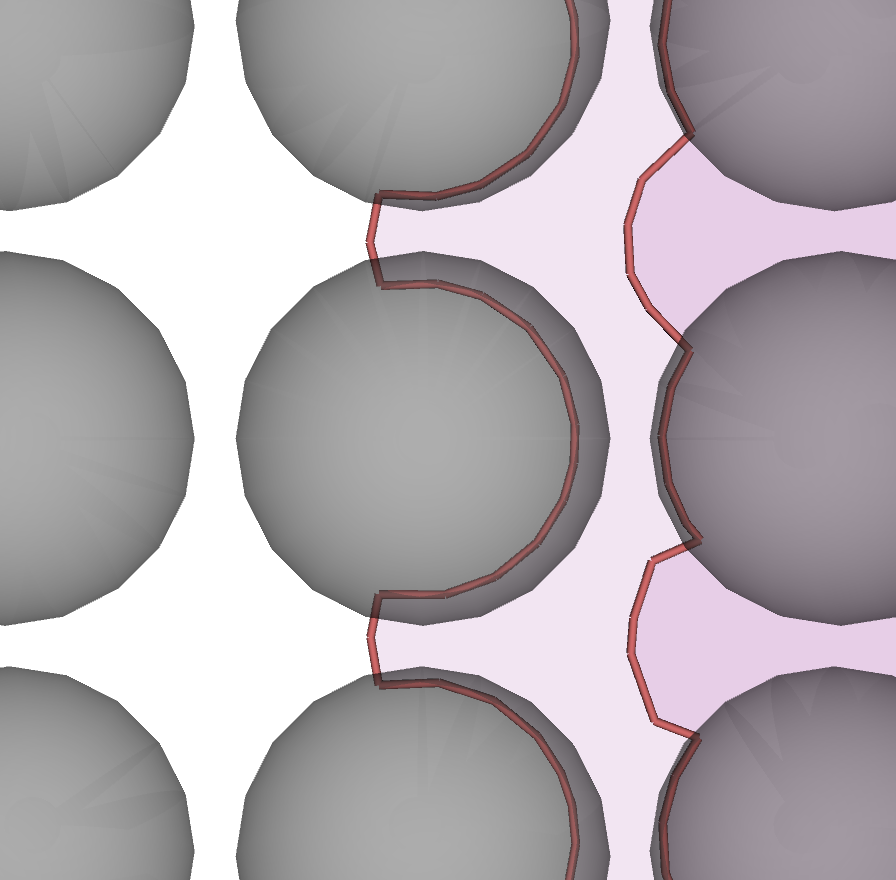}
     \caption{}
     \label{fig:ss_r100_e-5}
\end{subfigure}
\caption{Effect of a misfit on the bypass stress in the shearing regime ($r=78.5 \, \text{nm}$ and $f=0.69$).  (\subref{fig:ss_r100_e5}) Critical configuration with a repulsive interaction due to $\delta=0.5\%$, showing the hybrid looping-shearing bypass  mechanism. (\subref{fig:ss_r100_e0}) Critical configuration with zero misfit strain (baseline), showing the shearing bypass mechanisms. (\subref{fig:ss_r100_e-5}) Critical configuration with an attractive interaction due to $\delta=-0.5\%$, showing  pinning of the trailing dislocation.}
\label{fig:tc_misfit_doubleShear}
\end{figure}

We first consider the shearing regime. We select a precipitates square lattice with parameters $r=78.5 \, \text{nm}$ and $f=0.69$ which lies in this regime, and study the bypass mechanisms and strength as a function of $\delta$. Our simulation results are reported in Fig.~ \ref{fig:tc_misfit_doubleShear}. For increasing repulsive interaction (increasing positive misfit), the bypass mechanism changes from shearing to the hybrid looping and shearing mechanisms discussed in the previous section. This is because the initial repulsive interaction between the leading dislocation on the enter-side of the precipitates prevents the leading from shearing the precipitates and facilitates the Orowan looping process. In addition, the opposite force on the exit side of the precipitate  facilitates the closure of the leading loop (Fig.~\ref{fig:ss_r100_e5}, $\delta=0.5\%$). The overall effect is a significant increase in the bypass stress. On the other hand, in the case of an attractive interaction, shearing is facilitated for both leading and trailing dislocations. However, the critical bypass configuration is controlled by the attractive interaction on the exit-side of the precipitates, which pins the trailing dislocation causing a significant increase in the bypass stress (Fig.~\ref{fig:ss_r100_e-5}, $\delta=-0.5\%$). Hence a lattice misfit in general increases the bypass stress in the shearing regime.

\begin{figure}[t!]
\centering
\begin{subfigure}{.45\textwidth}
   \centering
    \scalebox{1.6}
    {\includegraphics[width=0.65\textwidth]{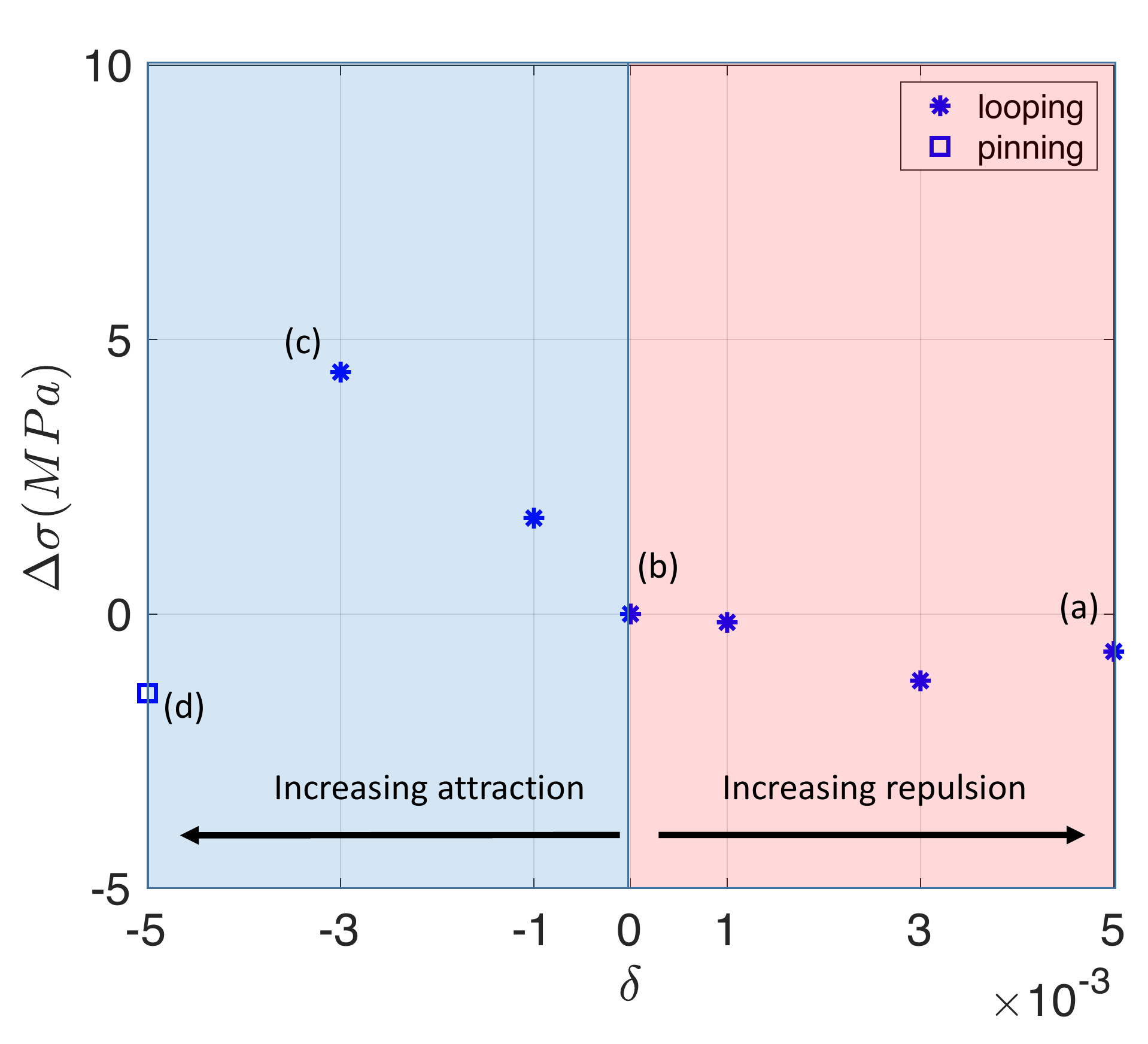}}
    \label{fig:tc_misfit_doubleLoop}
\end{subfigure}%
\hspace{0.5cm}
\begin{subfigure}{\misfitSnapWidth\textwidth}
   \centering
    \includegraphics[width=\misfitSnapWidthInside\textwidth]{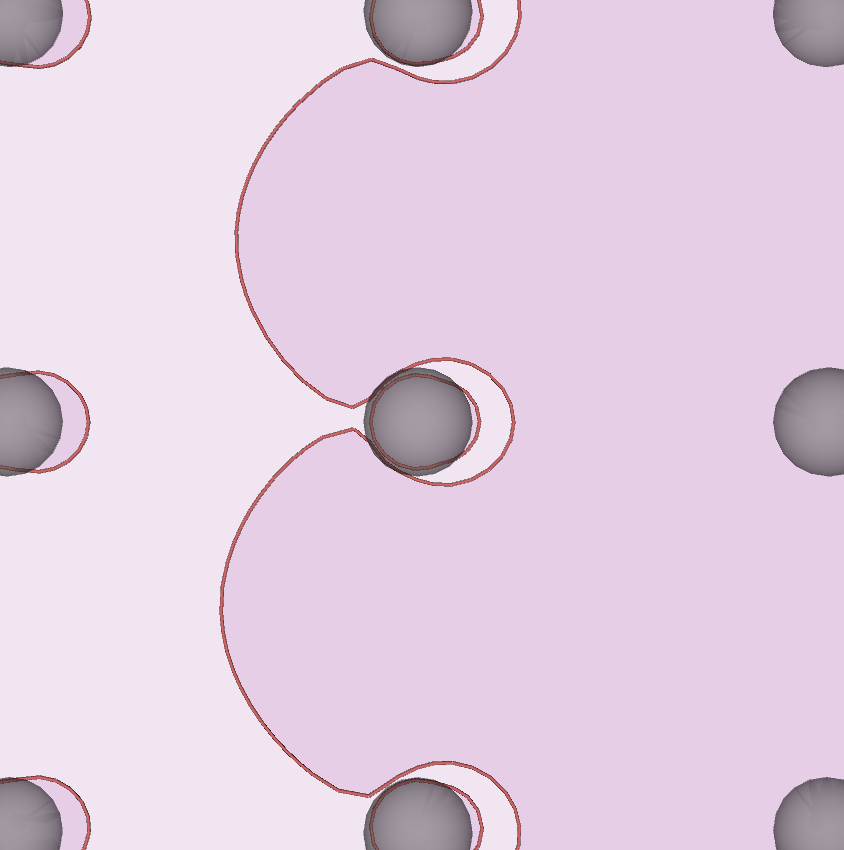}
     \caption{}
     \label{fig:ll_r300_e5}
\end{subfigure}

\vspace{0.5cm}

\centering
\begin{subfigure}{\misfitSnapWidth\textwidth}
   \centering
    \includegraphics[width=\misfitSnapWidthInside\textwidth]{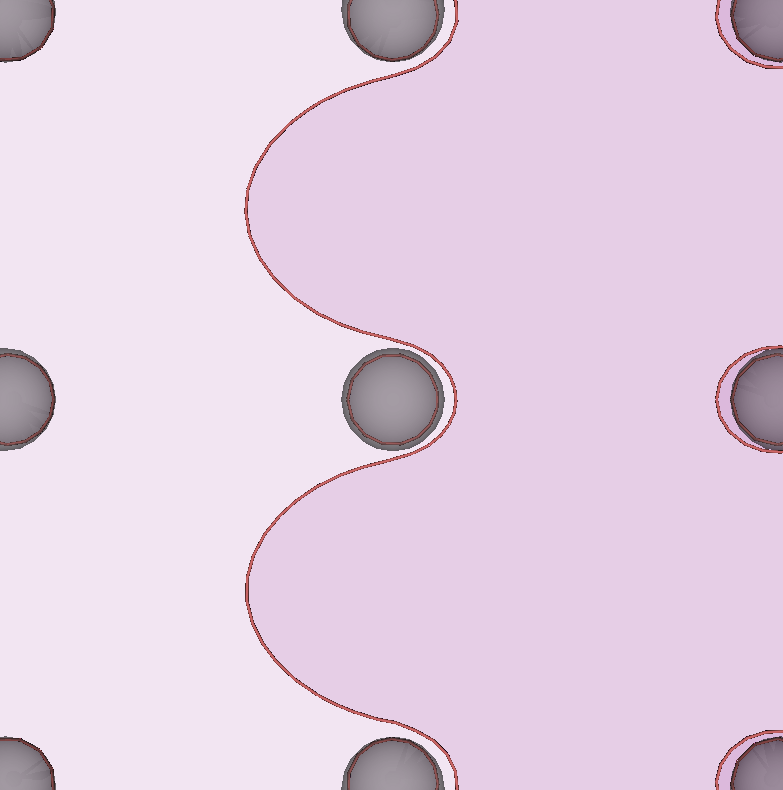}
     \caption{}
     \label{fig:ll_r300_e0}
\end{subfigure}
\hspace{0.5cm}
\begin{subfigure}{\misfitSnapWidth\textwidth}
   \centering
    \includegraphics[width=\misfitSnapWidthInside\textwidth]{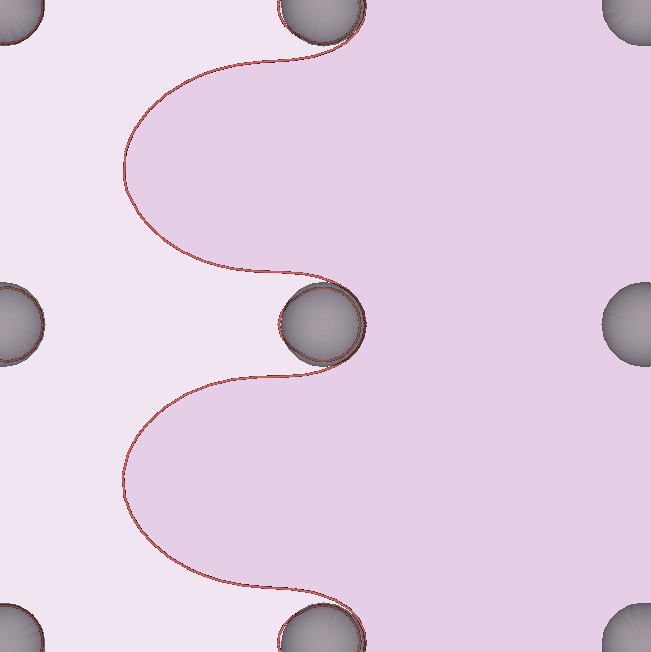}
     \caption{}
     \label{fig:ll_r300_e-3}
\end{subfigure}
\hspace{0.5cm}
\begin{subfigure}{\misfitSnapWidth\textwidth}
   \centering
    \includegraphics[width=\misfitSnapWidthInside\textwidth]{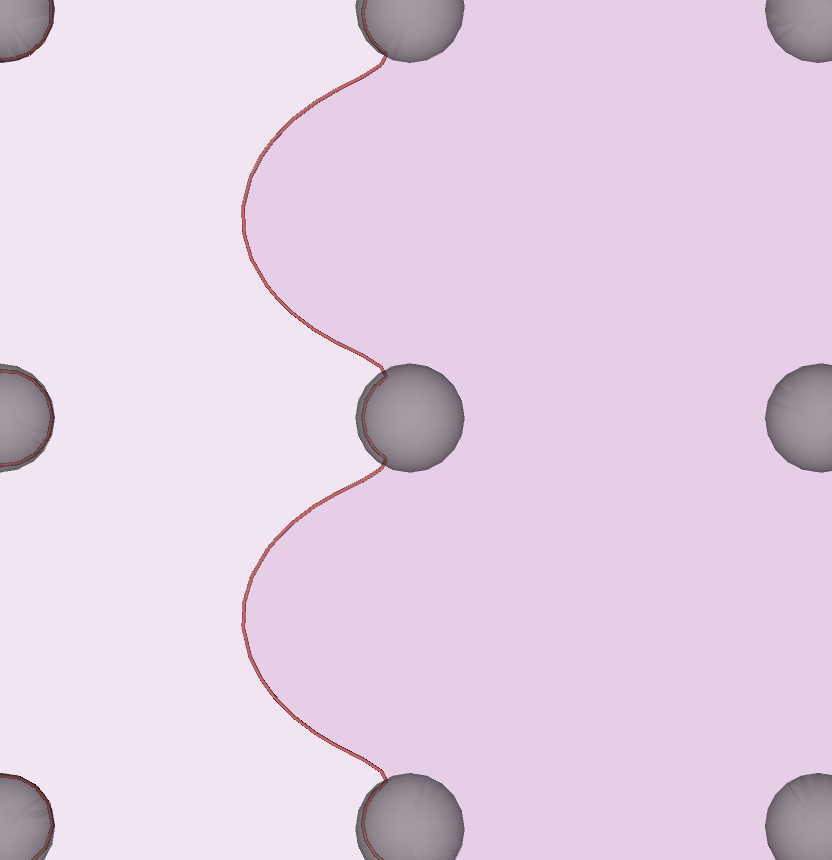}
     \caption{}
     \label{fig:ll_r300_e-5}
\end{subfigure}
\caption{Effect of a lattice misfit on the bypass  stress in looping regime ($r=235.62 \, \text{nm}$ and $f=0.06$). (\subref{fig:ll_r300_e5}) Critical configuration with a repulsive interaction due to $\delta=0.5\%$, showing the looping  mechanism. (\subref{fig:ll_r300_e0}) Critical configuration with zero misfit strain (baseline), showing the looping mechanism. (\subref{fig:ll_r300_e-3}) Critical configuration with an attractive interaction due to $\delta=-0.3\%$, showing the looping mechanism. 
(\subref{fig:ll_r300_e-5}) Critical configuration with an attractive interaction due to $\delta=-0.5\%$, showing the pinning of the trailing dislocation.}
\label{fig:misfit_doubleLoop}
\end{figure}

Next we consider the looping regime, with representative precipitate square lattice parameters  $r=235.62 \, \text{nm}$ and $f=0.06$. Simulation results are summarized in Fig.~\ref{fig:misfit_doubleLoop}. For a repulsive interaction (positive misfit here), the bypass mechanism is fundamentally unaltered, but the final state of the system may be significantly different. The critical stress still corresponds to the configuration where the trailing dislocation becomes unstable as it loops around the precipitates surrounded by the loops left behind by the leading dislocation. However, the positive misfit creates attractive forces  on the back side of the precipitates (Fig. \ref{fig:ll_r300_e5},  $\delta=0.5\%$), which help in completing the looping process of the trailing dislocation. Compared to an absence of misfit  (Fig. \ref{fig:ll_r300_e0}), the shape of the bowed-out trailing dislocation reflects the attraction on the back side of the precipitates. Hence a positive misfit results in a decrease of the bypass stress, although this effect is very mild. The most striking difference, however, is that upon completion of the loop process, the attractive stress on the back side of the precipitate may be sufficient to cause the collapse of both loops. Note that there are no concentric loops in the rightmost column of precipitates in Fig.~\ref{fig:ll_r300_e5}.
For intermediate attractive interaction (Fig. \ref{fig:ll_r300_e-3}, $\delta=-0.3\%$), there is a repulsive stress on the back side of the precipitate, which opposes the advancement of the trailing dislocation and causes an increase of the stress necessary to close the trailing loop. Even in this case, once the trailing loop is able to close, the two concentric loops  may collapse due to the attractive misfit stress on the entry side of the precipitate. Note the absence of concentric loops in the rightmost column of precipitates in Fig.~\ref{fig:ll_r300_e-3}. For even larger attractive interaction (Fig.~\ref{fig:ll_r300_e-5}, $\delta=-0.5\%$), the attractive stress on the entry side of the precipitates is sufficient to collapse the leading loop before the trailing dislocation completes the looping process. As a consequence, the critical stress reduces and the bypass process becomes similar to the   hybrid looping and shearing mechanism observed in the absence of a misfit. However, due to the high attractive stress the trailing dislocation remains pinned at the precipitate on the exit side, and this is the configuration corresponding to the critical bypass stress. Note that the effect of misfit stress is mild in this regime due to the presence of large channel width between the precipitates, which mitigates the effect of the attractive/repulsive interactions.

\begin{figure}[!h]
\centering
\begin{subfigure}{.45\textwidth}
   \centering
    \scalebox{1.6}
    {\includegraphics[width=0.65\textwidth]{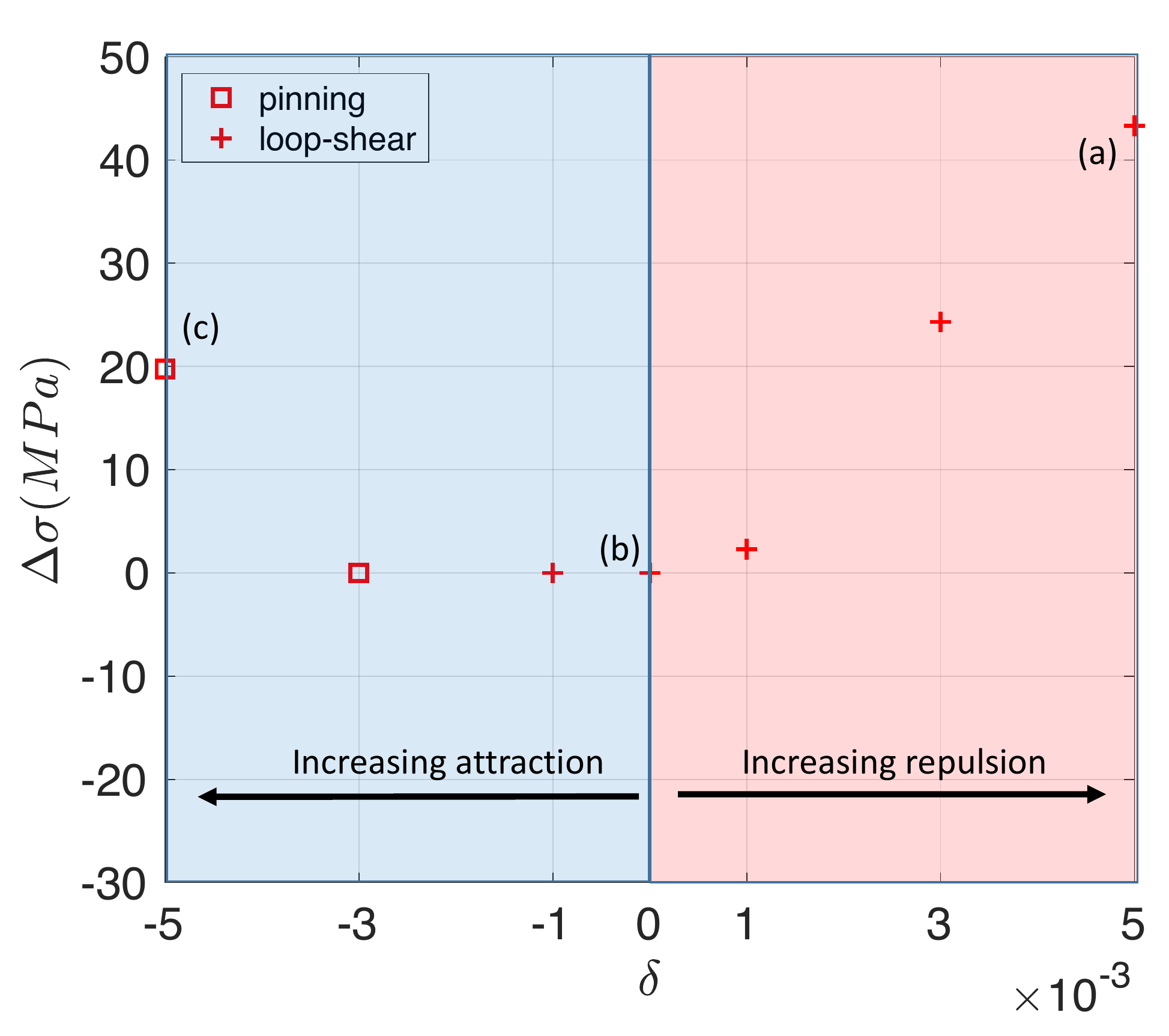}}
\end{subfigure}%

\vspace{0.5cm}

\centering
\begin{subfigure}{\misfitSnapWidth\textwidth}
   \centering
    \includegraphics[width=\misfitSnapWidthInside\textwidth]{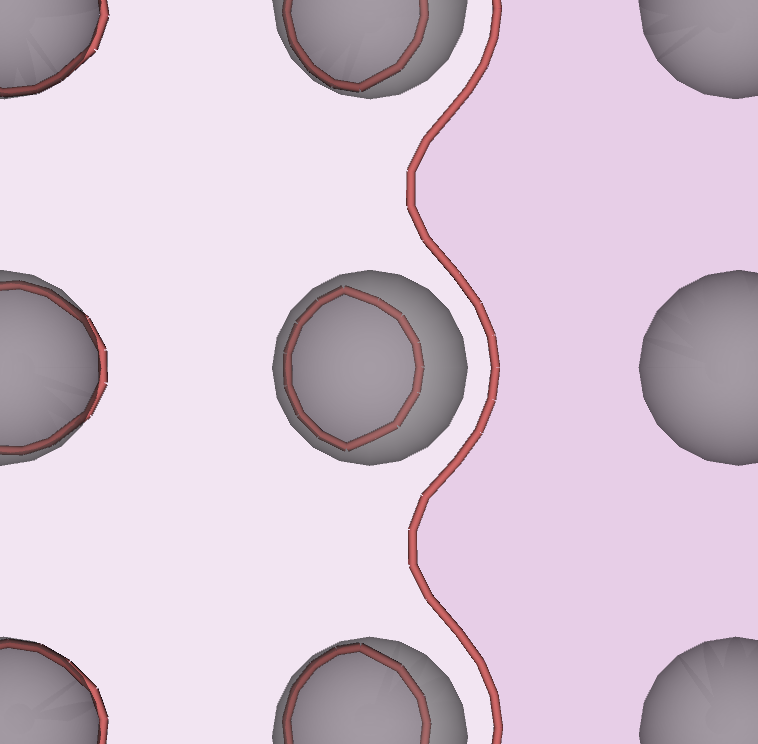}
     \caption{}
     \label{fig:ls_r100_e5}
\end{subfigure}
\hspace{0.5cm}
\begin{subfigure}{\misfitSnapWidth\textwidth}
   \centering
    \includegraphics[width=\misfitSnapWidthInside\textwidth]{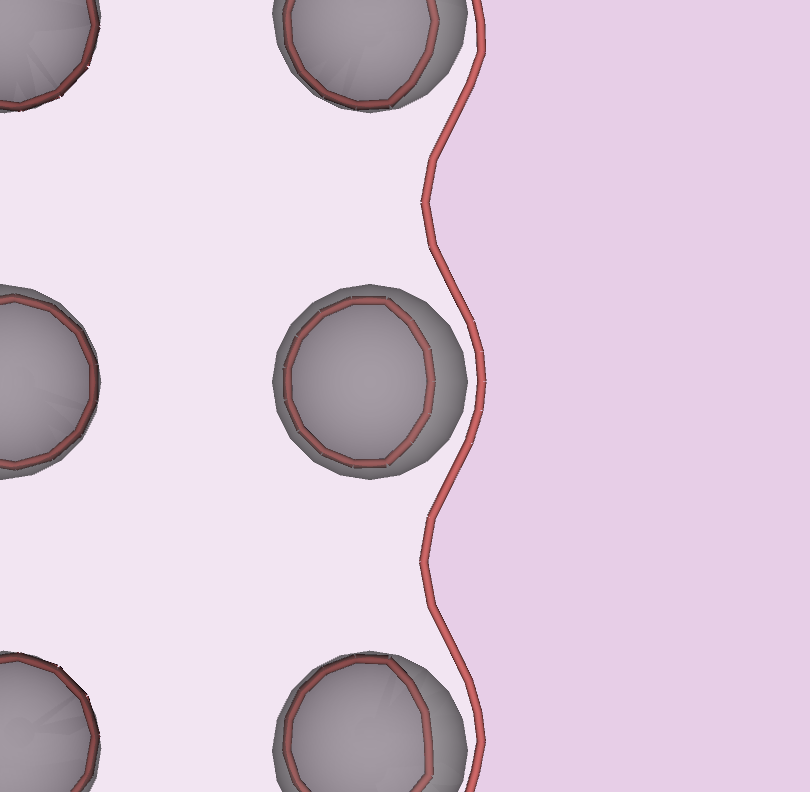}
     \caption{}
     \label{fig:ls_r100_e0}
\end{subfigure}
\hspace{0.5cm}
\begin{subfigure}{\misfitSnapWidth\textwidth}
   \centering
    \includegraphics[width=\misfitSnapWidthInside\textwidth]{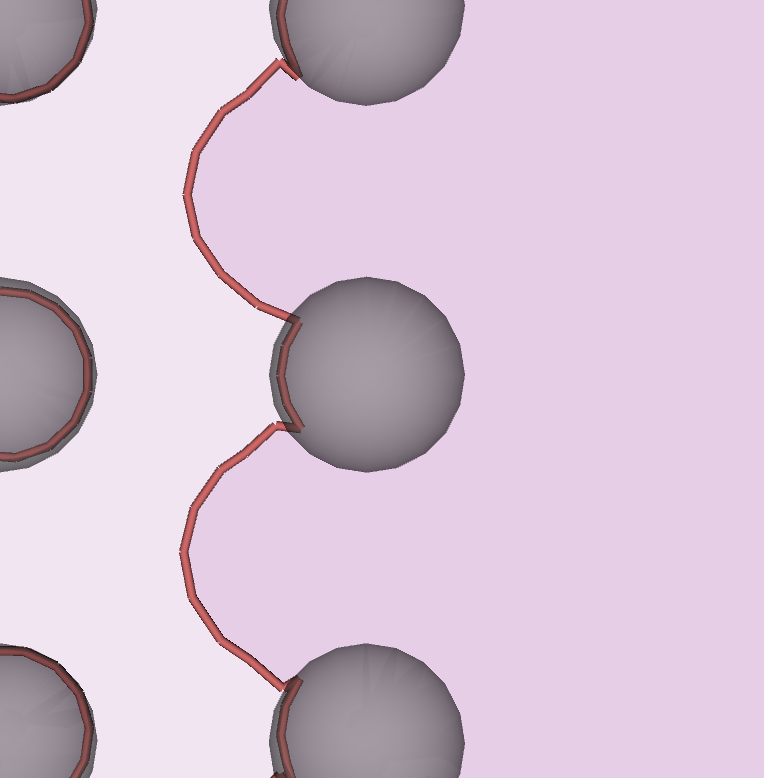}
     \caption{}
     \label{fig:ls_r100_e-5}
\end{subfigure}
\caption{Effect of a lattice misfit on the bypass stress in the hybrid looping-shearing regime for ($r=78.53 \, \text{nm}$ and $f=0.24$). (\subref{fig:ls_r100_e5}) Critical configuration with a repulsive interaction due to $\delta=0.5\%$, showing the hybrid looping-shearing   mechanism. (\subref{fig:ls_r100_e0}) Critical configuration with zero misfit strain (baseline), showing the hybrid looping-shearing mechanism. (\subref{fig:ls_r100_e-5}) Critical configuration with an attractive interaction due to $\delta=-0.5\%$, showing the pinning of the trailing dislocation. }
\label{fig:tc_misfit_loopShear}
\end{figure}
Finally, we consider the hybrid looping and shearing regime, with  selected representative precipitate square lattice parameters  $r=78.53 \, \text{nm}$ and $f=0.24$. For a repulsive interaction (positive misfit in Fig.\ref{fig:tc_misfit_loopShear}), the bypass stress  increases with increasing misfit. This is because the trailing dislocation experience a repulsive force on the entry side, and because the loop left behind by the leading dislocation is also shifted  towards the entry side causing an even greater repulsion (see Fig. \ref{fig:ls_r100_e5},  $\delta=0.5\%$).  An attractive interaction, on the other hand, helps collapsing the leading loop. The effect on the bypass stress is mild until the attractive stress on the exist side pins the trailing dislocation and becomes the dominant factor in determining the critical configuration (see Fig. \ref{fig:ls_r100_e-5},  $\delta=-0.5\%$). Also in this regime, therefore, a lattice misfit increases the bypass stress. 

In summary, for intermediate to high volume fraction, as is the case in the shearing and hybrid looping-shearing regimes,  dislocation motion is hindered by a lattice misfit, independently of the type of interaction (attractive or repulsive). This phenomenon was also observed by \cite{liu_misfit_pdd} who reported hardening in the stress-strain response in the presence of  precipitates with a lattice misfit. In a random distribution of precipitates, dislocations encounter an equal number of attractive and repulsive precipitate interactions. Hence, the presence of pinned dislocations at precipitates becomes a characteristic signature of strengthening by a lattice misfit. A similar strengthening by dislocation pinning was also proposed in the high-temperature creep literature \citep{arzt2001interface}, when dislocation climb becomes the dominant bypass mechanisms around precipitates.

\section{Summary and Conclusions\label{summary}}

In this work we have developed a new method to compute the forces exerted by generalized stacking faults on dislocations within the DDD framework. The method is based on a concept borrowed from complex analysis, and known as the \textit{winding number} of a closed curve about a point. Based on this quantity, the slip vector an any point on a slip plane can be computed if the loop topology of the dislocation network is preserved during the simulation. The method is general, and it applies seamlessly to Shockley partial dislocations in the fcc $\gamma$ matrix,  to superpartial dislocations in the $\text{L}1_2$ $\gamma'$ phase, or to  more complex situations such as $a\langle112\rangle\{111\}$ dislocation ribbons  observed in Ni-based superalloys. 

We have applied this method to investigate the mechanisms of bypass of a square lattice of spherical $\gamma'$ precipitates by pairs of $a/2\langle110\rangle\{111\}$ edge dislocations. The possible bypass mechanisms  were explored as a function of two microstructural parameters, namely the average precipitate radius on the glide plane $r$, and the precipitate volume fraction $f$.  DDD simulations were performed on a grid of points with $25\text{nm}\le r\le 400\text{nm}$ and $0.015\le f \le 0.687$. A map of bypass mechanism was produced as a function of these two parameters. Consistently with the general theory of precipitation strengthening by athermal bypass mechanisms, the low $r$ and high $f$ regions on this map are controlled  by APB shearing, while the high $r$ and low $f$ region is controlled by the Orowan looping process. However, our simulations indicate that the shearing and looping regime are separated by a  regime where a different ``hybrid" (looping-shearing) bypass mechanisms controls the bypass stress. This mechanism operates in the way described by  \cite{nabarro1995physics}, and it was indirectly observed by \cite{viswanathan2005investigation} and  \cite{unocic2008mechanisms} in superalloy Ren\'e 88 DT.  In this hybrid mechanism   the leading dislocation forms  stable loops around  precipitates while the approaching trailing dislocation drives these loops into the precipitates, forming  APB, and then quickly shears them removing the APB. The analysis of the critical dislocation configuration  during the bypass process also allowed us to derive an expression for the strength of this mechanism, which is in good agreement with the results of our DDD simulations.


We also studied the effect of a $\gamma/\gamma'$ lattice misfit on the bypass mechanisms by introducing a precipitate stress according to Eshelby's inclusion theory. Because dislocations statistically intercept precipitates on planes which are not equatorial planes, they experience either attractive or repulsive  interactions with the spherical precipitates. We observed that  both attractive and repulsive interactions lead to an increased   bypass stress in the intermediate/high volume fraction regime, where some type of precipitates shearing is involved. In the low volume fraction regime, the large channel width mitigates the effects of misfit stresses. It was also found that a large lattice misfit may significantly alter the mechanisms of precipitates bypass, and hence the  dependence of the bypass stress on the microstructural parameters $r$ and $f$. In particular, for strongly-attractive interactions, the critical bypass configuration is often controlled by \textit{pinning} of the trailing dislocation at the precipitates. In a random distribution of precipitates, dislocations encounter a statistically equal number of attractive and repulsive precipitate interactions. Hence, the presence of pinned dislocations at precipitates becomes a characteristic signature of strengthening by a lattice misfit. Interestingly, a similar strengthening mechanism by dislocation pinning is also reported in the high-temperature creep literature \citep{arzt2001interface}, when dislocation climb becomes the dominant bypass mechanisms around precipitates. It is therefore possible that, for sufficiently large lattice misfit, pinning becomes the dominant strengthening mechanism independently of the bypass mechanism (shearing, looping or climb) operating in different regimes of creep deformation. 

\section*{Acknowledgments}
 G. Po and S. Chatterjee acknowledge the U.S. Advanced Research Projects Agency–Energy (ARPA-E) under award number DE-SC0019157 with UCLA, and the U.S.  Department of Energy, Office of Fusion Energy Sciences (FES), under award number DE-SC0019157 with UCLA,  and related sub-awards with the University of Miami.

\appendix

\section{Calculation of \texorpdfstring{$\gamma$}{} surfaces}\label{app:calc_gamma_surf}
\setcounter{table}{0}
\setcounter{figure}{0}

In this work we use an analytical expression for the generalized stacking fault energy in the form
\begin{align}\label{eq:gammaSurf}
\gamma(\bm s)=\sum_{i}\sum_j S_i \, \sin(\bm k_i \cdot \bm M_j\bm s) + C_i \, \cos(\bm k_i \cdot \bm M_j\bm s), 
\end{align}
where $\bm s$ is a local (two-dimensional) slip vector on the glide plane, the $\bm k_i$'s are wave vectors, and the $\bm M_j$ are matrices representing the symmetries of the lattice. Let $\bm A$ be  the $2\times2$ matrix having in columns the basis vectors $\bm a_1$ and $\bm a_2$ of the glide plane lattice. Then the matrix $\bm B=2\pi \bm A^{-T}$ is the matrix having as columns the basis vectors $\bm b_1$ and $\bm b_2$ of the corresponding reciprocal lattice. The Born-Von Karman periodicity condition implies that the wave vectors $\bm k$'s have the form $\bm k=\bm B\bm n$, where $\bm n$ is a vector of integers.

\begin{figure}[!h]
\centering
\begin{subfigure}{.4\textwidth}
  \centering
  \includegraphics[width=\linewidth]{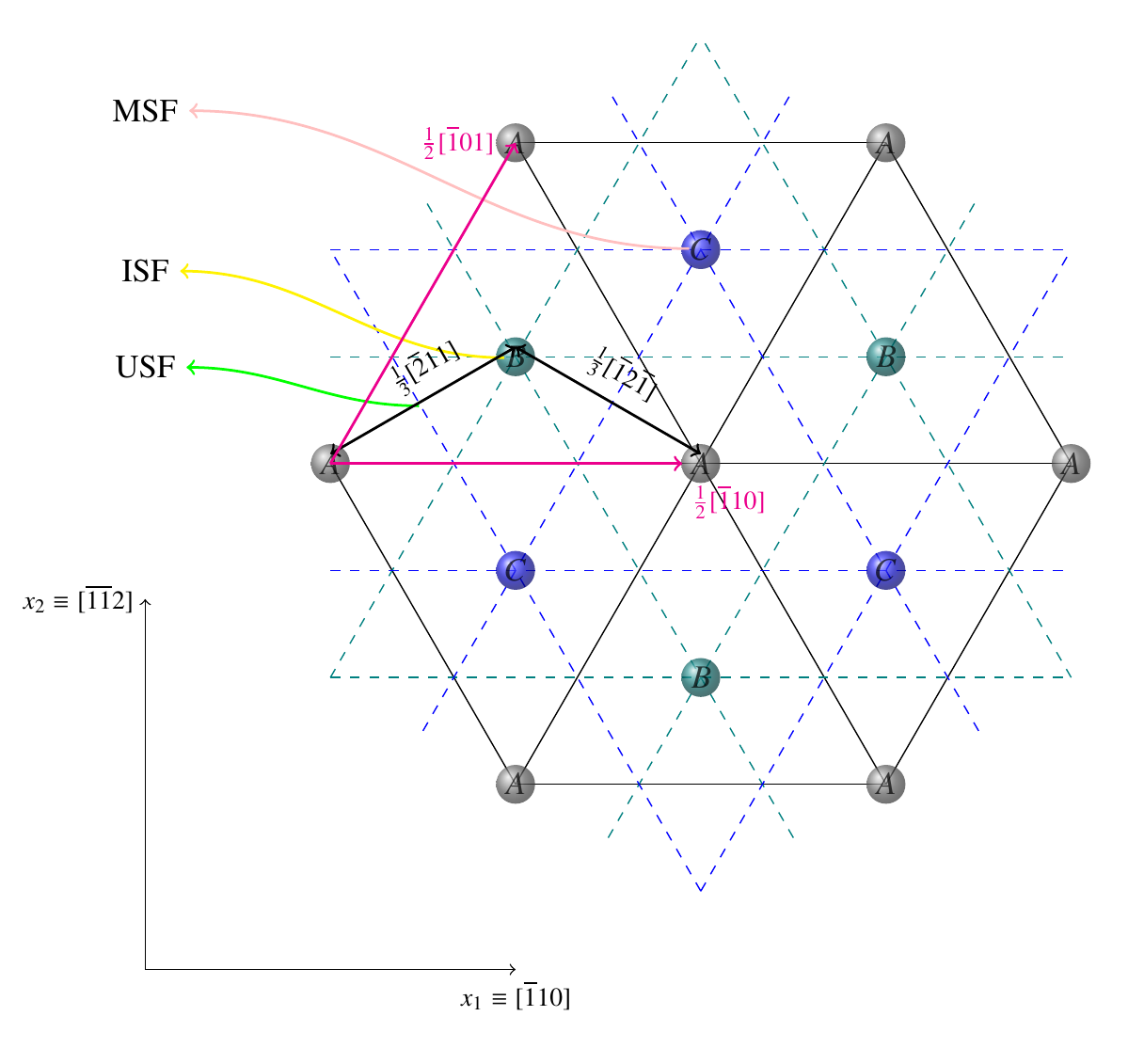}
  \caption{}
  \label{fig:fcc_ni}
\end{subfigure}%
\hfill
\begin{subfigure}{.32\textwidth}
\centering
\begin{tabular}[H]{|c|c|}
\hline
$\bm s_k$ &   $\gamma_k$ \\
\hline
$(0,0)$     &  $0$  \\
$(0.25,\frac{\sqrt{3}}{12})$     &  $\gamma_{USF}$  \\
$(0.5,\frac{\sqrt{3}}{6})$     &  $\gamma_{ISF}$  \\
$(1.0,\frac{\sqrt{3}}{3})$     &  $\gamma_{MSF}$  \\
\hline
\end{tabular}
	\caption{}
	  \label{fig:fcc_ni_conditions}
\end{subfigure}\hfill
\begin{subfigure}{.2\textwidth}
  \centering
  \includegraphics[width=\linewidth]{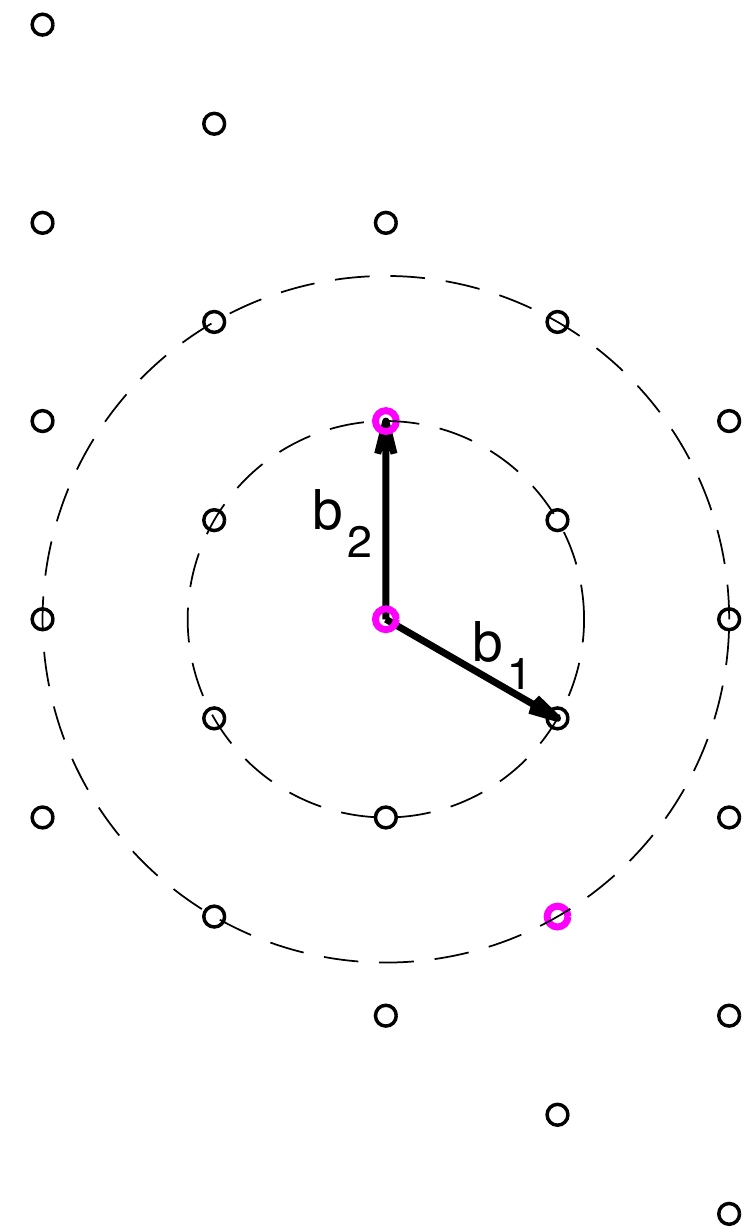}
  \caption{}
  \label{fig:fcc_ni_kpts}
\end{subfigure}\\
\begin{subfigure}{.48\textwidth}
\centering
	\begin{tabular}[H]{|c|c|c|}
		\hline
		wave vector $\bm n$ & $S$ [mJ\, m$^{-2}$]& $C$ [mJ\, m$^{-2}$]\\
		\hline
		$(0,0)$ & 0& 449.1 \\
		$(0,1)$ & -511.5&-326.1 \\
		$(1,-1)$ &0&-122.2 \\
		\hline
	\end{tabular}
	\caption{}
	  \label{fig:fcc_ni_SC}
\end{subfigure}\hfill
\begin{subfigure}{.48\textwidth}
  \centering
  \includegraphics[width=\linewidth]{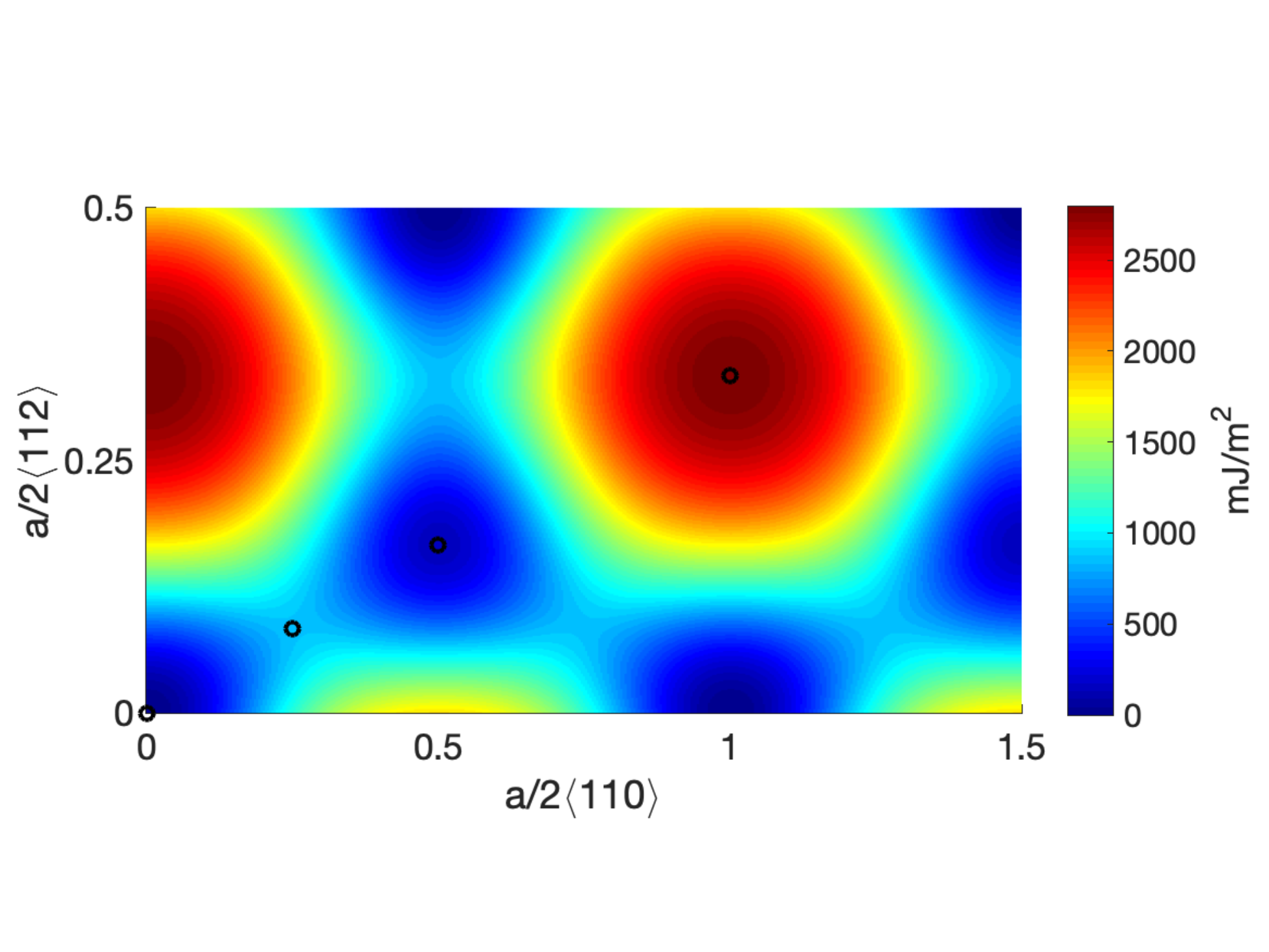}
  \caption{}
  \label{fig:gammaSurf_gamma}
\end{subfigure}%
\caption{Stacking fault energy for the $\gamma$ phase. (\subref{fig:fcc_ni}-\subref{fig:fcc_ni_conditions}) Structure of the $\{111\}$ plane,
and imposed fitting conditions  The value $\gamma_{ISF}=142~mJ/m^2$ is in good agreement with experimental measurements. The values $\gamma_{USF}=863~mJ/m^2$ and $\gamma_{MSF}=2200~mJ/m^2$ are estimated from atomistic simulations (\cite{wei_sfe}), which ensures the accurate description of the entire $\gamma$ surface. (\subref{fig:fcc_ni_kpts}-\subref{fig:fcc_ni_SC}) Selected wave vectors (magenta circles) in the reciprocal lattice of the $\{111\}$ plane, and corresponding fitting constants. (\subref{fig:gammaSurf_gamma}) The resulting $\gamma$ surface.
}
\end{figure}

\begin{figure}[!h]
\centering
\begin{subfigure}{.4\textwidth}
  \centering
  \includegraphics[width=\linewidth]{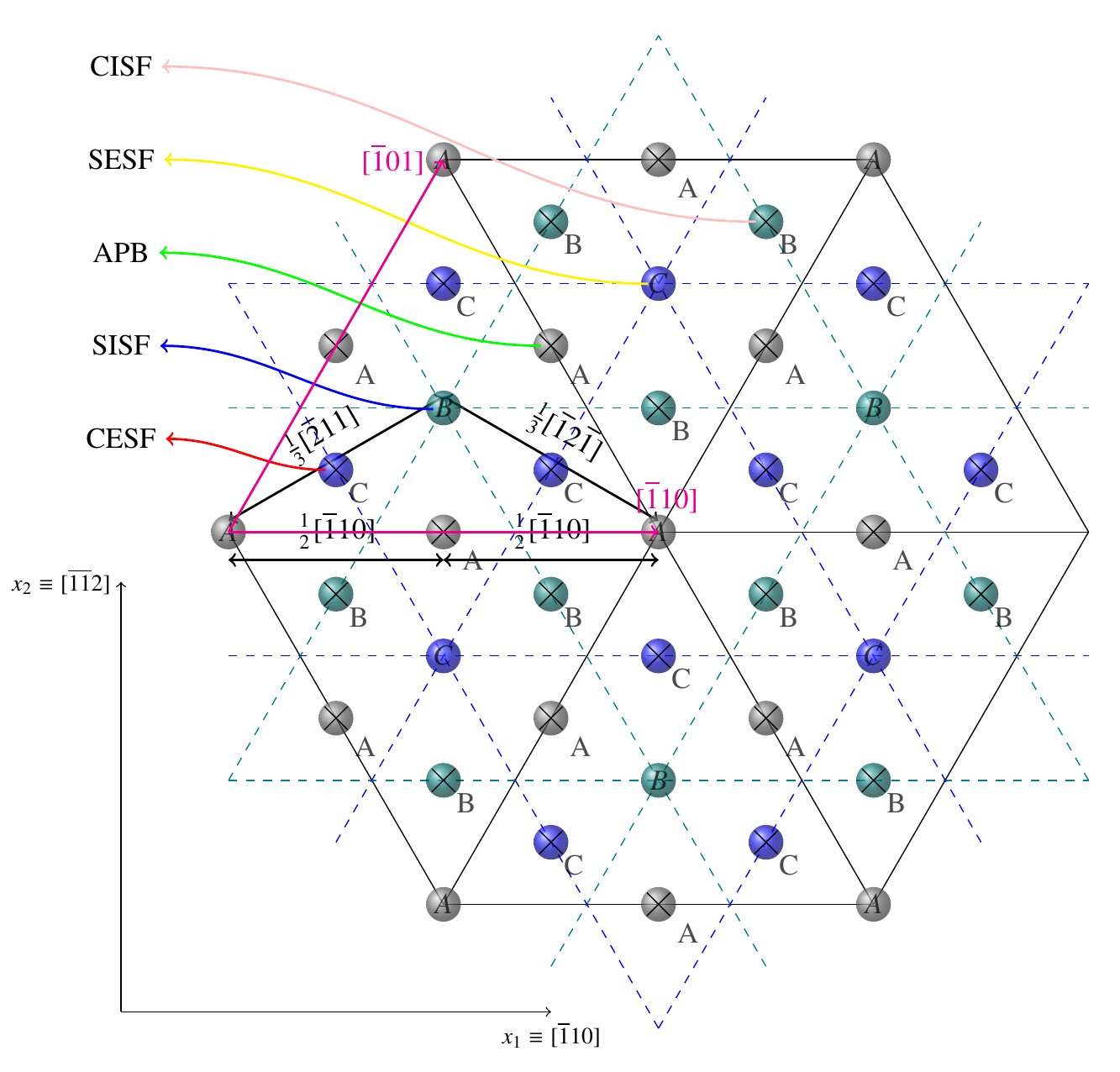}
  \caption{} 
  \label{fig:fcc_ni3al}
\end{subfigure}\hfill
\begin{subfigure}{.3\textwidth}
\centering
    \begin{tabular}[H]{|c|c|}
\hline
$\bm s_k$ &   $\gamma_k$  \\
\hline
$(0,0)$     &  $0$ \\
$(0.5,\frac{\sqrt{3}}{6})$     &  $\gamma_{CESF}$\\
$(1,\frac{\sqrt{3}}{3})$     &  $\gamma_{SISF}$\\
$(1.5,\frac{\sqrt{3}}{2})$     &  $\gamma_{APB}$\\
$(2,\frac{2 \sqrt{3}}{3})$     &  $\gamma_{SESF}$\\
$(2.5,\frac{5 \sqrt{3}}{6})$     &  $\gamma_{CISF}$\\
\hline
\end{tabular}
\caption{}
\label{fig:fcc_ni3al_conditions}
\end{subfigure}
\hfill
\begin{subfigure}{.2\textwidth}
  \centering
  \includegraphics[width=\linewidth]{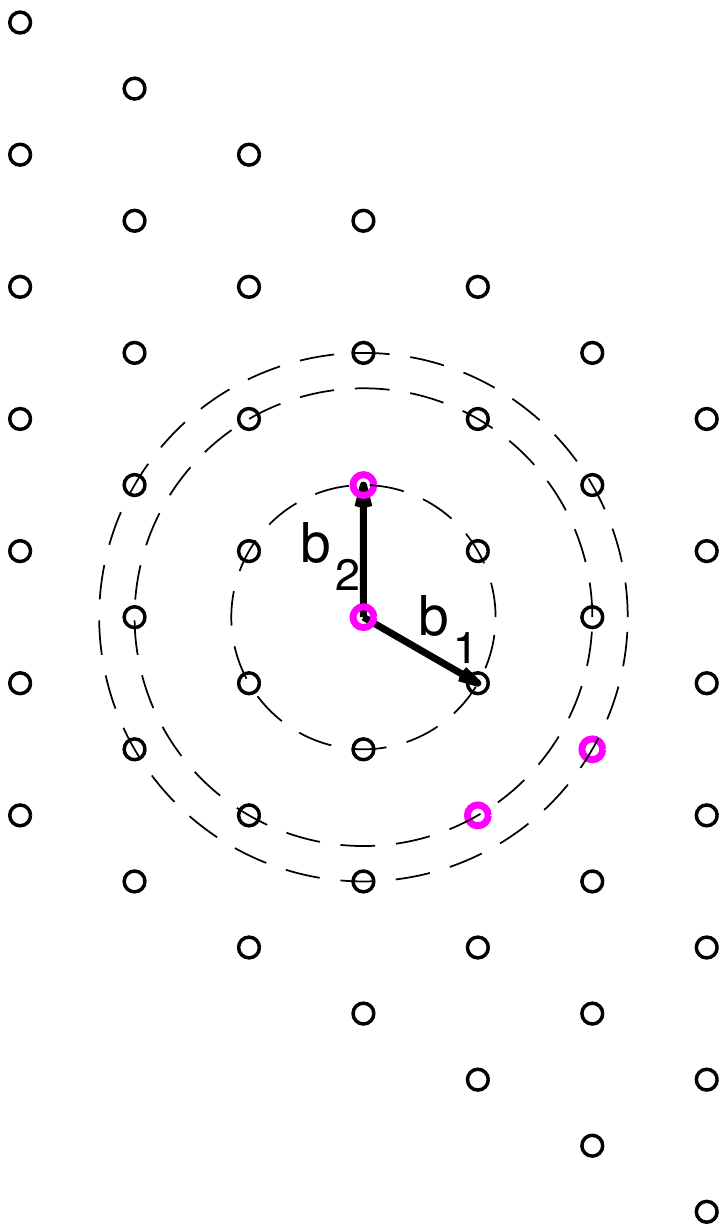}
  \caption{} 
  \label{fig:fcc_ni3al_kpts}
\end{subfigure}\\
\begin{subfigure}{.48\textwidth}
	\begin{tabular}[H]{|c|c|c|}
		\hline
		wave vector $\bm n$ & $S$ [mJ\, m$^{-2}$]& $C$ [mJ\, m$^{-2}$]\\
		\hline
		$(0,0)$ & 0& 58.61 \\
		$(0,1)$ & -3.608& 5.417 \\
		$(1,-1)$ &0&-49.17 \\
				$(2,0)$ &8.901&-14.86 \\
		\hline
	\end{tabular}
	\caption{}
	\label{fig:fcc_ni3al_SC}
\end{subfigure}
\hfill
\begin{subfigure}{.48\textwidth}
  \centering
  \includegraphics[width=\linewidth]{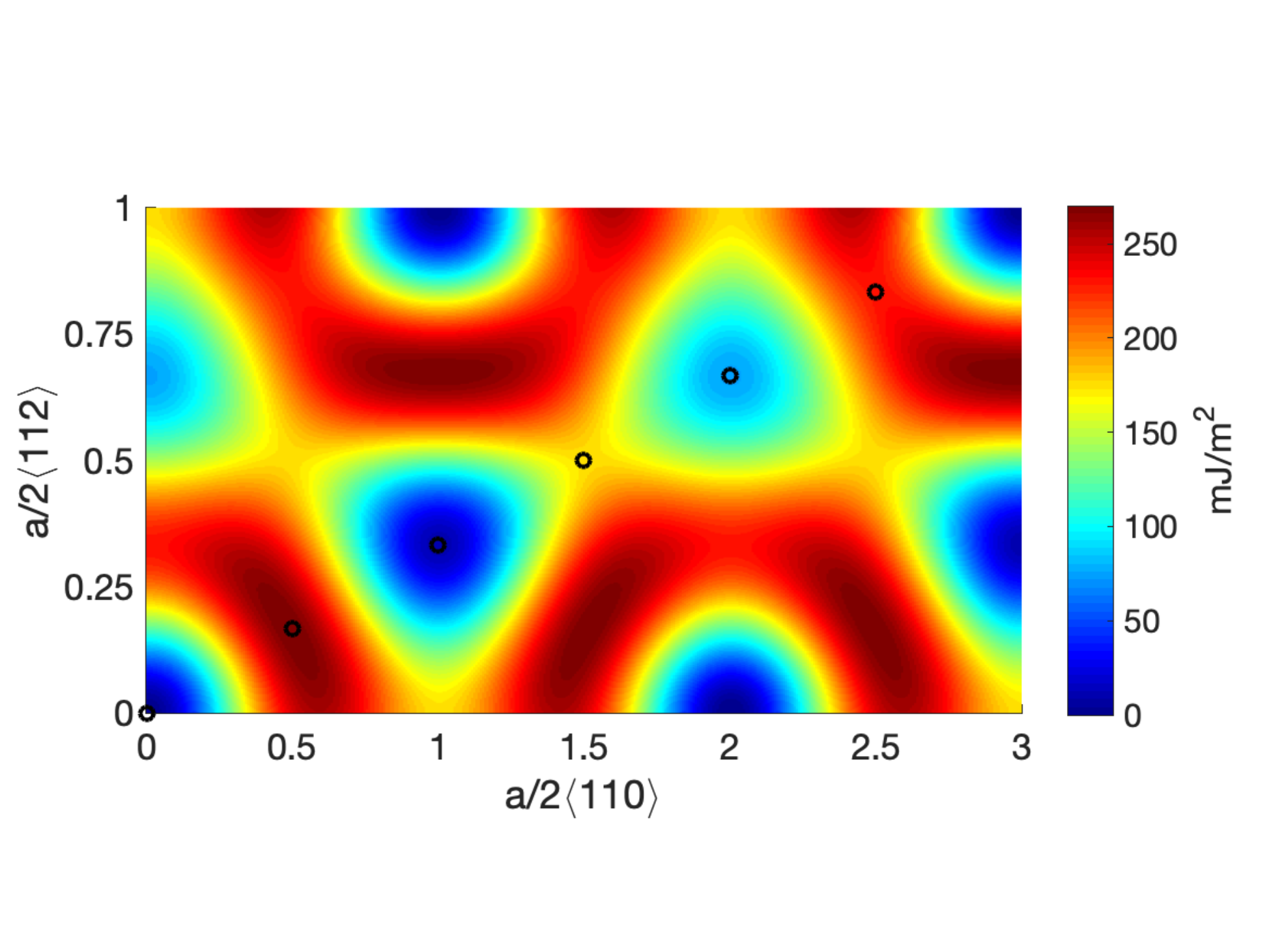}
  \caption{} 
  \label{fig:gammaSurf_gammaprime}
\end{subfigure}
\caption{Stacking fault energy for the $\gamma'$ phase. (\subref{fig:fcc_ni3al}-\subref{fig:fcc_ni3al_conditions}) Structure of the $\{111\}$ plane,
and imposed fitting conditions. The values $\gamma_{APB}=175 mJ/m^2$, $\gamma_{SISF}=10 mJ/m^2$, $\gamma_{CESF}=270 mJ/m^2$, $\gamma_{CISF}=230 mJ/m^2$ and $\gamma_{SESF}=75 mJ/m^2$ have been obtained from \cite{vorontsov_2012}, which ensures the accurate description of the entire $\gamma$ surface. (\subref{fig:fcc_ni3al_kpts}-\subref{fig:fcc_ni3al_SC}) Selected wave vectors (magenta circles) in the reciprocal lattice of the $\{111\}$ plane, and corresponding fitting constants. (\subref{fig:gammaSurf_gammaprime}) The resulting $\gamma$ surface.}
\end{figure}

For the $\{111\}$ plane in  the matrix fcc $\gamma$ phase,  we have
\begin{align}
\bm A_\gamma=b\left[\begin{array}{cc}
1&\frac{1}{2}\\
0&\frac{\sqrt{3}}{2}
\end{array}
\right]
&&\text{and}&&
\bm B_\gamma=\frac{2\pi}{b} \left[\begin{array}{cc}
1&0\\
-\frac{\sqrt{3}}{3}&\frac{2\sqrt{3}}{3}
\end{array}
\right]\, ,
\end{align}
where $b$ is the magnitude of the Burgers vector in the $\gamma$ phase. For the $\{111\}$ plane in  the precipitate  L$1_2$ $\gamma'$ phase we have 
\begin{align}
\bm A_{\gamma'}=2b\left[\begin{array}{cc}
1&\frac{1}{2}\\
0&\frac{\sqrt{3}}{2}
\end{array}
\right]
&&\text{and}&&
\bm B_{\gamma'}=\frac{\pi}{b} \left[\begin{array}{cc}
1&0\\
-\frac{\sqrt{3}}{3}&\frac{2\sqrt{3}}{3}
\end{array}
\right]\, ,
\end{align}
The three-fold rotation symmetry of the $\{111\}$ planes in both phases imply that the matrices $\bm M_j$'s are
\begin{align}
\bm M_0=\left[\begin{array}{cc}
1&0\\
0&1
\end{array}
\right]\, ,
&&
\bm M_1=\left[\begin{array}{cc}
-\frac{1}{2}&-\frac{\sqrt{3}}{2}\\
\frac{\sqrt{3}}{2}&-\frac{1}{2}
\end{array}
\right]
\, &&\text{and}
&&
\bm M_2=\left[\begin{array}{cc}
-\frac{1}{2}&\frac{\sqrt{3}}{2}\\
-\frac{\sqrt{3}}{2}&-\frac{1}{2}
\end{array}
\right]\, .
\end{align}
In each phase, the fitting constants  $S_i$ and $C_i$ are found by imposing a set of conditions 
\begin{align}\label{eq:gammaCond1} 
\gamma(\bm x_k)= \gamma_k 
\end{align}
where the pairs $(\bm s_k,\gamma_k)$ are obtained from ab-initio results from the literature. The specific conditions and choice of wave vectors for the  $\gamma$ and $\gamma'$ phases are detailed in Fig.~\ref{fig:fcc_ni} and  Fig.~\ref{fig:fcc_ni3al}, respectively.


\section{Critical bypass configuration in the looping regime}\label{app:comp_bks}

\begin{figure}[t!]
\centering
\begin{tikzpicture}
\node[anchor=south west,inner sep=0] (image) at (0,0) {\includegraphics[width=0.4\textwidth]{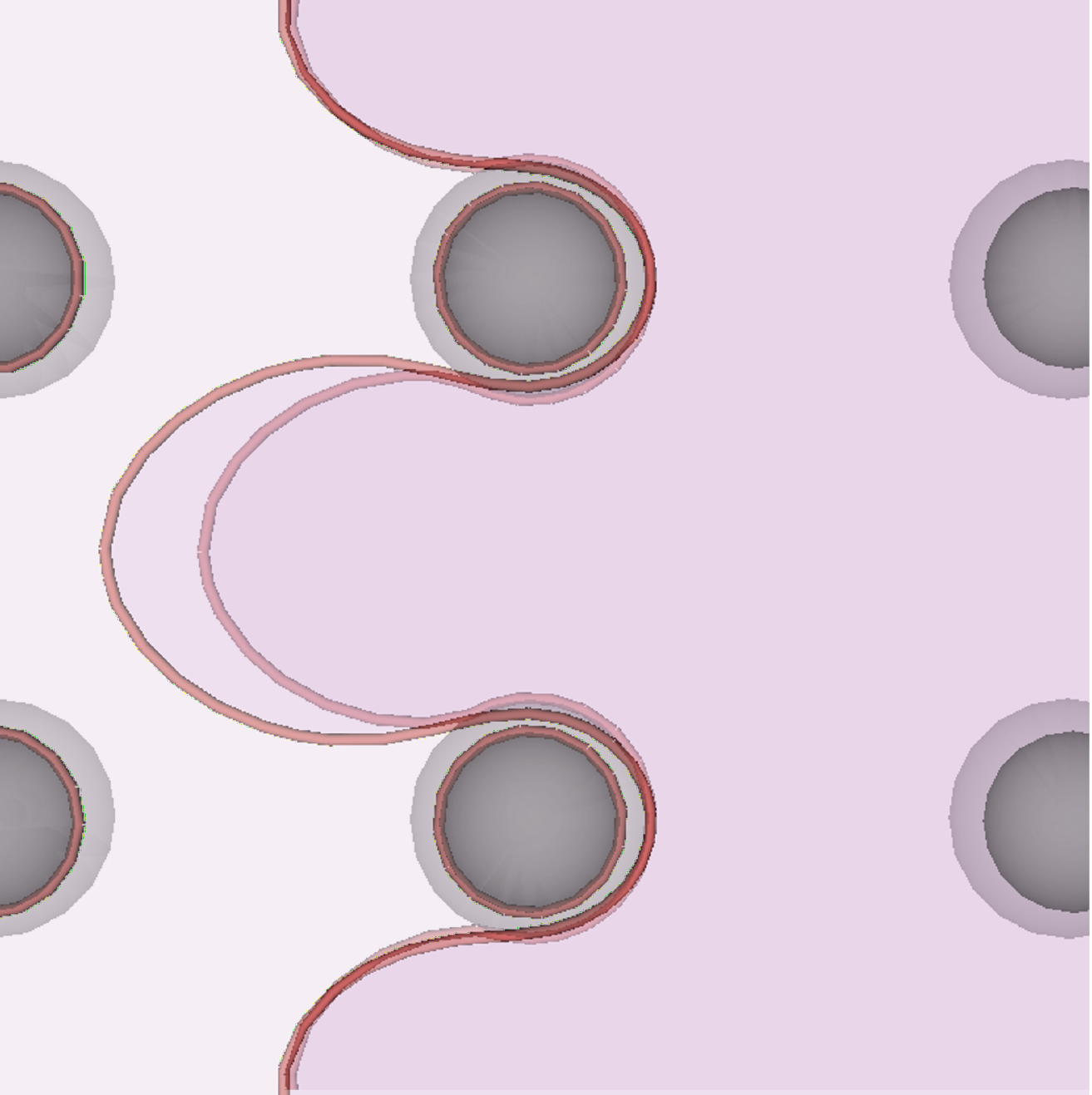}};
\begin{scope}[x={(image.south east)},y={(image.north west)}]
\draw [->,thick,black] (0.56,0.83) -- (0.68,0.88) 
node[right,black]{\small  $r=400 \, nm$}; 
\draw [->,thick,black] (0.55,0.7) -- (0.65,0.63) 
node[right,black]{\small  $r=300 \, nm$}; 
\draw [->,thick,black] (0.2,0.55) -- (0.3,0.55) 
node[black,right,xshift=-0.2cm]{\small  \begin{tabular}{l}
    $single$ \\
    $dislocation$ \\ 
    \end{tabular}}; 
\draw [->,thick,black] (0.15,0.4) -- (0.28,0.42) 
node[black,right]{\small $trailing ~ dislocation$}; 
\draw [->,thick,black] (0.57,0.27) -- (0.68,0.3) 
node[right,black,xshift=-0.2cm]{\small  \begin{tabular}{l}
    $leading$ \\
    $loop$ \\ 
    \end{tabular}}; 
\end{scope}
\end{tikzpicture}
\caption{Comparison of critical configurations in looping of a precipitate of $r=300 \, nm$ by a  dislocation pair and looping of a larger precipitate with $r_e=1.33\times300=400 \,nm$ by a single dislocation. The trailing dislocation experiences repulsion from the leading loop on the exit side. This repulsion forces the trailing dislocation to protrude further between the precipitates compared to the single dislocation encountering a precipitate of equivalent radius. This effect causes an increased bypass stress.
}
\label{fig:comp_bks}
\end{figure}

We discussed dislocation pair bypass by the looping mechanism in Section \ref{sec:result_dislocation_pair}. In Fig.~\ref{fig:looping}, we observed that the bypass stress in the DDD simulation is  larger than the BKS estimate (Eq.~\eqref{tauBKS}) by a factor $\alpha\approx2.2$, even if   a larger effective radius $r_e=1.33r$ is considered \cite{queyreau_Orowan}. Here, we provide a justification for this observation. In Fig.~\ref{fig:comp_bks}  we overlay two critical configurations extracted from the DDD simulations. 
In the foreground, the critical  configuration showing the trailing dislocation bypassing a precipitate of radius $r=300\, nm$ and surrounded by a loop left behind by the leading dislocation. More faded in the background, the critical configuration of a single dislocation bypassing a
larger precipitate of radius $r=1.33\times 300=400 \, nm$. 
It can be seen that these two critical configurations are not equivalent. In the case of the dislocation pair, the repulsive stress due to the leading loop on the back side of the precipitate hinders the closure of the trailing loop, and the trailing dislocation must protrude much deeper between the precipitates between closure can take place. This effect results in an increased bypass stress compered to the BKS model, even when an effective radius $r_e$ is considered. 


\section{Calculation of \texorpdfstring{$\ell(X)$}   {}}\label{app:lx} 
\begin{figure}[H]
\centering
\begin{minipage}{.5\linewidth}
  \centering
  \includegraphics[width=\linewidth]{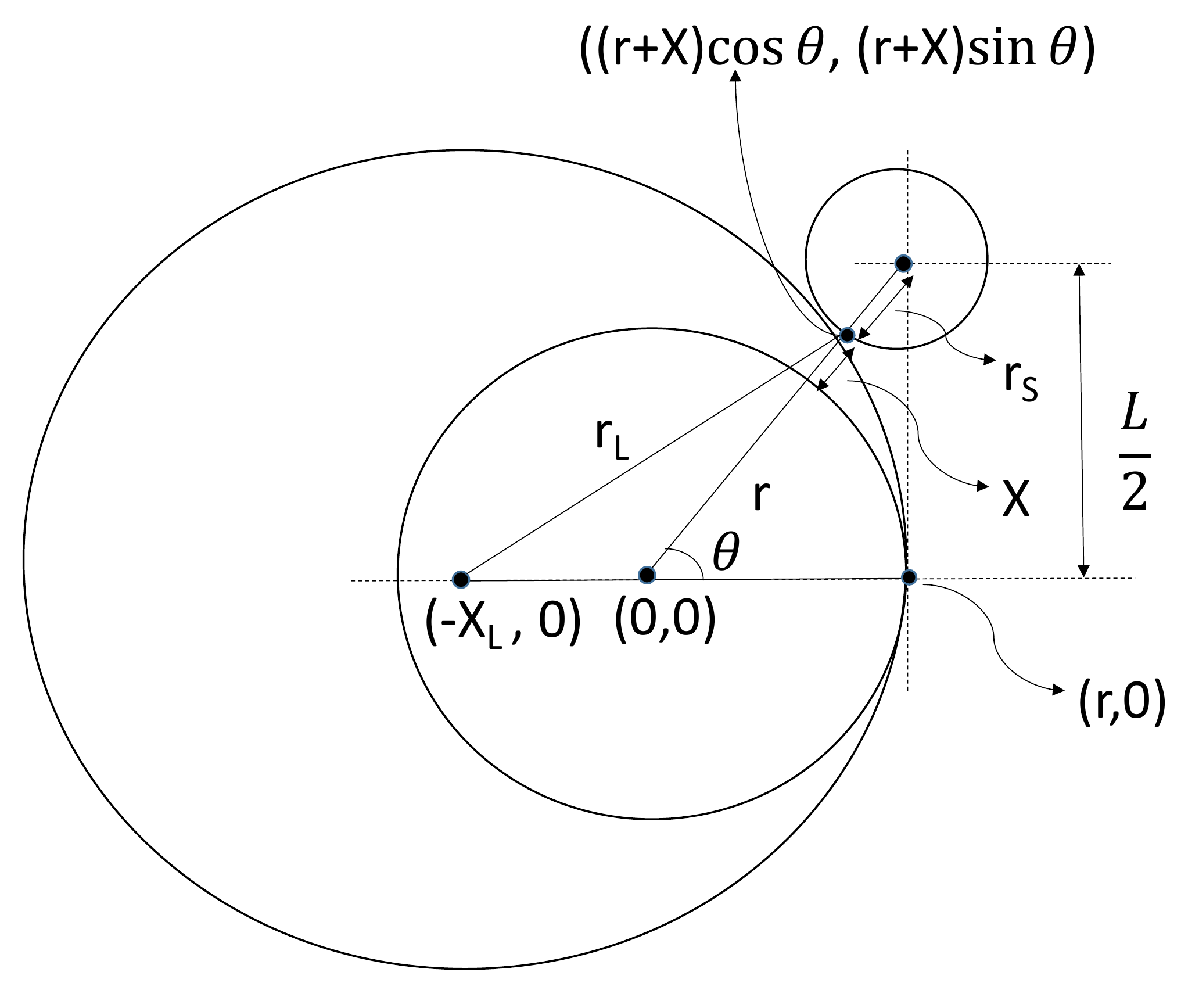}
  \caption{Calculation of $r_L(X)$}
  \label{fig:RL}
\end{minipage}%
\begin{minipage}{.4\linewidth}
  \centering
  \includegraphics[width=\linewidth]{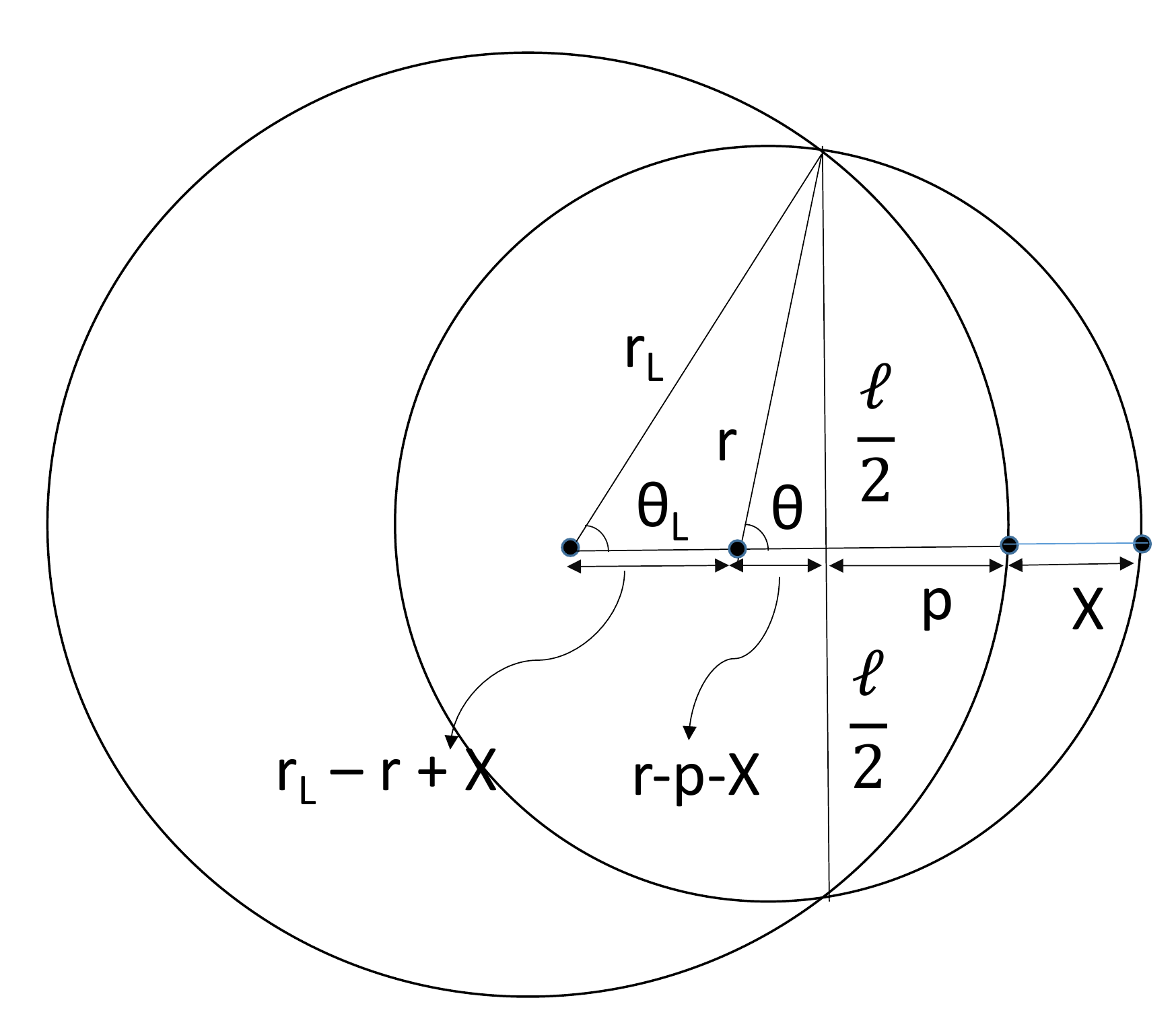}
  \caption{Estimation of $\ell(X)$.}
  \label{fig:Ax}
\end{minipage}
\end{figure}

The critical configuration for the hybrid looping and shearing mechanism was shown in Fig. \ref{fig:critical_config_loopShear}. The radius of curvature of the leading dislocation is $r_L$ such that $r_L > r$. 
Let us assume that the center of the precipitate be the origin. The leading dislocation is inside the precipitate by a distance $X$. Let us assume that the leading dislocation is shifted by $X$ to the right. Then it just touches the precipitate of radius $r$ as shown in Fig.~ \ref{fig:RL}. The circle, of which the leading dislocation is an arc, is assumed to be centered at $(-X_L,0)$ and is of radius $r_L$. Simple calculations lead to 
\begin{align}\label{eq:r_L}
r_L(X)= \frac{2rX + X^2}{2[r - (r+X)\cos\theta]} + r\, , && \text{where} && \cos\theta= \frac{r}{\sqrt{r^2 + {\frac{L^2} {4}}}}\, .
\end{align}

Now we shift the leading dislocation by $X$ to the left, back to its original position ( Fig.~\ref{fig:critical_config_loopShear}) as shown in Fig. \ref{fig:Ax}. Using simple geometric calculations, we obtain the length  $l(X)$ as 
\begin{align}\label{eq:l_x}
\ell(X)= 2 \sqrt{r^2 - (r-p-X)^2}\, , && \text{where} && p= \frac{2 r X - X^2}{2(r_L - r+X)}\, .
\end{align}

\section{Measurement of the bypass stress}\label{app:bypass_stress}

\begin{figure}[t!]
\centering
\begin{subfigure}{.4\textwidth}
   \centering
    \includegraphics[width=\linewidth]{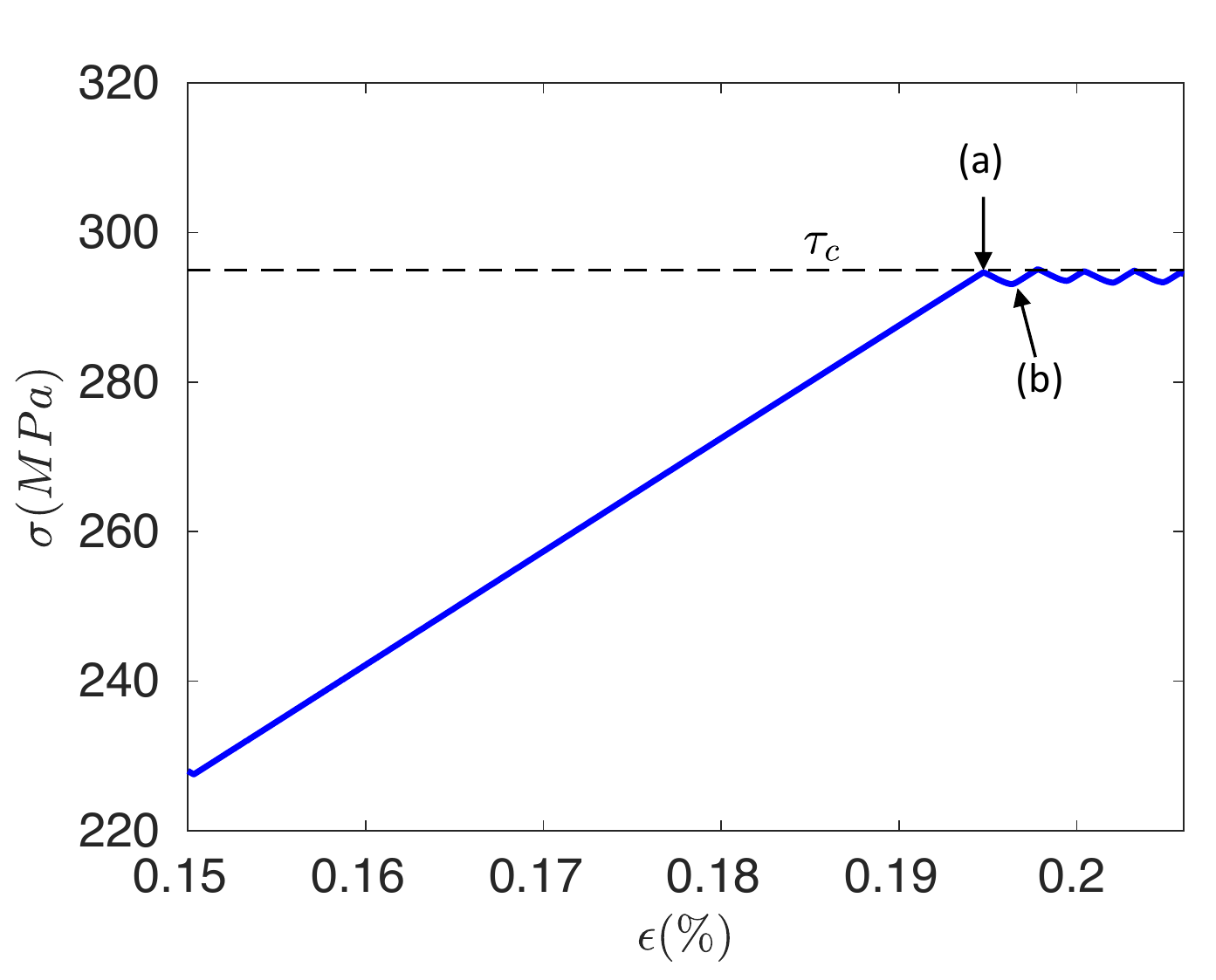}
\end{subfigure}
\hspace{0.5cm}
\begin{minipage}{0.22\textwidth}
\centering
\begin{subfigure}{\linewidth}
  \centering
    \includegraphics[width=0.9\linewidth]{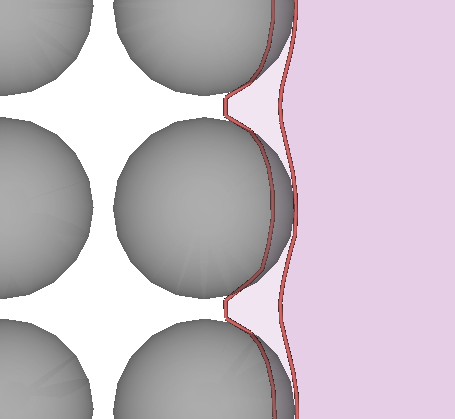}
     \caption{}
     \label{fig:before_cut}
\end{subfigure}

\vspace{0.5cm}

\begin{subfigure}{\linewidth}
  \centering
    \includegraphics[width=0.9\linewidth]{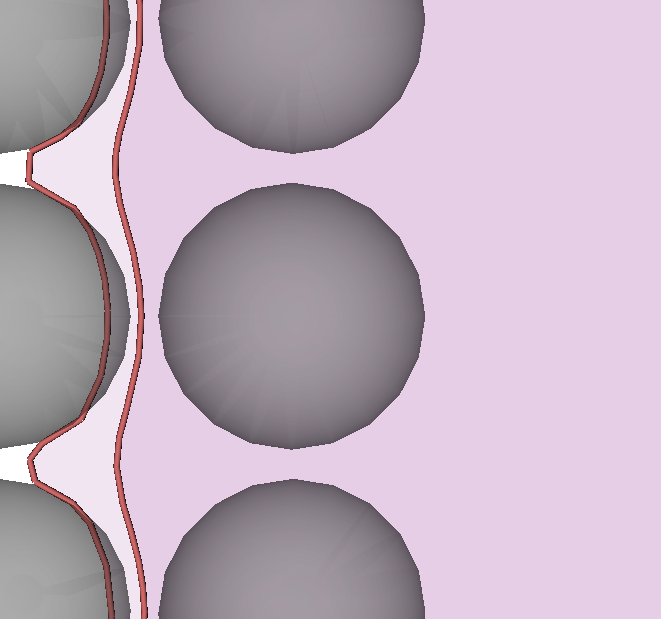}
     \caption{}
     \label{fig:after_cut}
\end{subfigure}
\end{minipage}
\caption{Stress-strain curve for $r=200 \mu m$ and $f=0.69$ under a constant applied strain rate. At point ({\subref{fig:before_cut}}),  a drop in stress is observed as the dislocation pair starts to shear the precipitate. This corresponds to the critical bypass stress (denoted by $\tau_c$). The drop continues till point ({\subref{fig:after_cut}}) at which the pair encounters  the next array of precipitates. ({\subref{fig:before_cut}}) The dislocation configuration corresponding to the critical bypass stress. ({\subref{fig:before_cut}}) The dislocation configuration corresponding to the state where the pair encounters  the next array of precipitates.}
\label{fig:bypass_stress}
\end{figure}

A typical stress-strain curve obtained from our simulations is shown in Fig.~{\ref{fig:bypass_stress}}. A dislocation pair is pinned at a precipitate until the stress state is sufficiently high for the pair  to bypass the precipitate (by shearing in this case). At constant applied strain rate, the bypass event corresponds to a drop in the stress (see point (a) in Fig.~{\ref{fig:bypass_stress}}). The value of stress at which this occurs  is the critical bypass stress which is recorded ($\tau_c$).

\section{Stress field of an isotropic spherical inclusion}
\label{EshelbyInclusion}
 The lattice misfit $\delta$ can be understood as the linear strain induced by the precipitate in the ``average" lattice. It corresponds to an eigenstrain or transformation-strain tensor
\begin{align}
    \bm\varepsilon^\star = \delta \bm I\, ,
\end{align}
where $\bm I$ is the identity tensor. In an isotropic material with shear modulus $\mu$ and Poisson's ratio $\nu$, the stress field induced by the inclusion can be written in terms of the auxiliary symmetric tensor
\begin{align}
\bm p =2\mu\left( \bm\varepsilon^\star+\frac{\nu}{1-2\nu}\text{tr}(\bm\varepsilon^\star)\bm I\right)\, .
\end{align}
The stress field  inside a spherical inclusion of radius $a$ is constant, and it reads \citep{mura2013micromechanics}
\begin{align}
\bm \sigma= 2 \mu \left[ \left(L + \frac{\nu}{1-2\nu}(3L+2M)\right) \, \text{tr}(\bm \varepsilon^\star) \bm I + 2 M \bm \varepsilon^\star \right] - \bm p\, , 
\end{align}
where $L=\frac{5\nu-1}{15(1-\nu)}$, and $M=\frac{4-5\nu}{15(1-\nu)}$. 
On the other hand, the stress field of a point $\bm x$ outside the inclusion is 
\begin{equation}    
\begin{split}
\bm \sigma(\bm x)= & \frac{a^3}{2\,(1-\nu)\,R^3} \Bigg[ \left(10(1-2\nu)+6 \frac{a^2}{R^2} \right)\frac{\bm p}{15} + \left(2\nu-2\frac{a^2}{R^2}\right)\frac{\left\{ (\bm p \bm r) \otimes \bm r + \bm r \otimes (\bm p \bm r) \right\}}{R^2} \\ 
  & + \left\{ \left(3\frac{a^2}{R^2}-5(1-2\nu) \right)\frac{\text{tr} (\bm p)}{15} + \left(1-2\nu-\frac{a^2}{R^2}\right)\frac{(\bm p \bm r) \cdot \bm r}{R^2} \right\} \bm I + \left\{ -\left(5-7\frac{a^2}{R^2}\right)\frac{(\bm p \bm r) \cdot \bm r}{R^4} + \left(1-\frac{a^2}{R^2}\right)\frac{\text{tr}(\bm p)}{R^2}  \right\} \bm r \otimes \bm r \Bigg],
 \end{split}
\end{equation}
where $\bm r=\bm x - \bm C$ where $\bm C$ is the center of the inclusion and $R=|\bm r|$.

\bibliographystyle{chicago}
\bibliography{testbibliog,biblio_GP}

\end{document}